\documentclass [preprint] {revtex4}
\usepackage[latin1]{inputenc}

\usepackage[dvips]{graphicx}
\usepackage{amssymb}

\def\dkkk{ D^+ \to K^- K^+ K^+}

\def\lp {\left( }
\def\rp {\right) }
\def\lb {\left[ }
\def\rb {\right] }
\def\lc {\left\{ }
\def\rc {\right\} }
\def\ra {\rangle }
\def\la {\langle }

\def\rar {\rightarrow}

\def\beq{\begin{equation}}
\def\eeq{\end{equation}}
\def\bea{\begin{eqnarray}}
\def\eea{\end{eqnarray}}
\def\nn {\nonumber}
\def\ni {\noindent}

\def\ct {\tilde{c}}

\def\Kb {\bar{K}}

\def\rtw {\sqrt{2}}
\def\rth {\sqrt{3}}
\def\rts {\sqrt{6}}
\def\sp {\!+\!}
\def\sm {\!-\!}

\def\mr2 {m_\rho^2 }

\def\cA {{\cal{A}}}
\def\cK {{\cal{K}}}

\def\cM {{\cal{M}}}

\def\cT {{\cal{T}}}

\def\a{\alpha}
\def\b{\beta}
\def\d{\delta}
\def\e{\epsilon}
\def\g{\gamma}

\def\D {\Delta}

\def\l{\lambda}
\def\m{\mu}
\def\n{\nu}

\def\O{\Omega}

\def\p {\pi}

\def\r{\rho}

\def\bQ {\mbox{\boldmath $Q$}}

\def\bI {\mbox{\boldmath $I $}}

\begin{document}


\title{Multibody decay anlyses - a new phenomenological model for meson-meson  subamplitudes}


\author{P.C. Magalh\~aes}
\affiliation{H.H. Wills Physics Laboratory, University of Bristol,  Bristol, BS8 1TLB, United Kingdom. }

\author{A.C. dos Reis}
\affiliation{Centro Brasileiro de Pesquisas F\'isicas, Rio de Janeiro, Brazil}

\author{M.R. Robilotta}
\affiliation{Instituto de F\'isica,  Universidade de S\~ao Paulo, S\~ao Paulo, Brazil}

\date{\today }

\begin{abstract}

Meson-meson amplitudes are important on their own and also play key roles in 
analyses of heavy-meson and tau decays.
In this work we propose a new phenomenological model suited to 
all $SU(3)$  mesonic two-body
final state interactions up to energies around 2 GeV.
It is aimed at replacing those entering the 
 old isobar model, produced in the 1960's,
long before the development of QCD.
The only similarity between our new proposal and amplitudes used in the isobar model
concern vector resonances in the elastic regime.
In  other situations, especially those involving scalar  resonances and coupled channels,
the isobar model is not compatible with post-QCD dynamics.
In order to support these claims convincingly and to motivate our approach, 
we consider applications to the 
$\pi \pi $ amplitude and compare our version with the isobar model in several 
different instances.
We also show that
the new model provides a clear indication of the 
mechanism responsible for the sharp rise observed in the $\p\p$ phase around
1\, GeV.
 The phenomenological amplitudes proposed here are suited to any number of
resonances in a given channel and rely just on masses and coupling constants as free parameters.
Concerning theory, they incorporate  chiral symmetry at low energies, include coupled channels and
respect unitarity whenever appropriate.

\end{abstract}

\pacs{...}

\maketitle

\section{motivation}

In the last decade, a considerable amount of precise data has been produced from BaBar, Belle, BES, LHCb  experiments on 
non-leptonic three-body decays of $D$ and $B$ mesons 
as well as on tau decays into pseudoscalars.
More comprehensive investigations can be done nowadays, 
using the very large and pure samples provided by the LHC experiments, 
and still more data is expected in the near future,  including neutral particles,  
with Belle II, BES III and LHCb (Run 2) experiments.

These decays involve two distinct sets of interactions. 
They begin with a primary vertex, in which the light $SU(3)$ quarks produced in the weak reaction 
disturb the surrounding QCD vacuum and give rise to an initial set of mesons.
This state then evolves by means of purely hadronic final state interactions (FSIs), 
whereby mesons rescatter many times before being detected. 
This rich hadronic final state structure is an important 
source of spectroscopic information about resonances and we recall
that the existence of the controversial  scalar states $f_0(500)$~\cite{sigma} 
 and  
$K^*_0$(700)~\cite{E791kappa} states was confirmed in three-body decays.
Final state interactions are also relevant  in the  study of  CP violation~\cite{CPviolation}. 

The analyses of non-leptonic three-body heavy-meson decays  is 
technically  involved and relies on models. 
The standard isobar model  (SIM) is by far the most popular choice 
amongst phenomenologists interested in resonance parameters.
It has been proposed in the early 1960s, long before the development of QCD, and 
fails to incorporate the new understanding of quark dynamics brought by the theory.
Its basic assumption is that a decay amplitude can be represented by a coherent sum of both 
non-resonant and  resonant contributions, with emphasis on the latter. The amplitude for the decay $ H(Q) \rar P_a(q_a) P_b(q_b) P_c(q_c) $, of a heavy meson $H$ 
into three pseudoscalars $P$ is denoted by $\cT$ and depends on the invariant masses
$m_{ab}^2=(q_a-q_b)^2$  and $m_{ac}^2=(q_a-q_c)^2$.
 \\
 What we define as standard isobar model assumes that $\cT$ can be written as:
\begin{eqnarray}
&& {\cT(m_{ab}^2, m_{ac}^2}) = c_{nr} \; \tau_{nr}(m_{ab}^2, m_{ac}^2) + 
\lb  \sum_k c_k \; \tau_k(m_{ab}^2)  + \sum_j c_j \; \tau_j(m_{ac}^2)  \rb \;,
\label{mot.1}
\end{eqnarray}
\ni where $k$ and $j$ are  resonances label that can be the same for a symmetric decay.
The first term in Eq.~(\ref{mot.1}) is non-resonant and that within square brackets implements the quasi-two-body, or ($2+1$),
approximation, in which only the interactions of a pair of particles matters and the third one, 
the bachelor, is just a spectator.  
 The $\tau_k(s)$ functions, for $s=m_{ab}^2, m_{ac}^2$,  represent dynamic two-body amplitudes 
and the complex coefficients $c_k=e^{i\theta_k}$
are fitting parameters. 
In the want of a theory,  the first term is usually taken to be $\tau_{nr}=1$.
For each resonance considered, one uses
$ \tau_k = [\mathrm{FF}] \times [\, \mathrm{angular} \; \mathrm{factor} \,] 
\times [\mathrm{line} \; \mathrm{shape} ]_k $,
where $[\mathrm{FF}]$ stands for form factors, 
[angular factor] is associated with spin
and $ [\mathrm{line} \; \mathrm{shape} ]_k$ represents a Breit-Wigner function 
depending on a mass $m_k$  and a width $\Gamma_k$, given by
\bea
 [\mathrm{line} \; \mathrm{shape} ]_k \;\; \rar \;\; 
 [\mathrm{BW}]_k = \frac{1}{[s- m_k^2 + i\,m_k\, \Gamma_k]}.
\label{mot.2}
\eea
For some states, variations such as the Flatt\'e or Gounaris-Sakurai are used. 
In applications,  both the qualities and quantities of resonances employed are regulated ad hoc and
the outcome of isobar model analyses are values for masses, widths, 
fit fractions and, sometimes, mixing couplings.
 Fit fractions, in particular, are associated with 
the complex parameters $c_{nr}$ and $c_k$,
which are neither directly related to an underlying dynamics nor allow the identification of
substructures.  
Important limitations of the isobar model are presented below.
\\[2mm]
{\bf 1.}  
Even if one overlooks  the
 problem of ascribing physical meanings to parameters extracted 
from the isobar model, there is another issue at stake.
Strictly speaking, their numerical values depend on the particular 
 assumptions underlying the use made of Eq.~(\ref{mot.1}), namely 
 the non-resonant term and the number and isospins of resonances employed.
 Therefore the numerical meaning
 of the parameters extracted remains 
 always attached to the specific reaction employed to derive them.
Final state interactions incorporated into the decay amplitude $\cT$ include both proper 
three-body interactions  and a wide range of elastic and inelastic two-body subamplitudes $\cA$
involving resonances and coupled channels, as we review in Sec.\ref{schem}.
In a given decay, the main information about resonances appears codified 	
in the $\cA$s and,
even if there are exceptions, it is important to distinguish them from $\cT$.
A  conspicuous difference between these amplitudes is that the latter 
includes  weak vertices and the former does not, but
this is sometimes bypassed  in the literature.
For instance, there is no justification for the assumption that the $\cA$s  are either identical or 
proportional to $\cT$,  as found in a partial-wave analysis of the $S$-wave $K^- \pi^+$ 
amplitude from the decay  $D^+ \rar K^- \pi^+ \pi^+$ produced some time ago~\cite{E791kappa}.
As a matter of fact, the empirical phase is different from that 
 produced by LASS for $K\pi$ scattering data~\cite{LASS}.
As expected, this discrepancy arose because one was comparing different objects and 
was later explained
by considering meson loops in the weak sector of $\cT$~\cite{BR,PatWV, kubisDKpp, satoshi}. 
\\[2mm] 
{\bf 2.} The extraction of information from the isobar model is hampered by 
the  presence of non-resonant terms.
An important message brought to hadron physics by QCD is that, provided  enough
energy is available,
the light quark condensate does show up and several pseudoscalars can be produced 
in a single vertex.
For instance,  the process $e^-\,e^+ \rar 4\,\p$ involves the multi-meson
matrix element  $\la \p\p\p\p|J_\g^\m|0\ra$, $J_\g^\m$ for the
electromagnetic current~\cite{EU}.
A similar matrix element, with the weak current $(V\sm A)^\m$, 
describes the decay $\tau \rar \n \, 4\p$~\cite{EU}.
In a recent work, we studied~\cite{PatDKKK} the doubly Cabibbo-suppressed 
decay $\dkkk $ departing from a non-resonant term based on the axial current matrix element 
$\la K^- K^+ K^+ |A^\m|\,0\,\ra$, describing the annihilation of the $D^+$
into a $W^+$ which subsequently hadronizes. 
  In that case 
 non-resonant terms and those involving resonances are 
entangled by a kind of diagramatic continuity.
\\[2mm]
{\bf 3.} In principle, the functions $\tau_k(s)$  in Eq.~(\ref{mot.1})  do
contain information about two-body interactions, but 
extracting it is difficult, for isospin channels are not clearly identified.
Scattering amplitudes $\cA$ depend 
on both the angular momentum $J$ and the
isospin $I$ of the channel considered, whereas just a $J$ dependence 
can be extracted from an empirical decay amplitude $\tau_k$.
Therefore, an attempt to extract $\cA^{(J,I)}$ from $\tau_k^{(J)}$ 
would amount to an artificial generation of physical content 
 from the reaction considered.
\\[2mm]
{\bf 4.}  For processes requiring several  resonances with the same quantum numbers,
SIM  amplitudes given by sums $\Sigma\, c_k\, \tau_k$ violate unitarity,
a criticism raised by many authors~\cite{unit, KpiUni, Ropertz2018pipi}. 
At present, there are solid conceptual techniques aimed at preserving unitarity 
in amplitudes involving several resonances~\cite{OOunit}, as  discussed in Sect.\ref{schem}.
Thus, nowadays,
the use of problematic guess functions based on 
sums of individual line shapes given by Eq.~(\ref{mot.2}) 
is difficult to be justified. 
\\[2mm]
{\bf 5.}  Meson-meson isoscalar amplitudes $\cA$ include important inelasticities  
due to couplings of intermediate states.
For instance, in $\p\p$ scattering the $K\Kb$ inelastic channel~\cite{Hyams} 
 opens  at $E\simeq 1\,$GeV.
So, this energy represents the upper bound for the validity of Eq.~(\ref{mot.1}), 
since there is no room in the BW-like representation of functions $\tau_k$, Eq.~(\ref{mot.2}), 
 for the  incorporation of  coupled channels.
In general, guess functions better suited for accommodating data should have structures similar
to those used in meson-meson scattering Ref.~\cite{Hyams, Bachir, Pelaez2019}.
 In the SIM, guess functions usually employed are not suited to
 accommodate coupled channels.
The role of resonances above inelastic thresholds is discussed in  Sect.\ref{resCOUP}.

All the problems of the standard isobar model mentioned above  
tend to corrode the physical meaning  of parameters  it yields from fits.
Since it was proposed,  more than half a century ago, many of the limitations pointed 
above were understood and tamed, 
especially owing to the formulation of  QCD. 
As a consequence, nowadays, serious flaws of the model are already rather clear, 
such as: it violates unitarity, it does not incorporate isospin and,
especially important, it is totally unsuited for dealing with coupled channels.
In the $SU(3)$ sector, scattering amplitudes for pions, kaons and etas are strongly
coupled and cannot be represented as sums of individual contributions. 
At present, as one knows,  QCD cannot be directly applied to heavy meson decays,
but their effective counterparts can.
Effective lagrangians rely just on  hadron masses and coupling constants,  
ensuring that the physical meaning of parameters is preserved 
from process to process.
Thus, guess functions for fitting heavy-meson decay data departing from lagrangians
deal with the same free parameters as employed in scattering amplitudes.
This makes the mutual comparison of their values meaningful.

This work is part of a program aimed at constructing guess functions for heavy-meson decays 
departing from effective lagrangians.
Here,  we concentrate on the two-body  scattering amplitudes 
 $\cA$, which are
directly  associated with observed quantities and also 
 important substructures
of decay amplitudes.
We depart  from a previous work on $\dkkk$ where a three-body amplitude 
was constructed based on  effective lagrangians with chiral symmetry and contained
unitarized scattering subamplitudes~\cite{PatDKKK}.
Although fits to Dalitz plots data were better than those based on 
the standard isobar model \cite{DKKK},
that work was performed in the $K$-matrix approximation. 
We draw attention to the fact that this $K$-matrix approximation 
is not the same thing as the K-Matrix approach \cite{Anisovich} used in some amplitude analyses.
Here, we propose a model which allows one to go beyond this approximation 
and discuss its implications.

Our presentation is organized as follows:
In  Sect.\ref{schem} we review how  heavy-meson decay amplitudes 
are related to weak vertices, scattering amplitudes and form factors.
This is intended to provide a broad conceptual framework for 
criticisms of the isobar model.
The full scattering amplitudes for the $SU(3)$ pseudoscalars
in the coupled channel formalism are presented in App.\ref{coupled}, combining 
interaction kernels and two-meson propagators given in Apps.\ref{omega} and \ref{kernel}.
In Sect.\ref{scatt} we present the full scattering amplitudes
 and specialize to 
the $\p\p$ amplitude, which is used as a standard 
for assessing the limitations 
of the isobar model.
In Sect.\ref{imUNC} we discuss  those limitations regarding
post-QCD physics and unitarity. 
In Sect.\ref{resCOUP} we discuss the impact of coupled channels into the problem 
and show that the meaning of a resonance as an independent contribution is lost 
in the inelastic region, supporting our claim that BW line shapes should not
be used above $1$ GeV.
We also compare coupled and uncoupled amplitudes and show that the impact of 
coupling is huge.
 In Sect.\ref{model} we present 
  our
  model for the real part of two-meson propagators which allows one to go beyond
 the $K$-matrix approximation.
 In Sect.\ref{extraR} we add  an extra resonance
 to each scalar channel using
 the methodology we developed and show the potentiality 
  of our model for extensions 
 to  higher energies.
Finally, in Sect.\ref{summ} we summarize our conclusions.

\section{schematic dynamics} 
\label{schem}

The theoretical description of a heavy-meson $H$ decay into three light pseudoscalars
$P_a\,P_b\,P_c$  
involves several classes of entangled problems and is necessarily rather complicated.
Below, we use simple topological arguments, based on hadronic degrees of freedom,
to classify these problems.
We rely on building blocks determined by {\em proper hadronic interactions}, 
defined as those associated with  diagrams 
that cannot be separated into two pieces by cutting  hadron
 lines only.
As one is dealing 
with with weak and strong interactions simultaneously,
it is convenient to isolate as much as possible these two sectors.

\begin{figure}[h] 
\includegraphics[width=.7\columnwidth,angle=0]{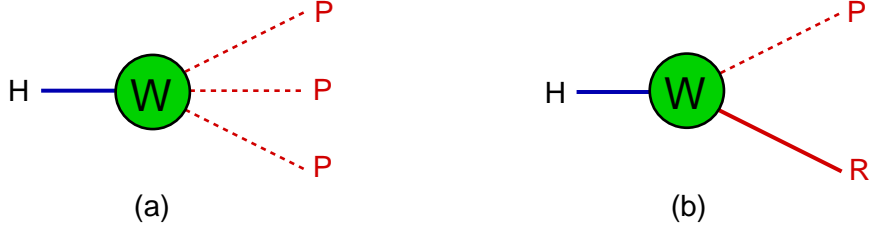}
\caption{Contributions to the primary weak vertex:
(a) $H \rar P_a\,P_b\,P_c$; (b) $H\rar P_a\,R_x$.}
\label{F-1}
\end{figure}

The basic weak interactions producing the decay of a heavy meson 
involve quarks in the QCD vacuum and were classified by  Chau~\cite{Chau}.
At the hadronic level,  the primary weak vertex 
 contains two kinds of 
proper Feynman diagrams, shown in Fig.~\ref{F-1},
describing the  processes $H\rar P_a\,P_b\,P_c$
and $H\rar P_a\, R_x$, where $R_x$ is a light resonance
which later decays as $R_x \rar P_b \, P_c$.
At this stage, this resonance is described by a bare pole
and does not have a width yet.
The green blob does not include hadronic degrees of freedom, but can
contain strong processes in the form of quarks and gluon exchanges.
In the literature the primary vertex is described by means of either 
factorization techniques~\cite{FacTec} or effective lagrangians~\cite{EffLag}.  
 
The mesons produced in diagram \ref{F-1}(a) can go directly to the detector 
 and give rise to a
 {\em non-resonant} contribution. 
Alternatively, 
it is possible that the hadrons produced in diagrams (a) and (b) 
have various forms of strong interactions before reaching the detector.
In this case, one talks about {\em final state interactions} (FSIs),
which are necessarily strong.

\begin{figure}[h] 
\includegraphics[width=1\columnwidth,angle=0]{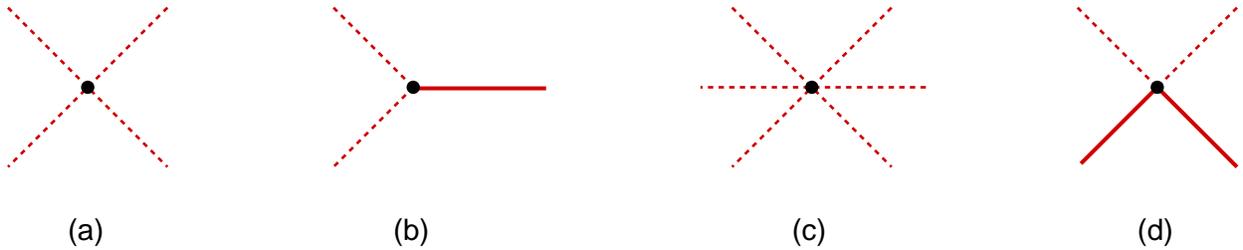}
\caption{Building blocks in the strong sector:
(a)  LO four-meson contact term; (b) NLO two-meson-resonance coupling;
(c) six-meson contact term; (d) two-meson-two-resonance coupling.}
\label{F-2}
\end{figure}

Nowadays, most approaches tend to organize the FSIs departing from 
chiral perturbation theory (ChPT).
Although Lattice QCD is improving \cite{LatticePiPi}, ChPT  still is
 the best available effective representation 
of QCD at low energies~\cite{WChPT,GL84,GL85} and can accomodate resonances~\cite{EGPR}.
As resonances correspond to nonperturbative states, predictions from ChPT are precise up 
to energies below the  $\rho(770)$ mass.
Beyond that point, one has to resort to extensions of ChPT, which may be performed
 by means of either
dynamical  models~\cite{PatDKKK, Bachir, Anisovich, DynMod, Ropertz2018} 
or dispersion relations~\cite{Pelaez2019,DisRel}. 
Here, we describe the basics of the former approach, which we find more suited to 
phenomenological studies of problems involving several resonances.
The idea is to define a few basic building blocks, 
as displayed in Fig.~\ref{F-2},
and to construct all relevant interactions departing from them.  
Diagram (a) represents a four-meson contact interaction, predicted by ChPT  to be 
the single leading contribution at low-energies and
 corresponds to an amplitude  given by a second order polynomial 
in momenta and meson masses.
Process (c) is a higher order term, describing a proper six-meson vertex.
Resonances are also included in the chiral formalism~\cite{EGPR} and diagrams 
(b) and (d) are associated with their decay and scattering amplitudes. 
 To our knowledge, diagram (c) has not yet been included into 
realistic calculations of heavy-meson decays,
whereas interactions described by diagram (d)  were  considered in a 
phenomenological description of 
the process $\sigma\sigma(\r\r)\to4\pi$ contributing to   $\pi\pi$ scattering~\cite{Ropertz2018}.

\begin{figure}[h] 
\includegraphics[width=1\columnwidth,angle=0]{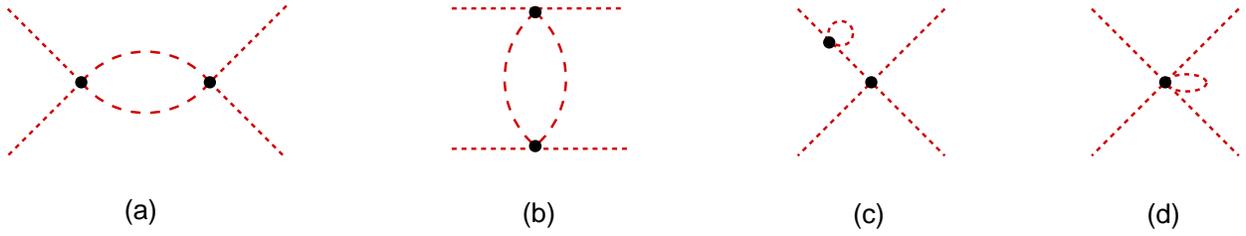}
\caption{One-loop corrections to the contact four-meson vertex:
(a) $s$-channel; (b) $t$- and $u$-channels;
(c) mass term; (d) vertex term.}
\label{F-3}
\end{figure}

The diagrams of Fig.~\ref{F-2} resemble interaction potentials $V$ in quantum mechanics 
and, to determine the full solution of a problem, one has to solve a dynamical
equation analogous to that of Lippmann-Schwinger.
This is not feasible in field theory and one has to resort to a piecemeal 
evaluation of perturbative corrections.
The procedure is similar to that used in quantum mechanics,
where full and free solutions are related by a series of the form
$1 + g\,V+ g\,V\,g\,V + \cdots$, $g$ being the free  propagator. 
In the present problem, one deals with relativistic propagators involving mesonic states,
denoted by $\O$.
In order to illustrate this procedure, in Fig.~
\ref{F-3} we show some perturbative corrections
involving a single {\em loop} to the four-meson contact term of Fig.~\ref{F-2}(a).
Diagrams (a) and (b) involve propagation between different points 
whereas (c) and (d) are local  and are
 incorporated into  actual values 
of masses and coupling constants.
Our main concern are diagrams (a) and (b).

A particularly important point in this constructive approach 
is that the $s$-channel contribution of  process (a) is complex
and one writes $\O(s)= \O^R(s) + i\, \O^I(s)$, where $\O^R$ and $\O^I$ are the real and imaginary parts.
The function $\O^I(s)$ is well behaved and underlies  imaginary contributions to the FSIs,
including resonance widths.
In field theory, this
 kind of imaginary components in some classes of propagators 
is of fundamental importance, for it is associated with unitarity.
A far reaching consequence is that 
 a reliable amplitudes must have a well defined balance 
between real and imaginary parts.
If this is not the case, they fail to  conserve probability, 
as in some instances of the isobar model. 
Concerning the real terms $\O^R$, explicit calculations show that they  
contain infinite contributions $\Lambda_\infty$.
Thus, formally, one has $\O^R = \O^R + \Lambda_\infty$, where  $\O^R$ is a 
 known regular function.
The elimination of $\Lambda_\infty$ requires renormalization, 
bringing unknown real constants into the problem. 
The model presented in this work regards 
$\O^R$, the real part of the two-meson propagator.

\begin{figure}[h] 
\includegraphics[width=1\columnwidth,angle=0]{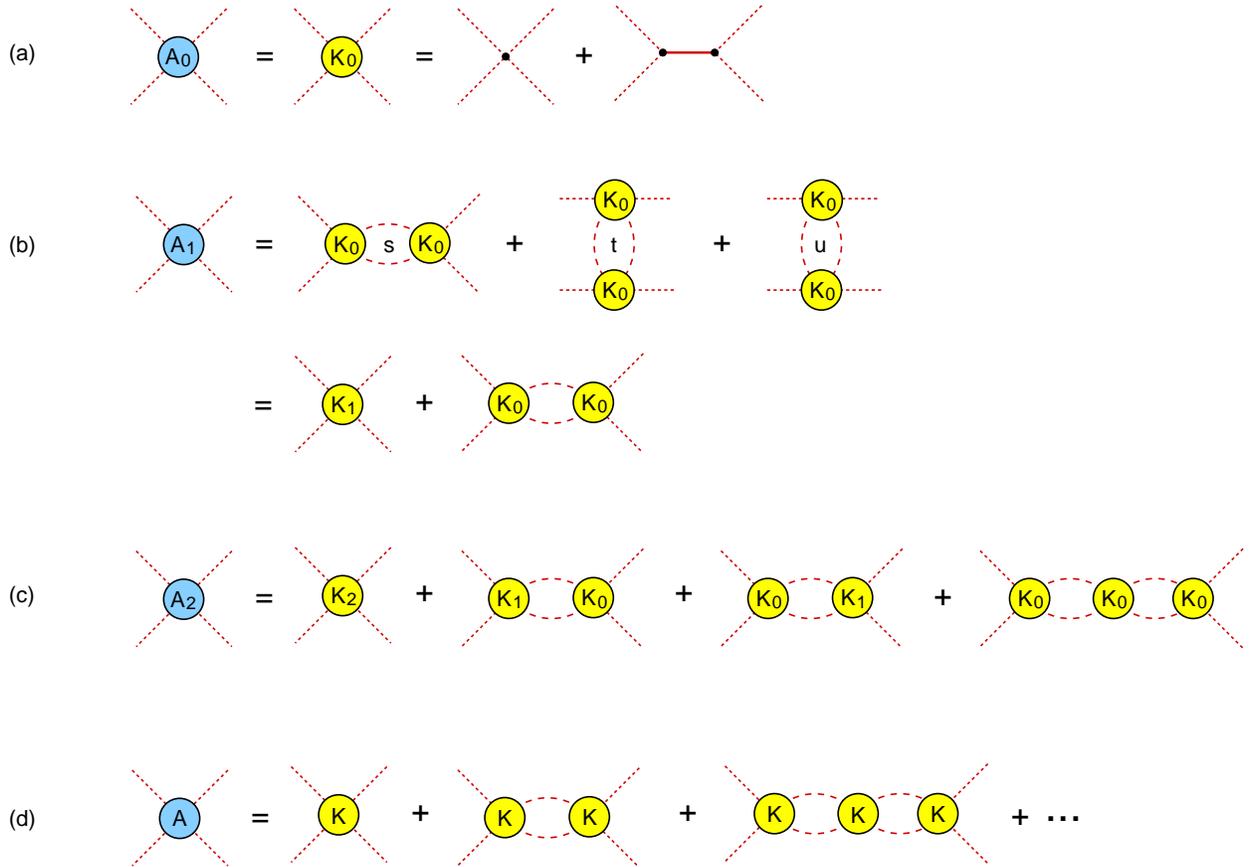}
\caption{Scattering amplitudes $\cA$ and kernels $\cK$:
(a) tree level;  (b) first perturbative correction; (c) second perturbative correction;
(d) full amplitude.}
\label{F-4}
\end{figure}

The study of FSIs in heavy-meson decays relies on non-perturbative 
amplitudes and their derivation 
requires the summation of infinite series of perturbative contributions. 
We exemplify this procedure 
in the case of a unitary meson-meson scattering amplitude,
denoting the full result by  $\cA$ and partial contributions with $n$ loops by $\cA_n$. 
We begin by defining a kernel $\cK_n$, as the part of $\cA_n$ that cannot be separated into two 
pieces by cutting {\em $s$-channel} two-meson loops only.  
The first kernel is $\cK_0$, associated with the tree processes displayed in Fig.~\ref{F-4} (a), and
it is a real function because, at this point we are still dealing with a bare resonance,
described by a pole at its mass. 
The tree amplitude is then given by  $\cA_0=\cK_0$.  

 The single-loop correction is shown in Fig.~\ref{F-4}(b) and involves three terms, in 
 $s$, $t$ and $u$ channels.
 The first one involves a two-meson $s$-channel propagator, whereas the last 
 two do not and are grouped into a new kernel $\cK_1$.
 The case of two loops is shown in Fig.~\ref{F-4} (c), where $\cK_2$ is a higher order kernel
 and the $s$-channel is represented by three successive $\cK_0$ interactions.
 Repeating this indefinitely and adding the results, we obtain a scattering amplitude of the form
\bea 
&& \cA =  \cK \times \lb 1 + (\mathrm{loop} \times \cK) + (\mathrm{loop}\times \cK)^2 
+ (\mathrm{loop}\times \cK)^3 + \cdots \rb \,,
\label{dyn.1}\\[2mm]
&& \mathrm{loop} = \O^R + i\, \O^I \;,
\label{dyn.2}\\[2mm]
&& \cK= \cK_0 + \cK_1 + \cK_ 2 + \cdots \;.
\label{dyn.3}
\eea
The geometric series in Eq.~(\ref{dyn.1}) can be summed and one has
\bea
&& \cA = \frac{\cK}{D}\;,
\label{dyn.4}\\[4mm]
&& D = 1- (\mathrm{loop} \times \cK)\;.
\label{dyn.5}
\eea
As discussed in the sequence, $1/D$  is the post-QCD version of the BW line shape,
Eq.~(\ref{mot.2}).

A very important feature of this result is that the amplitude $\cA$ is unitary, provided $\cK$ is real.
This property is quite general and derives from the structure of the denominator $D$,
which is suitably complex owing to the well defined imaginary function 
$\O^I$ in Eq.~(\ref{dyn.2}).
The forms adopted for both $\O^R$ and $\cK$
 are irrelevant for this 
  property of $\cA$, as discussed in Sect.\ref{imUNC}. 
This justifies the widespread use of the $K$-matrix approximation, which is implemented 
by neglecting $\O^R$ and writing 
\bea
K\mathrm{-matrix} \;\;\rar \;\;  \mathrm{loop} = 0 +  i\, \O_I \;.
\label{dyn.6}
\eea

\begin{figure}[h] 
\includegraphics[width=1\columnwidth,angle=0]{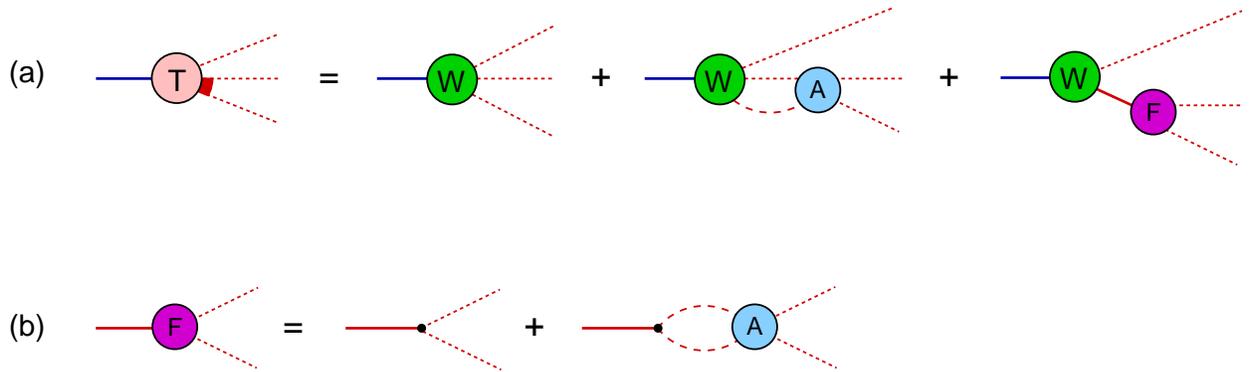}
\caption{(a) Decay amplitude in the $2+1$ approximation;
(b) form factor.}
\label{F-5}
\end{figure}

The amplitudes $\cA$ are key elements
 in the description of heavy-meson decays, for 
they are present in the  FSIs which supplement the weak process of Fig.~\ref{F-1}.
Strong interactions involving three bodies can be very complicated.
The simplest class of FSIs corresponds to the $(2+1)$ approximation,  represented in Fig.~\ref{F-5}, 
in which the first diagram in (a) represents the non-resonant contribution and the other two include 
particle interactions 
in the presence of a final meson acting as a spectator.
Structure (a) represents the heavy meson decay amplitude in the $(2+1)$ approximation 
and the blob indicated by $F$ is usually called {\em form factor},
which many authors take as the single contribution to the decay~\cite{FacTec}. 
It is isolated in Fig.~\ref{F-5}(b) and, denoting by $g$ the resonance-pseudoscalar coupling constant,  
the function $F$ can be related to the meson-meson scattering amplitude by
\bea
F = g\, \lb 1 +  (\mathrm{loop} \times A) \rb = g\; \frac{1}{D}\;,
\label{dyn.7}
\eea
where $D$ is the denominator given in (\ref{dyn.5}).
The imaginary part of $D$ gives rise to a finite width to the resonance.

\begin{figure}[h] 
\includegraphics[width=1\columnwidth,angle=0]{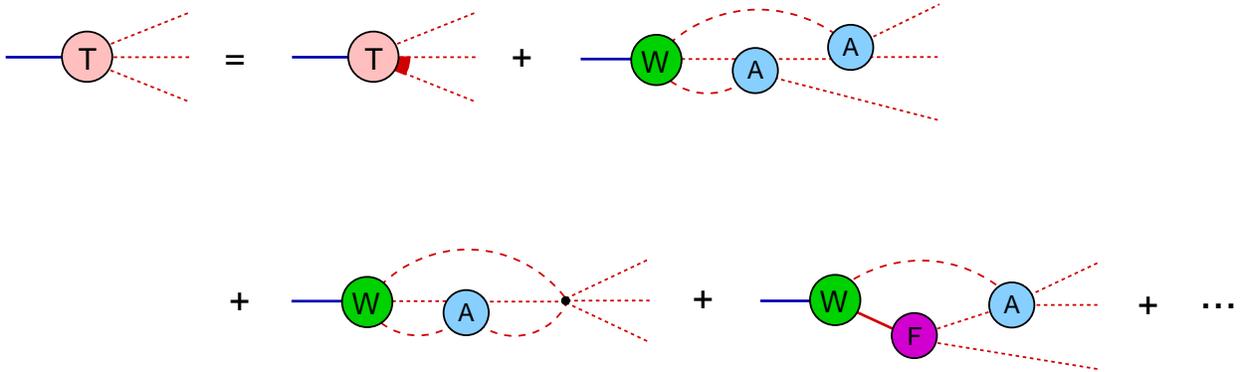}
\caption{Decay amplitude: $2+1$ approximation, supplemented by three-body interactions.}
\label{F-6}
\end{figure}

In order to go beyond the $(2+1)$ approximation, 
one would need to tackle a rather complicated three-body problem,
which involves both multiple scattering series and proper three-body interactions, 
as indicated in Fig.~\ref{F-6}.
It is worth stressing that these FSIs are not a matter of choice, 
since they are compulsory contributions to the problem.
Part of this sector can be tackled by means of Fadeev techniques~\cite{BR} or 
 the  Khuri-Treiman formalism \cite{kubisDKpp, mousallam_eta3pi} 
but this kind of effort 
to describe the full dynamics of heavy mesons nonleptonic decays is still incipient.

In summary, the decay of a heavy meson into three light mesons involves two 
distinct sectors, a weak primary vertex and a structure of final state strong interactions. 
Although the former is not simple, the latter may be expected to be much more complicated
and progress in the area depends on the definition of a  hierarchy among strong problems.
The simplest subset of problems is provided by the $(2+1)$ approximation
 depends on meson-meson scattering amplitudes.
 Nowadays
 even these two-body 
interaction are not sufficiently well known for systems involving pions, kaons and etas,
within the phase space provided by $D$ and $B$ decays.  
\\

\section{scattering amplitudes}
\label{scatt}

In this work we present a practical model for the inclusion of any number
of resonances in phenomenological meson-meson scattering amplitudes, so that they 
can be used as trial functions in more complicated reactions, such as 
 heavy-mesons or $\tau$ decays.  
Instead of presenting the model in its full complexity at once, we choose to construct it gradually,
so as to emphasize possible points of contact with the isobar model and point out
limitations of the latter.

The scattering amplitudes $ A_{(k\ell|ab)}^{(J,I)} $ for the process $P_k \, P_\ell \rar P_a \,P_b$
in a channel with spin $J$ and isospin $I$ are given in App.\ref{coupled} and involve three kinds
of conceptual ingredients, namely:
\\
{\bf a.  coupled channels - }  this sector of the problem is rather standard and model independent.
In our notation, the coupling among the various channels is implemented by the 
mixing matrices $ M_{ab}^{(J,I)}$ given by Eqs.~(\ref{c.1})-(\ref{c.4}). 
\\
{\bf b. multi-resonance dynamics - } the dynamical content of meson-meson ($PP$) 
interactions is incorporated 
into the kernels $\cK_{(k\ell|ab)}^{(J,I)}$ given in App.\ref{kernel}, 
which are real functions of masses and coupling constants.
While in kernels, resonances have no widths and are characterized just by their poles.
The inclusion of several resonances is performed by adding these
 poles
and the reader may want to inspect Eqs.~(\ref{k21})-(\ref{k26}) for an 
example. 
\\
{\bf c. unitarization - }  we neglect four-meson intermediate states and
the unitarization of amplitudes is directly associated with the
$s$-channel two-meson propagators $\O$ that occur in the full 
scattering amplitude.
These functions, described in App.\ref{omega}, contain real and imaginary parts:
$\O = \O^R + i\, \O^I$.
The latter, given by Eqs.~(\ref{a.13})-(\ref{a.14}), are free from ambiguities and 
constitute the only source of imaginary terms in the  amplitudes $ A_{(k\ell|ab)}^{(J,I)} $.
In particular, resonance widths are necessarily proportional to $\O^I$. 
The real component of $\O^R$ has infinite components which
are replaced by renormalization constants.
The form of this component in the case of several resonances is the object of this work.

At this point it is worth stressing that the model dependence incorporated in 
the amplitudes $ A_{(k\ell|ab)}^{(J,I)} $, given in App.\ref{coupled},
is restricted to the kernels $\cK$, which depends on dynamical assumptions, 
and to the real part $\O^R$ of two-meson propagators,
to be discussed in Sect.\ref{model}.
As the the imaginary part $\O^I$ is unambiguous, 
the scattering amplitudes are unitary and comply exactly with coupled channel requirements
for any choices made for $\cK$ and $\O^R$.  
In this sense, the approach tames model dependence as much as possible.

 In order to turn the  discussion more concrete, we concentrate on 
the case of $\p \p$ scattering, described by 
the amplitudes $ A_{(\p\p|\p\p)}^{(J,I)} $ 
with $(J,I=1,1)$ and  $(J,I=0,0)$, for comparisons with the isobar model and 
discussion of the main features of our model.
The extension to other channels is straightforward.
Using Eqs.~(\ref{c.9}) and (\ref{c.14}) we have
\bea
 A_{(\p\p|\p\p)}^{(1,1)} &\!=\!& \frac{(t-u)}{D^{(1,1)}}
\lc \lb 1 \sm M_{22}^{(1,1)} \rb \, \cK_{(\p\p|\p\p)}^{(1,1)} + M_{12}^{(1,1)} \, \cK_{(KK|\p\p)}^{(1,1)}  \rc \;,
\label{scat.1}\\[4mm]
 A_{(\p\p|\p\p)}^{(0,0)}  &\!=\!& \frac{1}{D^{(0,0)}}
\lc \lb \lp 1\sm M_{22}^{(0,0)}\rp \lp 1\sm M_{33}^{(0,0)}\rp \sm M_{23}^{(0,0)} \,  M_{32}^{(0,0)} \rb\, 
\cK_{(\p\p|\p\p)}^{(0,0)}
\right.
\nn\\[2mm]
&\! + \!& \left. \lb M_{12}^{(0,0)} \, \lp 1\sm M_{33}^{(0,0)} \rp \sp M_{13}^{(0,0)} \,  M_{32}^{(0,0)} \rb  \, 
\cK_{(KK|\p\p)}^{(0,0)}
\right.
\nn\\[2mm]
&\! + \!& \left.
\lb M_{13}^{(0,0)} \, \lp 1\sm M_{22}^{(0,0)} \rp \sp M_{12}^{(0,0)}\,  M_{23}^{(0,0)} \rb \, 
\cK_{(88|\p\p)}^{(0,0)} \rc  \;,
\label{scat.2}
\eea
 where the $\eta$ is represented by 8.
In these results, the complex mixing matrices, given by Eqs.~(\ref{c.1}) and (\ref{c.4}), 
have the general structure $M=\cK\times\O$.
The denominators $D$ contain the pole structure of the theory and have the form
\bea 
&& D^{(1,1)}= \lb 1\sm M_{11}^{(1,1)} \rb \,  \lb 1\sm M_{22}^{(1,1)} \rb - M_{12}^{(1,1)}  M_{21}^{(1,1)} \;,
\label{scat.3}\\[4mm]
&& D^{(0,0)} =\lb 1\sm M_{11}^{(0,0)}\rb \, \lb 1\sm M_{22}^{(0,0)} \rb \, \lb 1 \sm M_{33}^{(0,0)} \rb 
-  \lb 1 \sm M_{11}^{(0,0)} \rb \,   M_{23}^{(0,0)} \,  M_{32}^{(0,0)}  
\nn\\[2mm]
&& \hspace{10mm} 
- \lb 1 \sm M_{22}^{(0,0)} \rb \,  M_{13}^{(0,0)} \,  M_{31}^{(0,0)}  
- \lb 1 \sm M_{33}^{(0,0)}\rb \,  M_{12}^{(0,0)} \,  M_{21}^{(0,0)} 
\nn\\[2mm]
&& \hspace{10mm}
-  \, M_{12}^{(0,0)} \, M_{23}^{(0,0)} \, M_{31}^{(0,0)}  
- M_{21}^{(0,0)} \,  M_{13}^{(0,0)} \, M_{32}^{(0,0)}  \;.
\label{scat.4}
\eea
At low energies,  $M_{ab}^{(J,I)} \rar 0 $  and the amplitudes (\ref{scat.1}) and (\ref{scat.2})
become the real functions
\bea
 A_{(\p\p|\p\p)}^{(1,1)} &\! \rar \!& (t-u) \, \cK_{(\p\p|\p\p)}^{(1,1)}  \rar \frac{(t-u)}{F^2} \;,
\label{scat.5}\\[2mm]
 A_{(\p\p|\p\p)}^{(0,0)}  &\!\rar \!& \cK_{(\p\p|\p\p)}^{(0,0)} \rar \frac{(2\,s - M_\p^2)}{F^2} \;,
\label{scat.6}
\eea
where $F$ is the pion decay constant.


\section{standard isobar model  - uncoupled channels}
\label{imUNC}

The form of expressions (\ref{scat.1}) and (\ref{scat.2}) is involved owing to channel coupling. 
In order to discuss their contact with the standard isobar model, in this section 
we pretend that the $\pi\pi$ state cannot couple to $K\Kb$ and $\eta \eta$.
Labeling  with $U$ the corresponding uncoupled amplitudes, we have 
\bea
&&  A_{(\p\p|\p\p)}^{U \,(1,1)}  = \frac{(t-u)\; \cK_{(\p\p|\p\p)}^{(1,1)}}
{1 + \cK_{(\p\p|\p\p)}^{(1, 1)} \, [ \O_{\p\p}^P /2 ]} \;,
\label{imu.1}\\[4mm]
&& A_{(\p\p|\p\p)}^{U\, (0,0)}  = \frac{\cK_{(\p\p|\p\p)}^{(0,0)}}
{1 +  \cK_{(\p\p|\p\p)}^{(0,0)} \, [\O_{\p\p}^S/2]} \;.
\label{imu.2}
\eea
where $\O$ are the two-pion propagators discussed in App.\ref{omega}.
The kernels are given by Eqs.~(\ref{k1}) and (\ref{k21}) and, in order to simplify the discussion, 
we assume the value $\e=0$ for the mixing parameter in Eqs.~(\ref{k19}) and (\ref{k20}).
Thus
\bea
&& \cK_{(\p \p|\p\p)}^{(1,1)} =  \frac{1}{F^2}  
- \lb \frac{2\,G_V^2}{F^4}\rb  \frac{s}{s-m_\rho^2} 
-\, \frac{s\;G_{(\rho'|\p\p)}^2}{s-m_{\rho'}^2} 
\label{imu.3}\\[4mm]
&& \cK_{(\p \p|\p \p)}^{(0,0)}  = \frac{(2 s - M_\p^2)}{ F^2} 
- \lb \frac{12}{F^4} \rb \frac{\lb \ct_d \, s - (\ct_d \sm \ct_m) \, 2M_\p^2\rb^2 }{s-m_{S1}^2}\,
\nn\\[4mm]
&& 
- \lb \frac{2}{F^4}\rb \frac{\lb c_d \, s - (c_d \sm c_m) \, 2M_\p^2 \rb^2}{s-m_{So}^2}\,
- \, \frac{G_{(f'|\p\p)}^2}{s-m_{f'}^2}
\label{imu.4}
\eea
where $G_V$, $\ct_d, \ct_m, c_d, c_m$ are coupling constants~\cite{EGPR} and $m_{So}$ and $m_{S1}$ are the $SU(3)$ octet and singlet  scalar resonances.
We further simplify these results by considering just a single resonance in each channel.
In the vector case, using the approximate identity $G_V=F/\sqrt{2}$, one recovers
the classic vector meson dominance~\cite{EGPR} result 
\bea
&&  \cK_{(\p \p|\p\p)}^{(1,1)} \rar  - \,\frac{m_\rho^2/F^2}{s-m_\rho^2} \,
\label{imu.5}
\eea
whereas, for the scalar, one writes
\bea
&& \cK_{(\p \p|\p \p)}^{(0,0)}  \rar -\, \frac{\Theta^2(s)}{s-m_{So}^2} \;,
\label{imu.6}\\[4mm]
&& \Theta^2(s)  = 
\lb -\, \frac{[2 s \sm M_\p^2]\,[s-m_{So}^2]} { F^2}   
+ \frac{2\, \lb c_d \, s - (c_d \sm c_m) \, 2M_\p^2 \rb^2}{F^4} \rb \;. 
\label{imu.7}
\eea

Using results (\ref{imu.5})-(\ref{imu.7}) into Eqs.~(\ref{imu.1})-(\ref{imu.2})  
and recalling that the imaginary parts of $\O_{\p\p}$ are given by (\ref{a.13}) and (\ref{a.14}), 
the uncoupled amplitudes can be expressed in terms of functions $\cM$ 
and $\Gamma$ that resemble masses and widths as
\bea
&& A_{(\p\p|\p\p)}^{U\, (1,1)} = -\, \frac{(t-u)\; m_\rho^2/F^2}
{s-\cM_V^2 + i\, \cM_V\, \Gamma_V} \;,
\label{imu.8}\\[4mm]
&& \cM_V^2  = m_\rho^2  - \frac{m_\rho^2 \, \O_{\p\p}^{P\,R}}{2\,F^2} \;, 
\label{imu.9}\\[4mm]
&& \cM_V\,\Gamma_V = \frac{m_\rho^2}{96\pi\, F^2} \, \frac{(s-4 M_\p^2)^{3/2}}{s^{1/2}} \;,
\label{imu.10}\\[4mm]
&& A_{(\p\p|\p\p)}^{U\, (0,0)} = -\, \frac{\Theta^2(s)}
{s-\cM_S^2 + i\, \cM_S\, \Gamma_S} \;,
\label{imu.11}\\[4mm]
&& \cM_S^2  = m_{So}^2  - \frac{\Theta^2(s) \, \O_{\p\p}^{S\,R}}{2} \;, 
\label{imu.12}\\[4mm]
&& \cM_S\,\Gamma_S = \frac{\Theta^2(s)}{32\pi} \, \frac{(s-4 M_\p^2)^{1/2}}{s^{1/2}} \;.
\label{imu.13}
\eea

These results illustrate a number of features from  constructive descriptions of 
resonances, namely:
\\
{\bf a.}  even if we begin with a bare resonance, it acquires a dynamical width
by means of interactions with pseudoscalars, whereas the $s$-channel pole  present in the kernel
becomes complex.
In the case of the $\rho$, Eq.~(\ref{imu.10}) yields  
 $\Gamma_P$ $ \rar$ $\Gamma_\rho\sim 145\, $Mev, close to
the PDG value~\cite{PDG}.
\\
{\bf b.}  the functions $\cM$ shift the resonance masses from their nominal $m$ values.
As indicated by Eqs.~(\ref{imu.9}) and (\ref{imu.12}) these are model dependent effects,
because the real parts $\O_{\p\p}^{J\,R}$ of the two-pion propagators contain 
undetermined free constants, remnants of renormalization.
A popular way to avoid this problem consists in using the $K$-matrix approach,
in which this function is set to zero by fiat.
In sect.\ref{model},  our alternative is presented.
\\
{\bf c.}  equations (\ref{imu.8}) and (\ref{imu.11}) resemble the Breit-Wigner line shapes
given by Eq.~(\ref{mot.2}),but superficially only.
In fact, they are rather different, because the $\cM$ and $\Gamma$ are running functions of $s$.
The usual BW expressions, on the other hand, employ masses $\cM_V^{BW} = m_\r^2$,
$\cM_S^{BW}= m_{So}^2$ and widths given by
\bea
&& \Gamma_V^{BW} = \frac{(s-4 M_\p^2)^{3/2}}{96\pi\, F^2} \;,
\label{imu.14}\\[4mm]
&& \Gamma_S^{BW} = C_S^{BW} \, \frac{(s-4 M_\p^2)^{1/2}}{32\pi} \;,
\label{imu.15}
\eea
where $C_S^{BW}$ is a coupling constant.
Comparing these expressions with Eqs.~(\ref{imu.10}) and (\ref{imu.13}),
we learn that the 
BW line shape is a good approximation for
vector but unsuited for scalar resonances. 
The fact that Eq.~(\ref{imu.10}) is identical with the classic Gounaris-Sakurai result, produced in
1968~\cite{GS}, indicates that the vector sector has been stable in the last 50 years.
However,  the scalar sector 
 is different, because the way one understands
it changed significantly after the development of QCD.
The ground state of the theory, its vacuum, is not empty and 
chiral perturbation theory implements this feature into low-energy physics.
In the present case, it gives rise to the incorporation of both contact interactions and
$s$-dependent couplings of scalar resonances
 to pseudoscalars~\cite{EGPR} into the function $\Theta(s)$.
In this exercise, even if we assume $C_S^{BW} =  \Theta^2(m_{So}^2)/m_{So}^2$, 
the BW approximation for scalars remains unsuited, for all the rich $s$-dependence of Eq.~(\ref{imu.7}) 
is lost.

A very important feature of Eqs.~(\ref{imu.1}) and (\ref{imu.2}) is that they are 
automatically unitary, irrespective of the features of the kernel $\cK$ employed,
provided it is real, and of the real part of the two-pion propagator $ \O_{\p\p}^{J\,R} $.
In practice, an easy way to check unitariy is to evaluate 
the inelasticity $\eta$,
using the non-relativistic amplitudes $ f_{(\p \p|\p\p)}^{(J,I)} $
given in App.\ref{nonrelat}.
Skipping labels, they are related to the $ A_{(\p \p|\p\p)}^{(J,I)} $ by
\bea
f = -\, \frac{\O^I}{2} \, A_{(\p \p|\p\p)}^{(J,I)}
\label{imu.16}
\eea
where the $ \O^I  $ are the imaginary parts of the two-pion propagator,
given by Eqs.~(\ref{a.13}) and (\ref{a.14}).
 Thus, one has the generic form
\bea
&& f= -\, \frac{1}{w +i}\;,
\label{imu.17}\\[2mm]
&& w = \frac{1 +  \cK \, \O^R /2}
{\cK \, \O^I/2 } \;.
\label{imu.18}
\eea
Unitarity is ensured because, for any function of the form (\ref{imu.17}),  
irrespective of the value of $w$, the inelasticity parameter, given by Eq.~(\ref{ps.8}),
is always $\eta=1$ , in the absence of other channels.
So, this is a model independent result, valid for any choices of $\cK$ and $\O^R$.

One now considers the case of several resonances in the same channel.
As shown in App.\ref{kernel}, the kernel for a channel containing $n $ resonances 
represented by individual terms $\cK_j$  is written as 
\bea
\cK = \cK_c + \cK_1 + \cdots + \cK_n \;,
\label{imu.19}
\eea
where $\cK_c$ is a contact term.
Using Eqs.~(\ref{imu.1}), (\ref{imu.2})  and (\ref{imu.16}), 
we  write  the non-relativistic amplitude as 
\bea
&& f = -\, \frac{[ \cK_c + \cK_1 + \cdots + \cK_n]\, \O^I/2}
{1 + [ \cK_c + \cK_1 + \cdots + \cK_n]\, [\O^R + i\, \O^I] /2}
= -\, \frac{1}{w +i}\;,
\label{imu.20}\\[4mm]
&& w = \frac{1 +  [ \cK_c + \cK_1 + \cdots + \cK_n]\, \O^R /2}
{ [ \cK_c + \cK_1 + \cdots + \cK_n]\, \O^I /2} \;.
\label{imu.21}
\eea
This amplitude is unitary because this property does not depend on the form of the kernel. 

In the standard isobar model, on the other hand, unitarized resonances are treated individually
and, for each of them, one would write
\bea
&&  f_i = -\, \frac{\cK_i \, \O^I/2} {1 +  \cK_i \, [\O^R + i\, \O^I] /2}
= -\, \frac{1}{w_j +i} \;,
\label{imu.22}\\[4mm]
&&  w_j = \frac{1 + \cK_j \, \O^R /2} { \cK_j \, \O^I /2} \;.
\label{imu.23}
\eea
These unitary terms are then added schematically as 
$ f_\mathrm{model}^\mathrm{isobar} = \a_c \, f_c + \a_1 \,f_1 + \cdots + \a_n\, f_n$,
where the $\a$ are complex functions of $s$.
Thus, one has $ f \neq f_\mathrm{model}^\mathrm{isobar}$ and learns that
the standard isobar model prescription for adding resonances is not compatible with unitarity.
This happens because it treats each resonance as an individual object 
whereas, in the amplitude, they are necessarily coupled among themselves by the intermediate states
they share.
Unitarity is a global property that cannot be split as sums of individual contributions.

In summary, addition of resonances and unitarization does not commute and, 
after QCD, the SIM structure is suited just for  the case of a single uncoupled vector resonance.


\section{resonances - coupled channels}
\label{resCOUP}

The qualitative features of coupled channels are discussed just in the case of 
the scalar-isoscalar amplitude $ A_{(\p\p|\p\p)}^{(0,0)} $, 
including $KK$ and $\eta\eta$ couplings, given by Eq.~(\ref{scat.2}) and cast in the form
\bea
&& A_{(\p\p|\p\p)}^{(0,0)} = \frac{N_{(\p\p|\p\p)}^{(0,0)}}{D^{(0,0)}} \;,
\label{coup.1}\\[5mm]
&& N_{(\p\p|\p\p)}^{(0,0)} = \cK_{(\p\p|\p\p)}^{(0,0)} 
+ C_{(\p\p|KK)}^{(0,0)} \, \O_{KK}^S/2 
+ C_{(\p\p|88)} \, \O_{88}^S/2
+ C_{(\p\p|KK|88)} \, \O_{KK}^S \, \O_{88}^S /4 \;,
\label{coup.2}\\[5mm]
&& D^{(0,0)} = 1 
+ \lb \cK_{(\p\p|\p\p)}^{(0,0)} \, \O_{\p\p}^S/2 + \cK_{(KK|KK)}^{(0,0)} \, \O_{KK}^S/2 
+ \cK_{(88|88)}^{(0,0)} \, \O_{88}^S/2 \rb 
\nn\\[3mm]
&& \;\;\; + \;  C_{(\p\p|KK)} \; \O_{\p\p}^S \, \O_{KK}^S /4
+ \;  C_{(\p\p|88}) \; \O_{\p\p}^S \, \O_{88}^S /4
+ \; C_{(KK|88)}  \; \O_{KK}^S \, \O_{88}^S /4
\nn\\[3mm]
&& \;\;\; + \; C_{(\p\p|KK|88)} \; \O_{\p\p}^S \, \O_{KK}^S \, \O_{88}^S/8 \;,
\label{coup.3}\\[5mm]
&&  C_{(\p\p|KK)} = \cK_{(\p\p|\p\p)}^{(0,0)} \, \cK_{(KK|KK)}^{(0,0)}  - \lb\cK_{(\p\p|KK)}^{(0,0)}\rb^2 
 \label{coup.4}\\[3mm] 
&&  C_{(\p\p|88)} =  \cK_{(\p\p|\p\p)}^{(0,0)} \, \cK_{(88|88)}^{(0,0)}   - \lb \cK_{(\p\p|88)}^{(0,0)} \rb^2 \;,
\label{coup.5}\\[3mm]
&& C_{(KK|88)} =  \cK_{(KK|KK)}^{(0,0)} \, \cK_{(88|88)}^{(0,0)} - \lb \cK_{(KK|88)}^{(0,0)}\rb^2 \;,
\label{coup.6}\\[3mm]
&& C_{(\p\p|KK|88)} =  \cK_{(\p\p|\p\p)}^{(0,0)} \, \cK_{(KK|KK)}^{(0,0)} \,\cK_{(88|88)}^{(0,0)} 
 - \cK_{(\p\p|\p\p)}^{(0,0)} \, \lb \cK_{(KK|88)}^{(0,0)} \rb^2 
 - \cK_{(KK|KK)}^{(0,0)} \, \lb \cK_{(\p\p|88)}^{(0,0)} \rb^2
 \nn\\[3mm]
 && \;\;\; - \; \cK_{(88|88)}^{(0,0)} \lb \cK_{(\p\p|KK)}^{(0,0)} \rb^2 
 + 2\, \cK_{(\p\p|KK)}^{(0,0)} \, \cK_{(\p\p|88)}^{(0,0)} \,\cK_{(KK|88)}^{(0,0)} \;,
 \label{coup.7}
\eea

\subsection{close to the poles}
\label{clopol}

The kernels $\cK_{(aa|bb)}^{(0,0)}$ involving three bare poles are displayed
in App.\ref{kernel} and a na\" ive inspection of Eqs.~(\ref{coup.2})-(\ref{coup.7}) 
could suggest that the amplitude (\ref{coup.1}) would be highly singular.
However, this is not the case.
In order to simplify the discussion,  we assume that the mixing angle $\epsilon=0$ in Eqs.~(\ref{k19}) and
(\ref{k20}) and, at the vicinity of a pole, be it  $S_o$, $S_1$ or $S'$, the kernels have the general structure
\bea
&& \cK_{(aa|bb)}^{(0,0)} \simeq - \frac{G_{aa}\, G_{bb}}{\D} - B_{aabb} \;,
\label{coup.8}\\[2mm]
&& \D = (s-m^2) \;,
\label{coup.9}
\eea
where the $B_{aabb}$ are finite backgrounds and redundant indexes  were skipped.
Below, we show that divergent terms proportional to $\D^{-2}$ and $\D^{-3}$
cancel out in both $N_{(\p\p|\p\p)}^{(0,0)}$ and $D^{(0,0)}$ and the amplitude 
$ A_{(\p\p|\p\p)}^{(0,0)}$ is finite at the pole.
Close to the pole, explicit calculation yields
\bea
&& N_{(\p\p|\p\p)}^{(0,0)} \simeq \frac{1}{\D} \, \lc - \lb G_{\p\p}^2 + \D\, B_{\p\p\p\p} \rb 
+  H_{(\p\p|KK)} \;  \O_{KK}^S /2 
+ H_{(\p\p|88)} \;  \O_{88}^S /2 
\right.
\nn\\[2mm]
&&  \;\; \left. - \; H_{(\p\p|KK|88)} \; \O_{KK}^S   \O_{88}^S /4  + \D \,[\cdots]\rc
\label{coup.10}\\[4mm]
&& D^{(0,0)} \simeq \frac{1}{\D} \, \lc (s-m^2) 
\right.
\nn\\[4mm]
&&\;\; \left. - \; \lp G_{\p\p}^2 +\D  B_{\p\p\p\p} \rp  \O_{\p\p}^S/2
- \lp G_{KK}^2 +\D  B_{KKKK} \rp  \O_{KK}^S/2
- \lp G_{88}^2 +\D  B_{8888} \rp   \O_{88}^S/2
\right.
\nn\\[2mm]
&&\;\; \left. + \; H_{(\p\p|KK)} \; \O_{\p\p}^S \O_{KK}^S /4 
+  H_{(\p\p|88)} \; \O_{\p\p}^S \O_{88}^S /4 
+  H_{(KK|88)} \;  \O_{KK}^S \O_{88}^S /4
\right.
\nn\\[2mm]
&& \;\; \left. - \; H_{(\p\p|KK|88)} \; \O_{\p\p}^S \, \O_{KK}^S  \, \O_{88}^S /8 + \D\, [\cdots] \rc
\label{coup.11}\\[4mm]
&& H_{(\p\p|KK)} =  G_{\p\p}^2  B_{KKKK} + G_{KK}^2 B_{\p\p\p\p} 
- 2 \, G_{\p\p} G_{KK}  B_{\p\p KK} \;,
\label{coup.12}\\[4mm]
&& H_{(\p\p|88)} =   G_{\p\p}^2  B_{8888} + G_{88}^2 B_{\p\p\p\p} 
- 2 \, G_{\p\p} G_{88}  B_{\p\p 88} \;,
\label{coup.13}\\[4mm]
&& H_{(KK|88)} =   G_{KK}^2 \, B_{8888} + G_{88}^2\, B_{KKKK} 
- 2\, G_{KK}\, G_{88} \, B_{KK88} \;,
\label{coup.14}\\[4mm]
&& H_{(\p\p|KK|88)} =   G_{\p\p}^2  \lp B_{KKKK} B_{8888} - B_{KK88}^2 \rp 
+ G_{KK}^2  \lp B_{\p\p\p\p} B_{8888} - B_{\p\p 88}^2 \rp 
\nn\\[2mm]
&&  \;\;  + \; G_{88}^2  \lp B_{\p\p\p\p}  B_{KKKK} - B_{\p\p KK}^2 \rp 
- 2\,  G_{\p\p}  G_{KK}  \lp  B_{8888}  B_{\p\p KK} - B_{\p\p 88}\, B_{KK88}  \rp 
\nn\\[2mm]
&& \;\;  
-\; 2 \, G_{\p\p}\, G_{88}  \lp  B_{KKKK} B_{\p\p 88} - B_{\p\p KK}\, B_{KK88}  \rp 
\nn\\[2mm]
&& \;\; - \; 2\, G_{KK}  G_{88}  \lp  B_{\p\p\p\p}  B_{KK88}- B_{\p\p KK}\, B_{\p\p 88}  \rp 
\label{coup.15}
\eea

These results show that, at the pole,  both $ N_{(\p\p|\p\p)}^{(0,0)}$  and $D^{(0,0)}$ diverge as $1/\D$ 
and yield, as expected, a finite amplitude. 
They also shed light on a conceptual limitation of the isobar model.
Since the functions $H$ involve products of coupling  constants $G$ 
and background contributions $B$ from other channels, 
resonances no longer  behave as individual objects.
This contradicts the tacit assumption underlying the isobar model,
namely that background terms can be neglected and resonances can be isolated. 

In order to check the importance of background terms,   
we consider the case of a hypothetical single octet  resonance of mass $m=1.05\,$GeV,
between the $KK$  and $88$
 thresholds, where the finite backgrounds 
are given just by the chiral LO contact terms in Eqs.~(\ref{k21}-\ref{k24}), 
 with opposite signs.
Using the coupling constants prescribed in Ref.~\cite{EGPR}, the non-vanishing contributions
come from $G_{\p\p}=8.06\, $GeV, $G_{KK}=10.76\, $GeV, 
$B_{\p\p\p\p} = - 252.69$, $B_{\p\p KK}= - 110.39$, $B_{KKKK}= - 191.21$, 
which yield $G_{\p\p}^2 = 64.93\, $Gev$^2$,
$G_{KK}^2=115.69\, $GeV$^2$ and $H_{\p\p KK}=-22\,513.61\, $GeV$^2$.  
We adopt the $K$-matrix approximation, that consists in setting $\O^R =0$ and keeping $\O^I$ only.
Using  $[\O_{\p\p}^S]^I = - 191.78\,\times\,10^{-4}$ and $[\O_{KK}^S]^I = - 67.69\,\times\,10^{-4}$,
one finds $N_{(\p\p|\p\p)}^{(0,0)} = \{ -\, 64.93 + [i\,  76.20] \}/\Delta $ 
and $D^{(0,0)} = \{ i\, 0.62 + i\, 0.39  + [0.73]\}/\Delta $,
where the contributions involving the background were indicated by $[\cdots]$.
They cannot be neglected, indicating that Breit-Wigner line shapes, Eq.~(\ref{mot.2}), 
are not suited for describing resonances above a crossing threshold.

\subsection{$K$-matrix results}
\label{Kmare}

As already stressed, the imaginary component $\O^I$ of 
the two-meson propagators $\O $ is fully  determined by theory.
In the widely used $K$-matrix approach, just this part is 
kept and the choice $\O^R=0$ amounts, 
in fact, to a disguised model for the real part.
In the case of uncoupled channels, this choice has the advantage of allowing a clear 
identification of the nominal value of the resonance mass. 
In this subsection, we present numerical studies for the scalar-isoscalar amplitude
$ A_{(\p\p|\p\p)}^{(0,0)} $ given by Eq.~(\ref{scat.2}) and rely on expressions for the kernel 
given in App.\ref{kernel},  with resonance masses 
$m_{fa}=1.37\,$GeV, $m_{fb}=0.98\,$GeV,
and coupling parameters fixed in Ref.~\cite{EGPR}.
Once the value of $\O^R$ is fixed, predictions depend just on models used for the interaction kernel.

\begin{figure}[h!] 
\includegraphics[width=.6\columnwidth,angle=0]{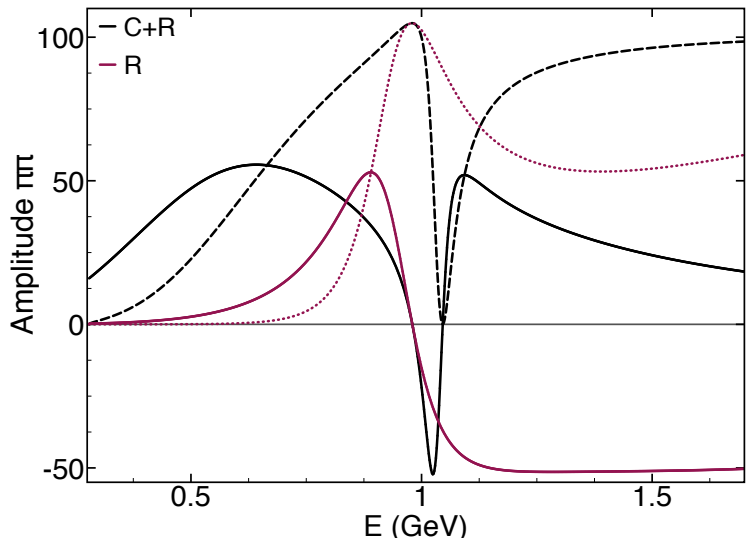}
\caption{Predictions for real (full curves) and imaginary (dashed curves) parts of 
the scalar-isoscalar $\p\p$ amplitude based on a single resonance (R) and 
the same resonance superimposed to a chiral contact term (C+R).}
\label{FK-1}
\end{figure}

In Fig.~\ref{FK-1}, we neglect $K\Kb$ and $\eta\eta$ couplings and compare results from 
two versions of 
Eq.~(\ref{k21}), both with $\epsilon=0$. One of them keeps
 just its third term, representing an octet resonance(R),
and the other also includes
 the first term, describing a contact chiral interaction(C+R),
which is one of the signatures of post QCD physics.
In the jargon of the isobar model, the
 resonant structure corresponds to 
a BW line shape, as discussed in Sect.\ref{imUNC}.
One notes that the contact term is rather important and the dominance of the resonance 
is restricted to a narrow band around its mass $m_{fb}$.
Close to threshold, 
the chiral contribution yields Eq.~(\ref{scat.6}) and give
 the correct magnitude for the scattering length.

\begin{figure}[h] 
\includegraphics[width=.7\columnwidth,angle=0]{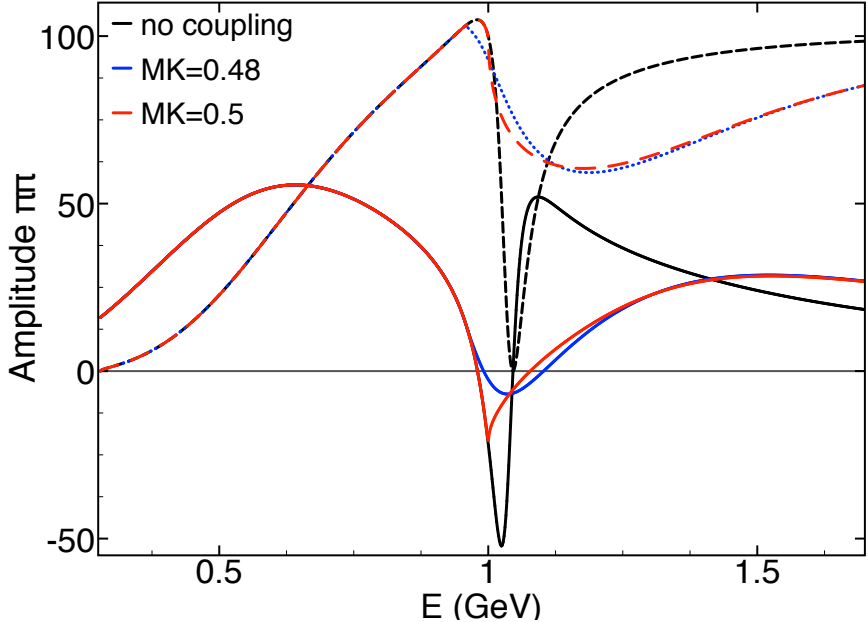}
\caption{Predictions for real (full curves) and imaginary (dashed curves) parts of 
the scalar-isoscalar $\p\p$ amplitude based on a single resonance superimposed 
to a non-resonant background (NR+R) 
for no coupled channels (black) and a coupled $K\Kb$ channel
with threshold below (blue) and above (red) the resonance mass.}
\label{FK-2}
\end{figure}

The opening of the $K\Kb$ channel is studied in Fig.~\ref{FK-2}, for the same C+R 
case considered before, 
 keeping the resonance mass fixed at $m_{fb}=0.98\,$GeV,
while adopting two fake values for $M_K$, namely $0.48$ and $0.50\, $GeV,
so as to have the $K\Kb$ threshold both below and and above it.
As expected, all curves coincide below the thresholds.
Above them, however, one learns that the impact of the coupling is important,
since the previous C+R form provides a very poor 
representation for the new results, irrespective of the value of $M_K$ chosen.
At threshold, one has a usual cusp in the real part of the amplitude for $m_{fb} < 2\,M_K$
and a discontinuity in its imaginary part for $m_{fb}> 2\, M_K$.
Beyond that point,  the real curves display the upward bending associated with the 
polynomial chiral background whereas usual connections between real and imaginary parts 
are lost, owing to inelastic effects.  
Altogether, the shift  in $M_K$ affects the amplitudes just
in a narrow region of about $200\,$MeV above threshold.

\begin{figure}[h!] 
\includegraphics[width=.7\columnwidth,angle=0]{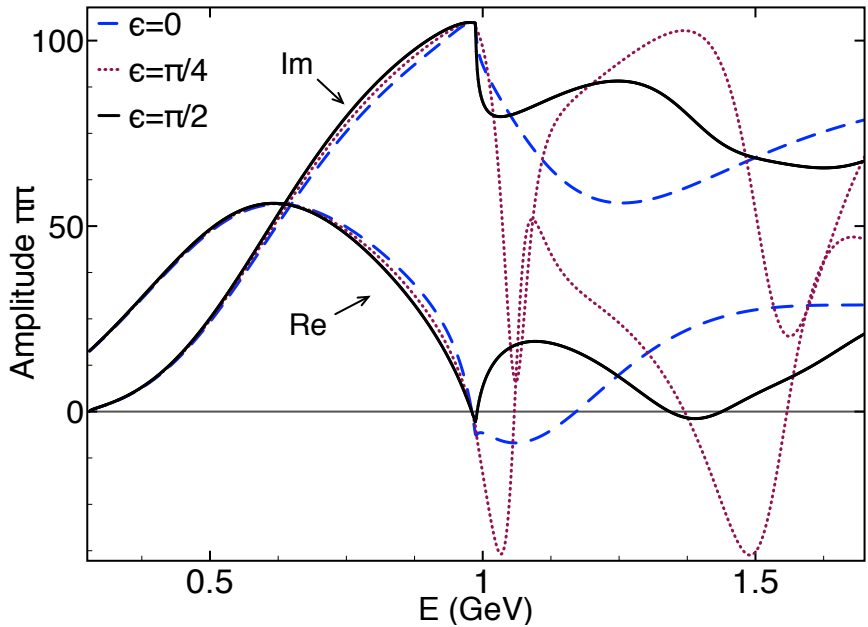}
\caption{ Predictions for real and imaginary  parts of 
the scalar-isoscalar $\p\p$  phase shift $\d^{(0,0)}$ (left)
and inelasticity parameter $\eta^{(0,0)}$ (right) based on two resonances  superimposed 
to a non-resonant background (C+$f_a$+$f_b$) with a coupled $K\Kb$ channel, 
for mixing parameters $\e=0$ (full lines), $\e=\p/4$ (dotted lines 
and $\e=\p$ (dashed lines).}
\label{FK-3}
\end{figure}

\begin{figure}[h!] 
\includegraphics[width=.45\columnwidth,angle=0]{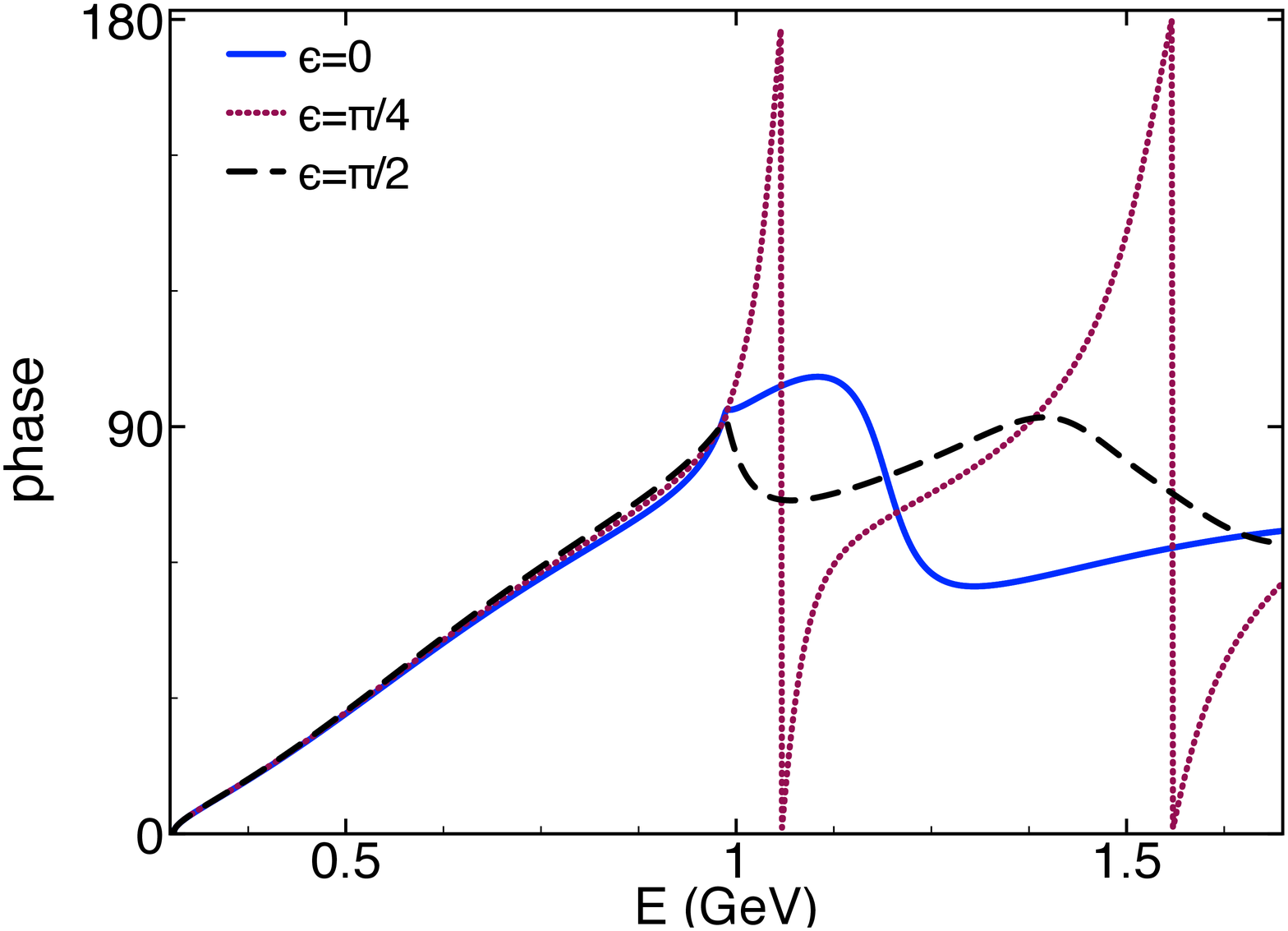}
\hspace{5mm}
\includegraphics[width=.45\columnwidth,angle=0]{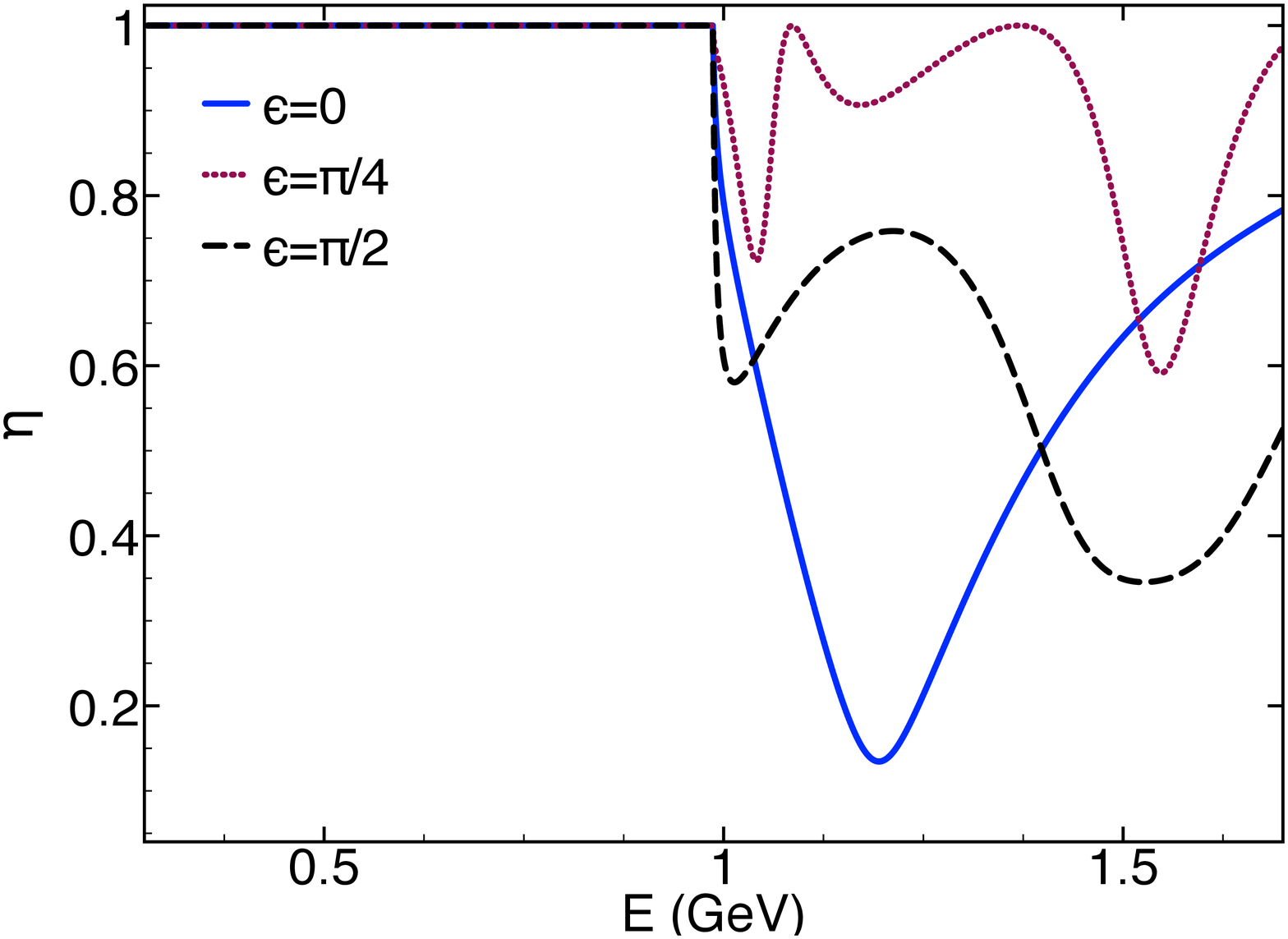}
\caption{ Predictions for real  and imaginary  parts of 
the scalar-isoscalar $\p\p$ amplitude based on two resonances  superimposed 
to a non-resonant background (NR+$f_a$+$f_b$) with a coupled $K\Kb$ channel, 
for mixing parameter $\e=0$ (full lines), $\e=\p/4$ (dotted lines) 
and $\e=\p$ (dashed lines).}
\label{FK-4}
\end{figure}

In the scalar-isoscalar sector,  $SU(3)$ gives rise to octet and singlet states $So$ and $S1$,
which can be combinations of the observed resonances $f_a=f(1370)$ and 
$f_b=f(980)$, with a mixing  angle $\e $ defined by Eqs.~(\ref{k19}) and (\ref{k20}). 
The influence of this parameter in the $\p\p$ amplitude is discussed in Fig.~\ref{FK-3}
for two resonances superimposed to the chiral background, adopting $\e = 0, \p/4, \p/2$.
All curves coincide up to $E=0.98\,$GeV, but become quite different afterwards,
the most striking feature being the change in the number of zeroes of the real part
over the energy range considered.
The influence of the mixing angle over the phase shift $\d^{(0,0)}$ and inelasticity
parameter  $\eta^{(0,0)}$ is presented in Fig.~\ref{FK-4}

\section{model for two-meson propagator}
\label{model}

The discussion presented here is general and applies to all meson-meson channels.
The amplitudes given in App.\ref{coupled} are model dependent both through 
the kernels $\cK$ 
and the real components $\O^R$ of the two-meson propagators $\O = \O^R + i\, \O^I$. 
The dependence on $\cK$ has a dynamical character, since it relies on parameters from lagrangians,
such as masses and coupling constants,
whereas the model for $\O^R$, discussed now, is kinematic.

The intermediate two-meson propagators for states $a$ and $b$ are given in App.\ref{omega},
Eqs.~(\ref{a.11}) and (\ref{a.12}),  and their complex forms for $J=0,1$ read
\bea
&& \O_{ab}^S  = - \, \frac{\Pi_{ab}(s)  }{16\p^2}  \;,
\label{m1}\\[4mm]
&& \O_{ab}^P =  -\, \frac{\l}{48\, \p^2\, s} \, \Pi_{ab}(s) \;.
\label{m2}
\eea 
where $\l$ is the  K\"all\'en function  whereas 
$\Pi_{ab}$  represents the regular parts of loop integrals,
that are determined by theory and shown in Eqs.~(\ref{a.4})-(\ref{a.9}).
Owing to renormalization, the real parts of the functions $\O$ must be supplemented 
by arbitrary constants, to be fixed by experiment and that is why a model dependence comes in.
In the framework of chiral perturbation theory, these constants are coefficients of 
polynomials on external momenta~\cite{Bachir}.

The model introduced here consists in a generalization scheme for Eqs.~(\ref{m1}) and (\ref{m2})  
and its explicit form depends on the number of resonances considered, 
which are denoted by $R_x, R_y, R_z \cdots$.
Their masses and coupling constants are taken as
 free parameters, so that
they can be fitted in phenomenological analyses.

In order to motivate the choices made, we consider the case $J=0$
and begin with the case of a single resonance, which is written as
\bea
&& \O_{ab}^S(s) \rar \frac{1}{16 \p^2} \lc \lb  F_x(s)\, \Pi_{ab}^R(m_x^2) \rb -  \Pi_{ab}(s) \rc\;,
\label{m3}
\eea 
where the term within square brackets is {\it real} and
 corresponds to a subtraction. 
It generalizes an expression employed earlier in the study of the $K \pi$ amplitude~\cite{PatWV}.
The function $F_x(s)$ is a form factor that satisfies the conditions:
\\
(a) $F_x(s) \rar 0$   for $s\rar 0$ - 
this is important to ensure that loop corrections do not spoil chiral symmetry results at low energies.
In that region, the symmetry predicts amplitudes to be proportional to the real contact terms 
present in the kernels given in App.\ref{kernel} and therefore the functions $\O$ cannot 
show up  there.
\\
(b) $F_x(s)= 1$ for $s = m_x^2$ - this condition implies that 
the real component satisfies
$\O_{ab}^{SR}(m_x^2) = 0$ and was 
chosen with practical purposes in mind, so that results coincide 
with those of the $K$-matrix approach at $s=m_x^2$.
In the case of uncoupled channels, this  allows the nominal mass of the resonance
to be identified with a zero of the real part of the scattering amplitude.
In the case of coupled channels, this property is preserved in the elastic regime below the first threshold 
but changes afterwards, as shown in Fig.~\ref{FK-3}.
The subtraction performed at the resonance mass is a conservative one, 
intended to prevent the increase of free parameters in the model.  
\\  
(c) $F_x(s)$ is finite  for $ s\rar \infty$ - 
chiral symmetry holds at low energies only, where it requires subtraction terms as polynomials in $s$.
However, these may become too important  at high energies, where the theory is no longer valid, and
this unwanted behavior is avoided by imposing  the form factor to be bound in that limit.
  
\begin{figure}[h] 
\includegraphics[width=.45\columnwidth,angle=0]{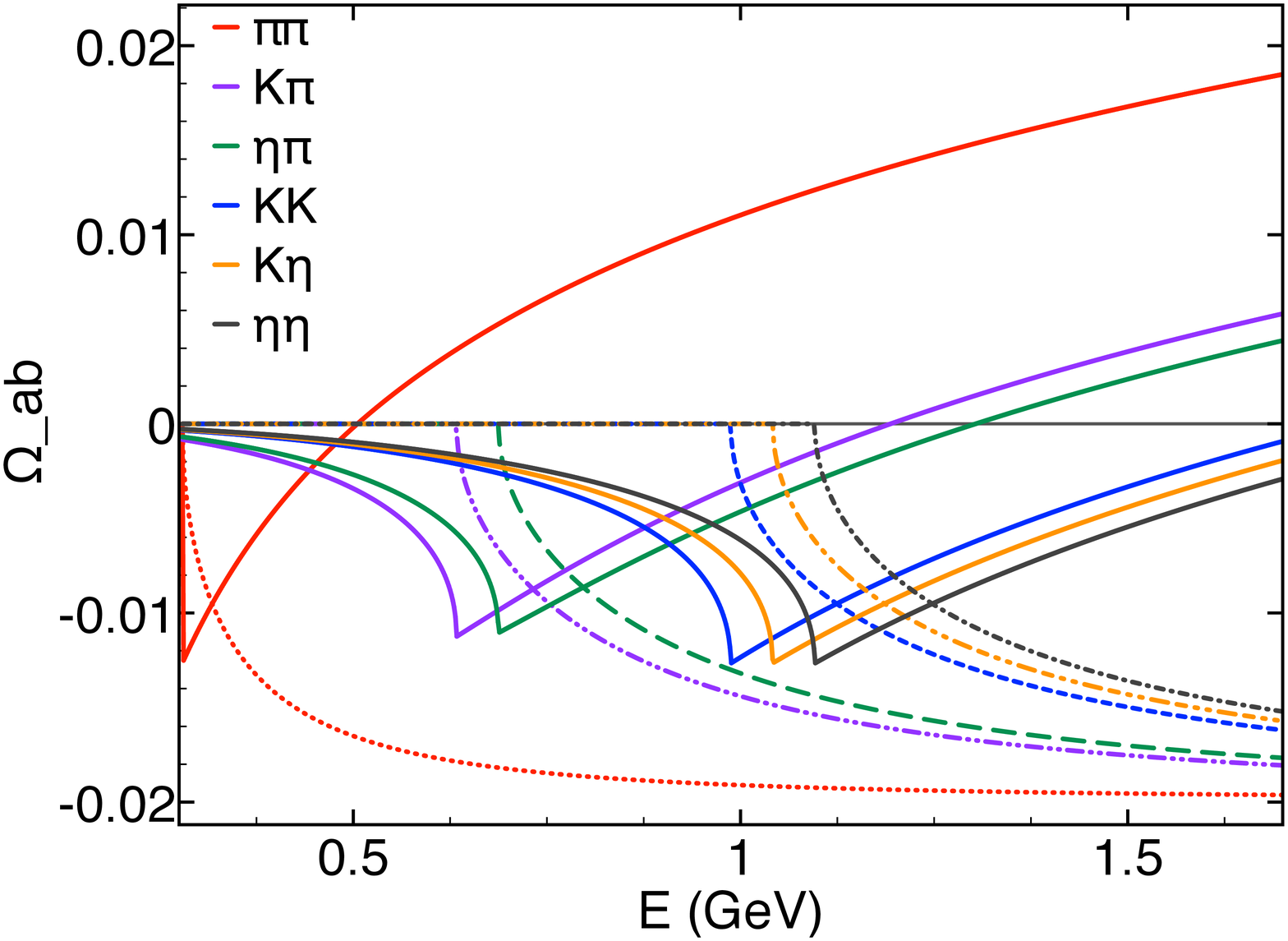}
\includegraphics[width=.45\columnwidth,angle=0]{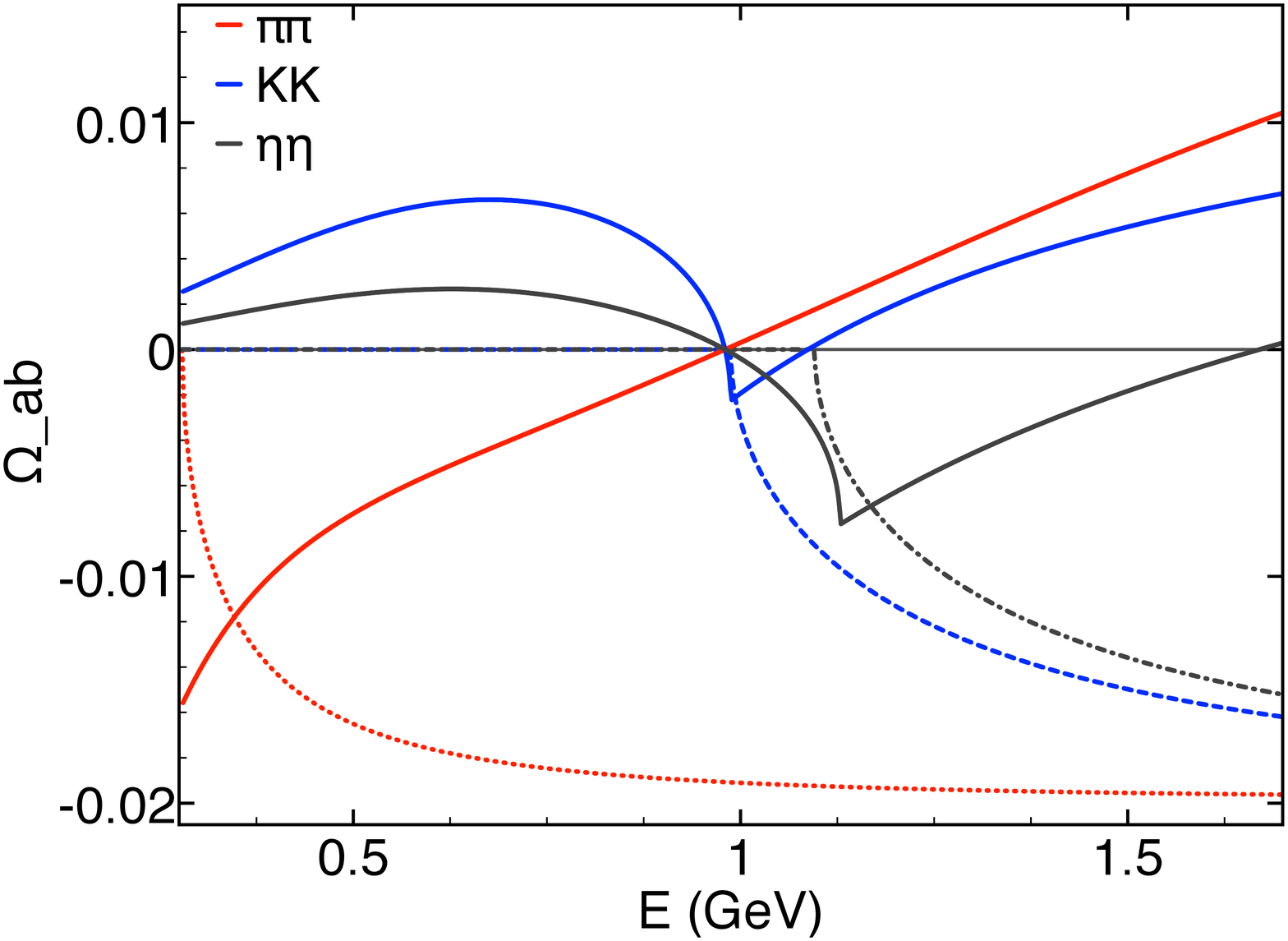}
\caption{Behaviour of the real (continuous lines) and imaginary (dashed lines) parts
of two-meson propagators: (left)  functions 
$\O_{\p\p}^S$, $\O_{K\p}^S$,  $\O_{\pi 8}^S$, $\O_{KK}^S$, $\O_{K8}^S$ and $\O_{88}^S$ from   Eq.~(\ref{m1});
(right) model predictions for the isospin 0 channel, based on a single $f_b$ resonance 
of mass $m_{f_b} = 0.98\,$GeV, from Eq.~(\ref{m3}). }
\label{Fm-1}
\end{figure}

The class of functions satisfying these criteria is, of course, very large and our choice is 
\bea
F_x(s) = \frac{4\,m_x^2\,s}{(s+m_x^2)^2}\;,
\label{m4}
\eea
which has a maximum at $s=m_x^2$.
In Fig.~\ref{Fm-1} we show, on the left, the energy dependence of 
the two-meson propagators 
for $\p\p$, $K\p$, $\pi \eta $, $K\Kb$,  $K\eta$  and $\eta \eta$ states given by Eq.~(\ref{m1}),
where it is possible to see the different scales associated with $SU(2)$  and $SU(3)$ sectors.
On the right, we present  model predictions based on Eq.~(\ref{m3})
for the isospin 0 channel, based on a single $f_b$ resonance 
of mass $m_{f_b} = 0.98\,$GeV.
We notice that the subtraction makes the real parts of $\O^S$ to vanish at the resonance mass 
and that the effects of the form factor $F_x(s)$ are more important at low energies, the very region 
where the functions $\O$ are less important owing to chiral symmetry.
These combined features suggest that the overall influence of the specific choice made 
in Eq.~(\ref{m4}) is expected to be small.

The extensions of Eq.~(\ref{m3}) to the case of two and three resonances read
\bea
&& \O_{ab}^S(s) \rar  \frac{1}{16\p^2} 
\lc F_x(s)\, \frac{(s-m_y^2) }{(m_x^2-m_y^2)} \, \Pi_{ab}^R(m_x^2) 
+ F_y(s) \, \frac{(m_x^2-s)}{(m_x^2-m_y^2)} \, \Pi_{ab}^R(m_y^2) - \Pi_{ab}(s)  \rc \;,
\label{m5}\\[4mm]
&& \O_{ab}^S(s) \rar \frac{1}{16 \p^2} 
\lc F_x(s)\, \frac{ (s-m_y^2) \, (s-m_z^2)}{ (m_x^2-m_y^2) \, (m_x^2-m_z^2)} \, \Pi_{ab}^R(m_x^2)
+ F_y(s)\, \frac{(m_x^2-s) \, (s-m_z^2)}{(m_x^2-m_y^2) \, (m_y^2-m_z^2)} \, \Pi_{ab}^R(m_y^2) 
\right.
\nn\\[4mm]
&& \left. +  
 F_z(s) \, \frac{(m_x^2- s) \, (m_y^2 - s)}{(m_x^2-m_z^2) \, (m_y^2-m_z^2)} \, \Pi_{ab}^R(m_z^2)  
 - \Pi_{ab}(s)
\rc \;.  
\label{m6}
\eea
The corresponding expressions for the $J=1$ case $\O_{ab}^P $ can be obtained from 
Eqs.~(\ref{m3}), (\ref{m5}) and (\ref{m6}) through multiplication by a factor $\l/3\,s$.

\begin{figure}[h!] 
\includegraphics[width=.45\columnwidth,angle=0]{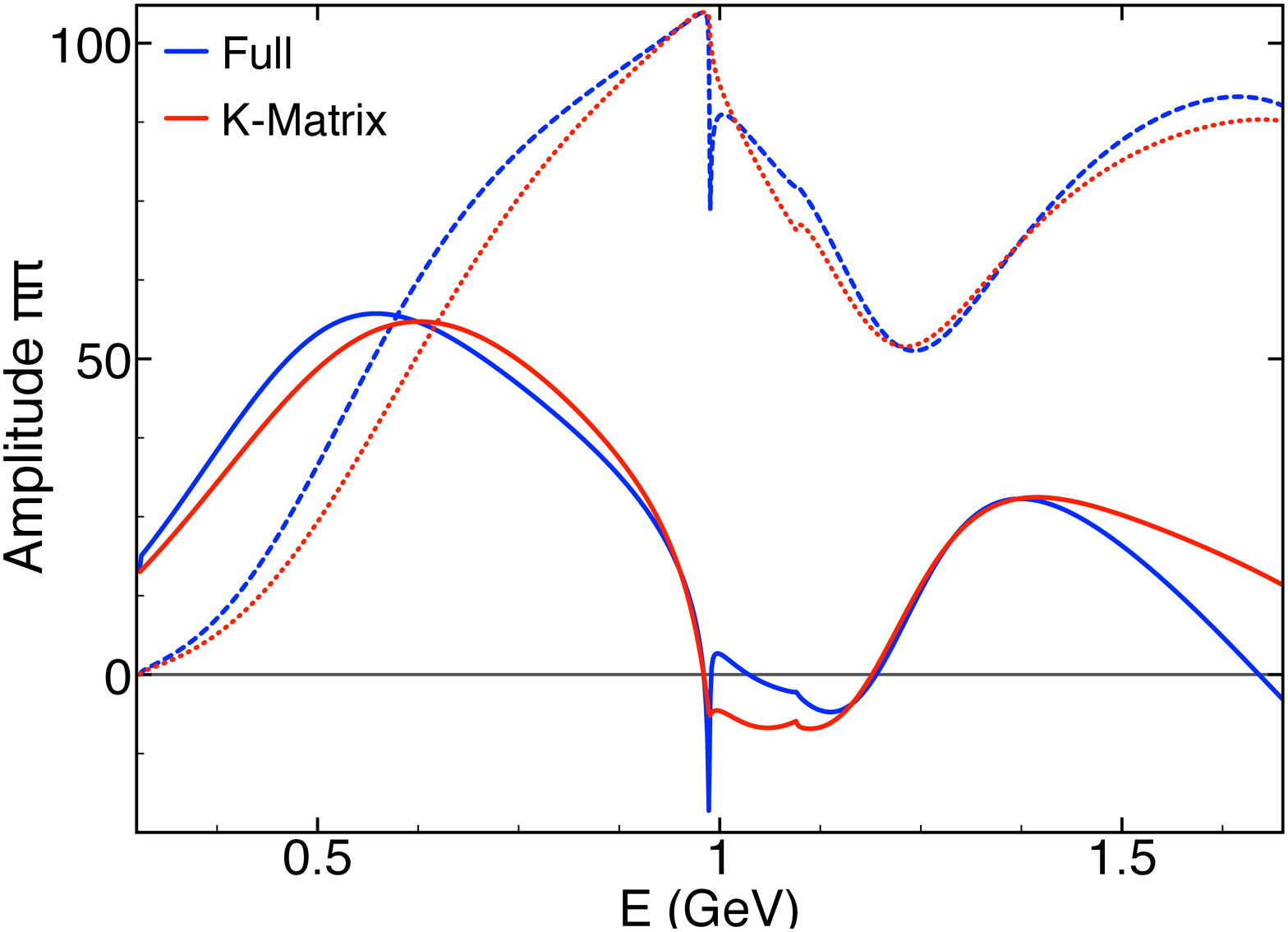}
\includegraphics[width=.45\columnwidth,angle=0]{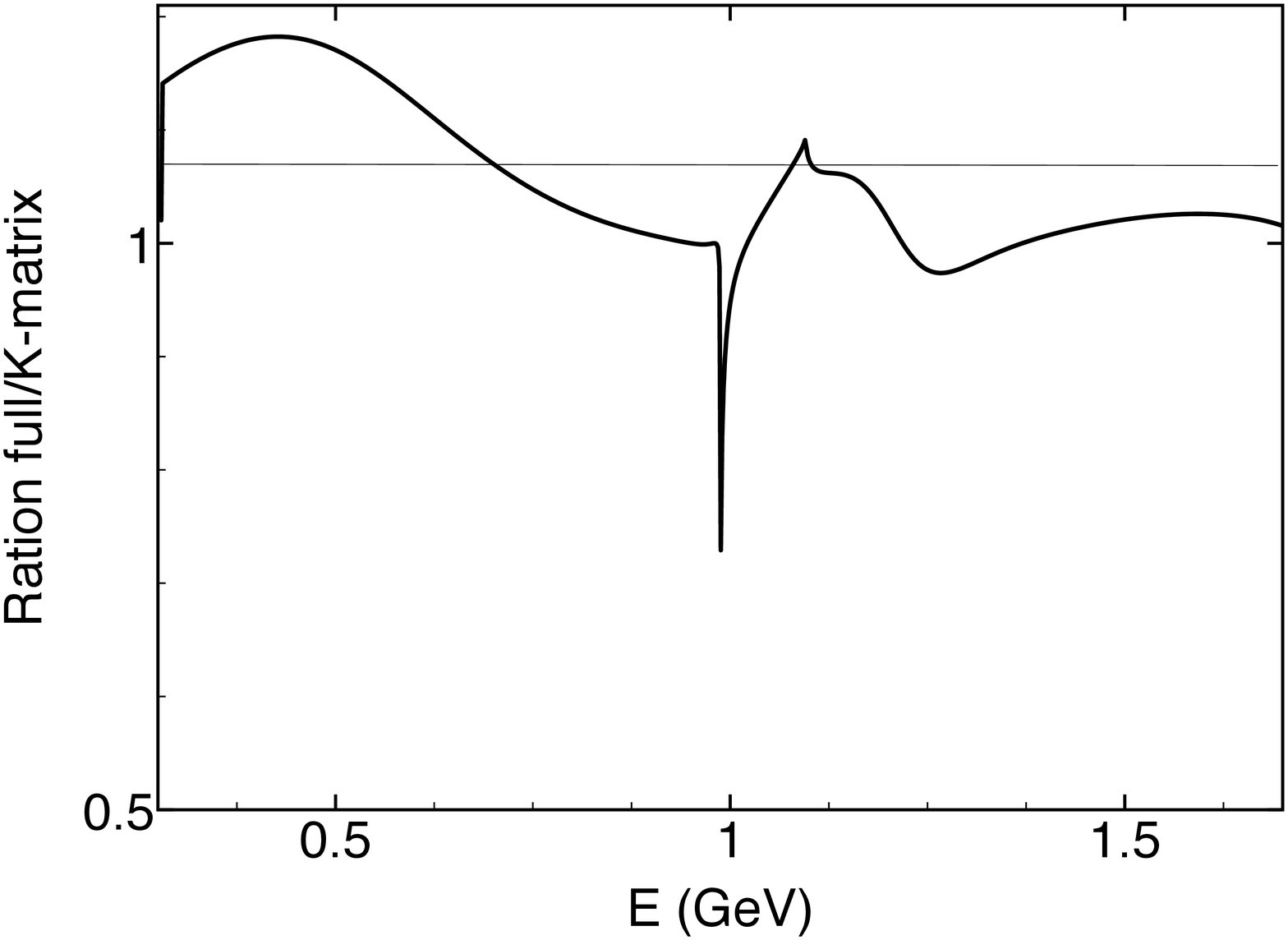}
\caption{ Predictions for the scalar-isoscalar $\p\p$ amplitude with
two resonances $f_a(1370)$ and $f_b(980)$, with $\e=0$, superimposed 
to a non-resonant background: (blue)  from the model,
Eq.~(\ref{m5})  and (red) the K-matrix;
left: (full curve) real and (dashed curve) imaginary parts;
right: ratio of magnitudes.}
\label{Fm-2}
\end{figure}

\begin{figure}[h!] 
\includegraphics[width=.45\columnwidth,angle=0]{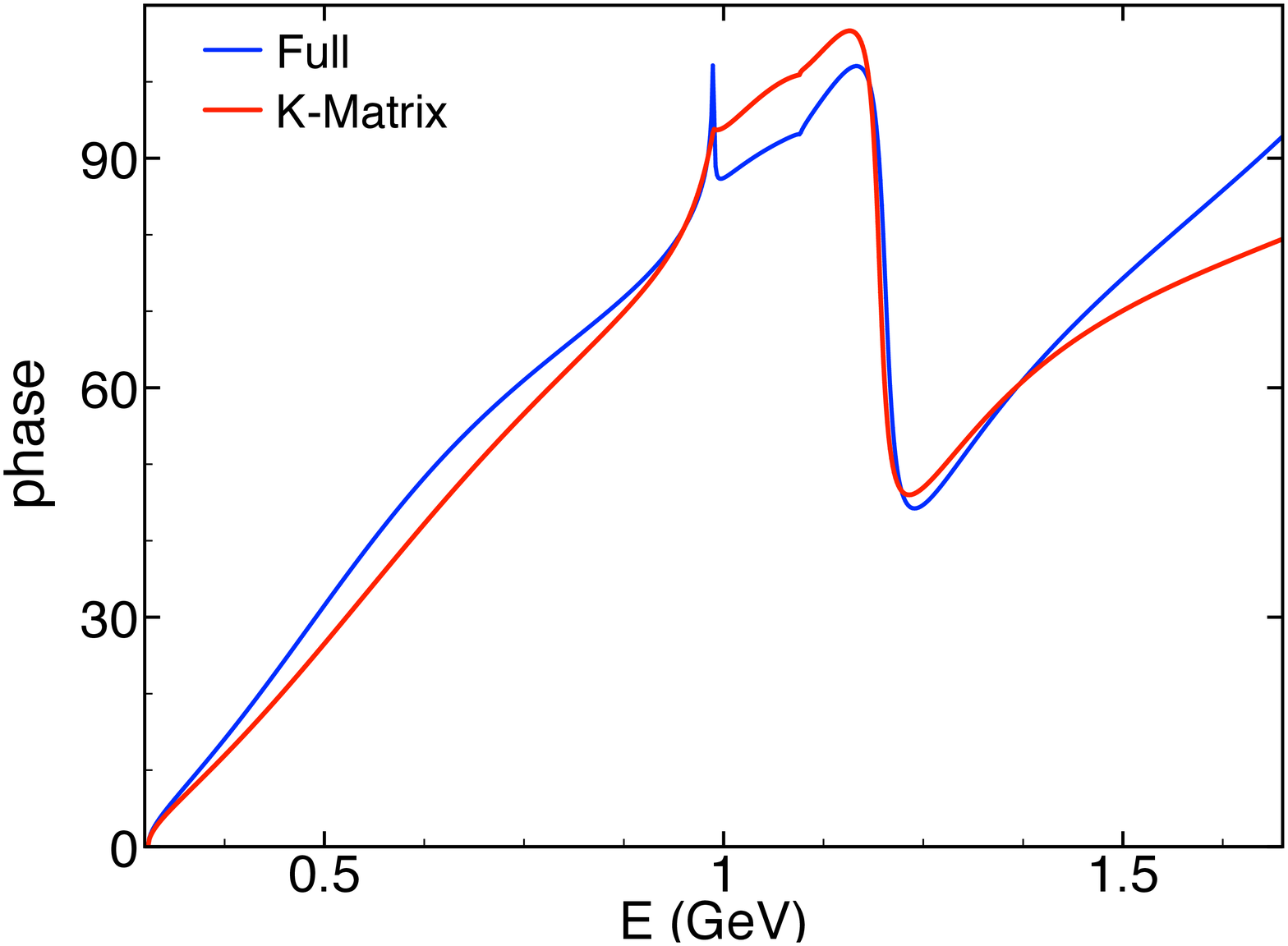}
\includegraphics[width=.45\columnwidth,angle=0]{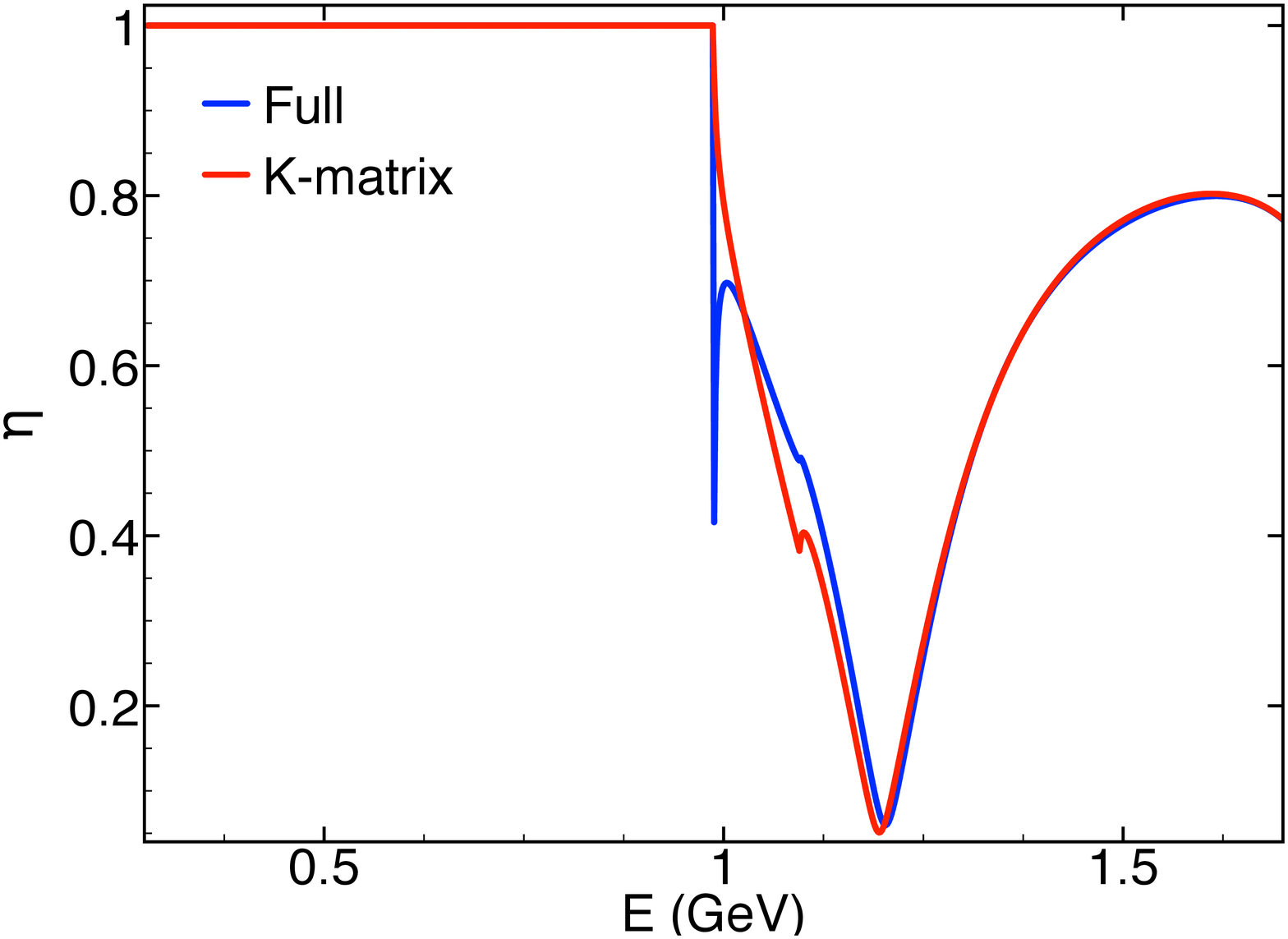}
\caption{  Predictions for phase shifts (left) and inelasticity parameter (right)
of the scalar-isoscalar $\p\p$ amplitude with
two resonances $f_a(1370)$ and $f_b(980)$, with $\e=0$, superimposed 
to a non-resonant background: (blue)  from the model,
Eq.~(\ref{m5})  and (red) the K-matrix.}
\label{Fm-3}
\end{figure}

\begin{figure}[h!] 
\includegraphics[width=.45\columnwidth,angle=0]{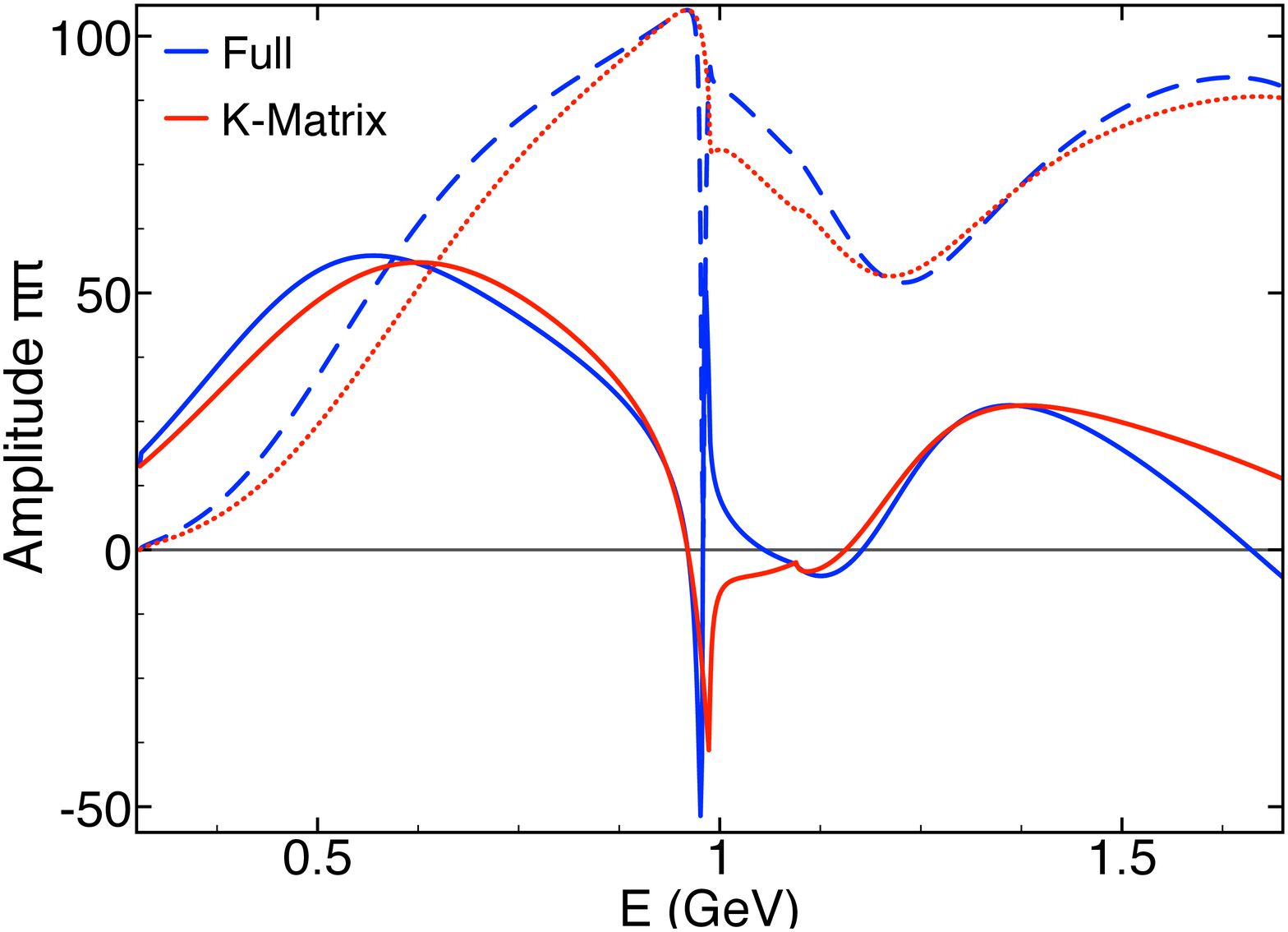}
\includegraphics[width=.45\columnwidth,angle=0]{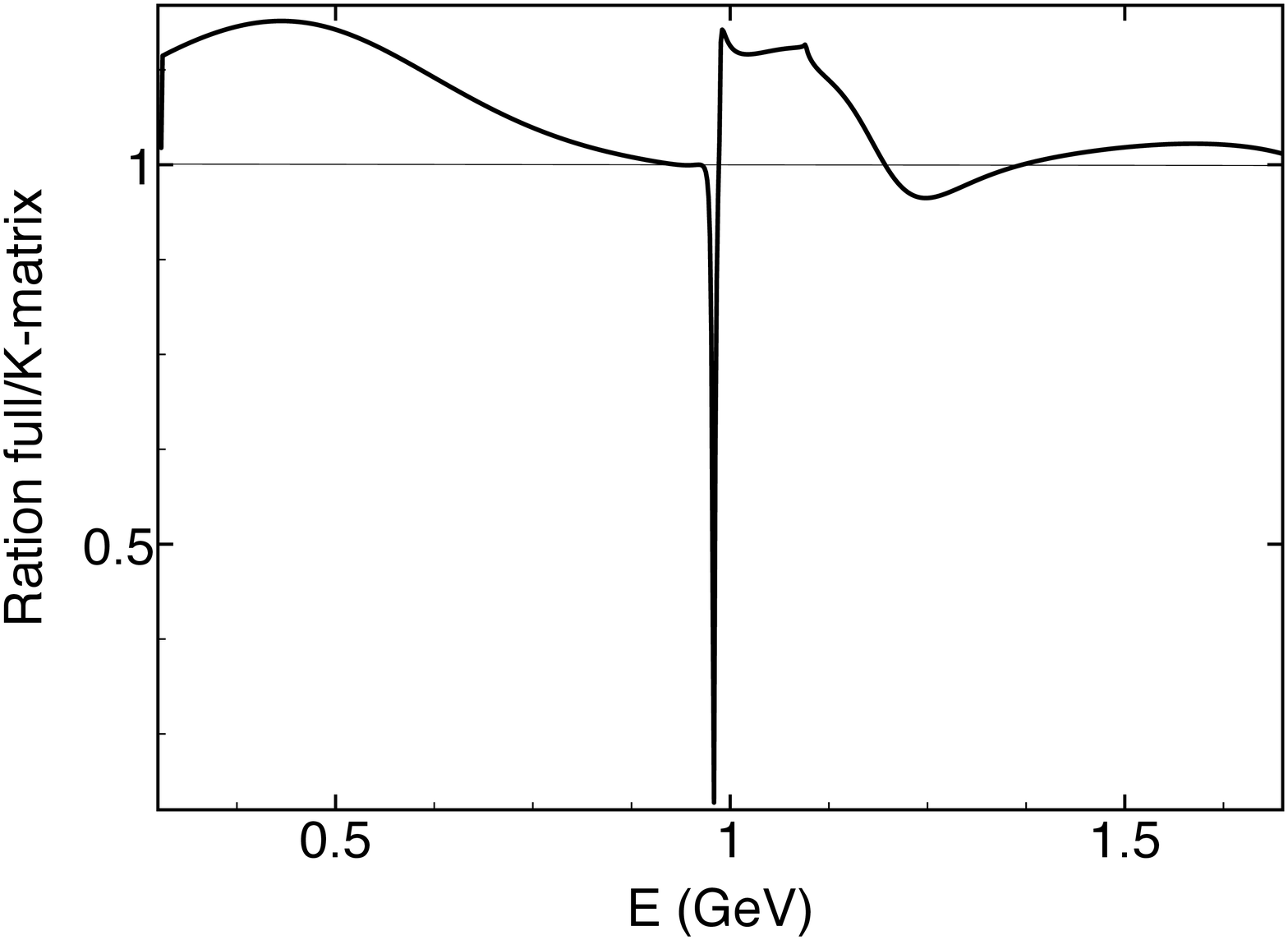}
\caption{ Predictions for the scalar-isoscalar $\p\p$ amplitude with
two resonances $f_a(1370)$ and $f_b(960)$, with $\e=0$, superimposed 
to a non-resonant background: (blue)  from the model,
Eq.~(\ref{m5})  and (red) the K-matrix;
left: (full curve) real and (dashed curve) imaginary parts;
right: ratio of magnitudes.}
\label{Fm-4}
\end{figure}

\begin{figure}[h!] 
\includegraphics[width=.45\columnwidth,angle=0]{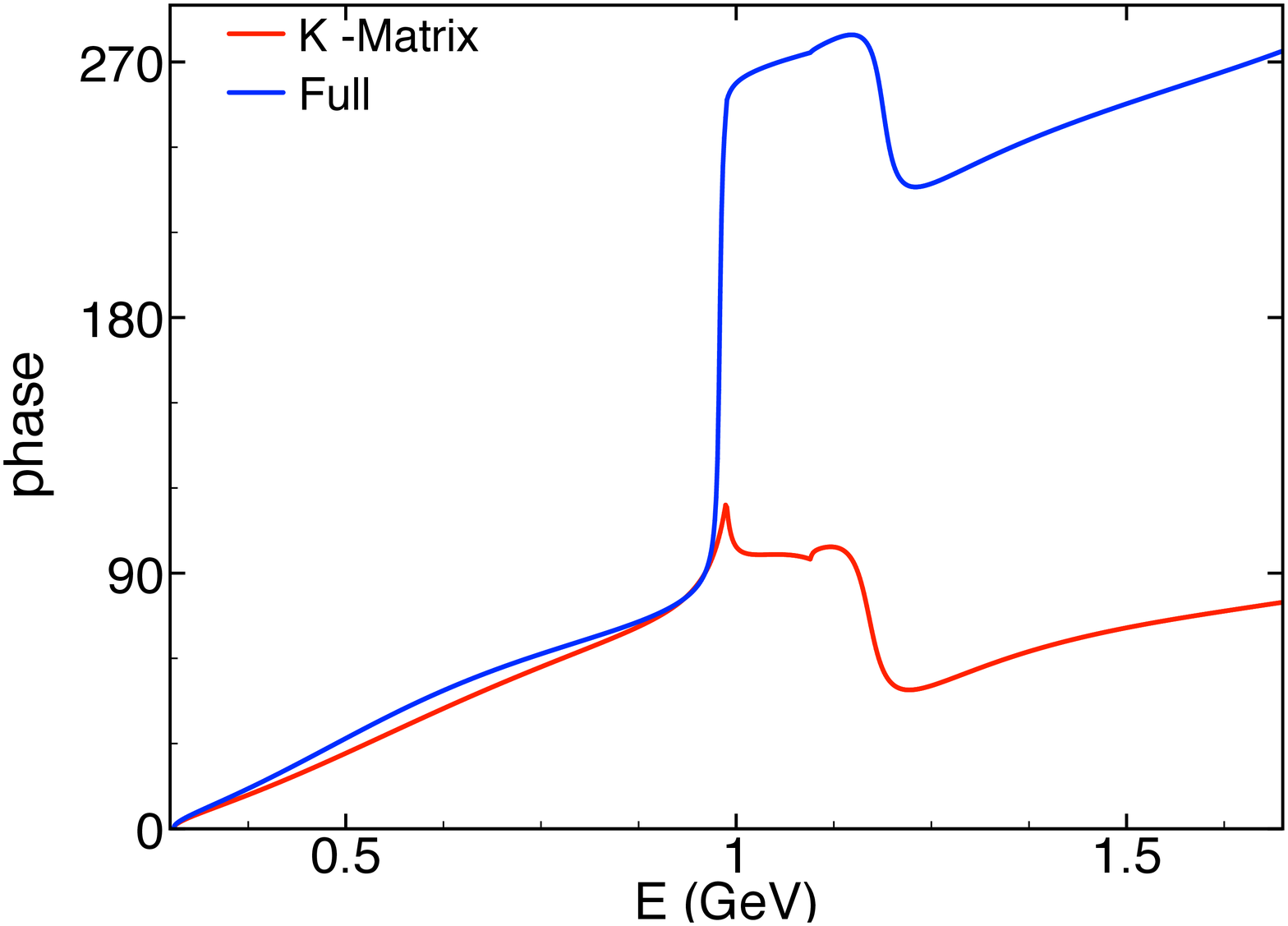}
\includegraphics[width=.45\columnwidth,angle=0]{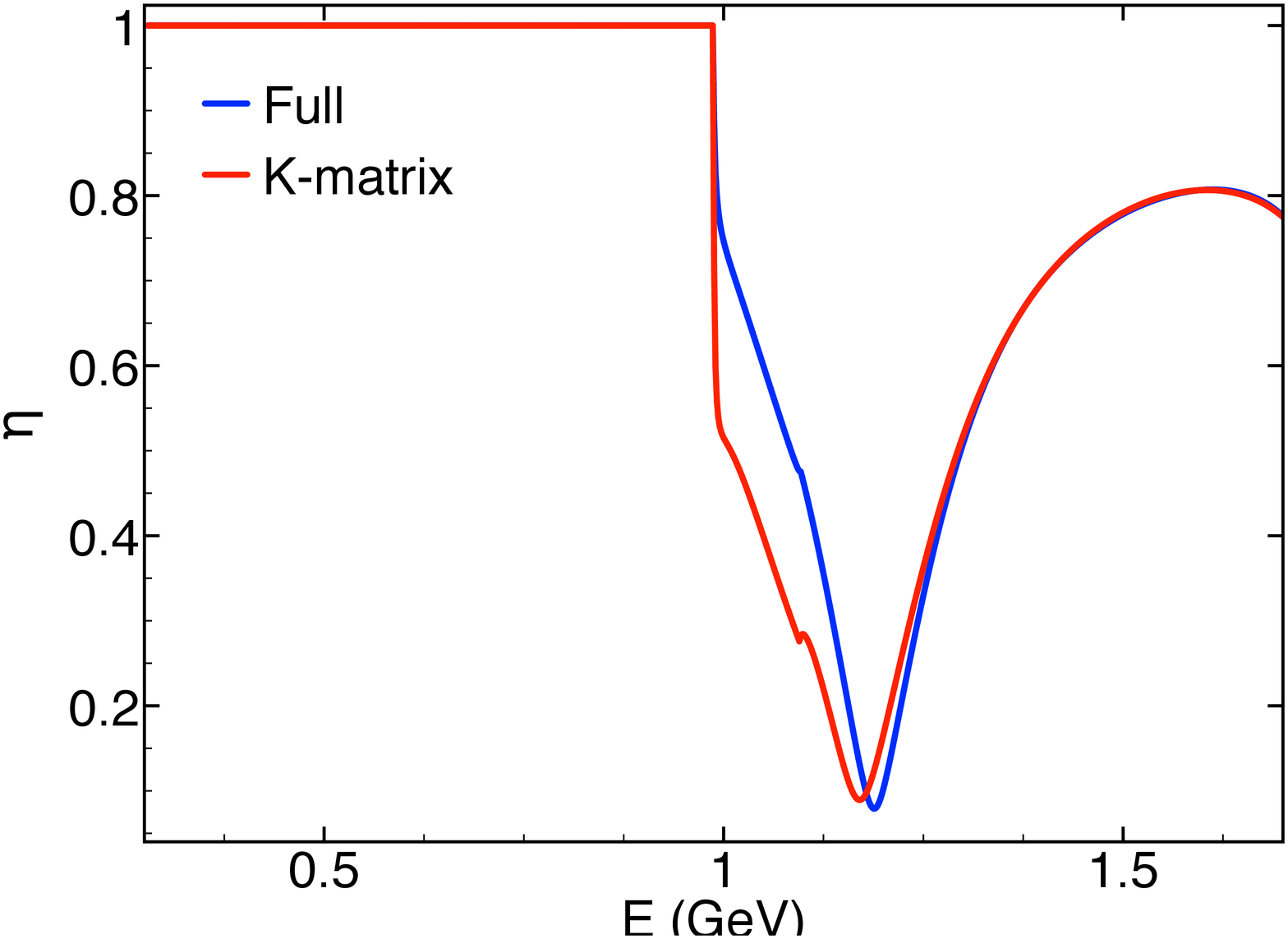}
\caption{  Predictions for phase shifts (left) and inelasticity parameter (right)
of the scalar-isoscalar $\p\p$ amplitude with
two resonances $f_a(1370)$ and $f_b(960)$, with $\e=0$, superimposed 
to a non-resonant background: (blue)  from the model,
Eq.~(\ref{m5})  and (red) the K-matrix.}
\label{Fm-5}
\end{figure}

We compare  predictions from the model and the K-matrix
for the scalar-isoscalar $\p\p$ amplitude
in Fig.~\ref{Fm-2}, for
the case of two resonances $f_a=f(1370)$ and $f_b=f(980)$ 
with the mixing parameter $\epsilon=0$. 
The corresponding phase shift and inelasticity parameter are shown in   Fig.~\ref{Fm-3}.
It is possible to notice that results from the model and K-matrix are qualitatively 
similar over the energy range considered, except for a small region around 1 GeV, 
where effects from the resonance $f_b$  and the opening of the $K\Kb$ channel compete.
This can be seen more clearly in the sharp peak in figure for the phase, whose 
tip occurs at threshold.
For slightly lower energies, the resonance tends to push the phase upwards, whereas
the coupled $K\Kb$ interaction does the opposite afterwards.
In order to explore this picture, we use  a   little
lower mass for the octet resonance, 
namely $f_b=0.96\,$GeV and the results of Figs.~\ref{Fm-4} and \ref{Fm-5} show 
that effects near threshold become much stronger.
The phase for the model, in particular, has a sharp rise around 1 GeV, 
as shown in fig.\ref{Fm-5} and also observed by experimet~\cite{Hyams}, 
but this does not happen for the $K$-matrix.
Another interesting feature of this channel concerns the second resonance $f_a(1370)$.
Inspecting Figs.~\ref{Fm-1}-\ref{Fm-4} around the corresponding energy, 
we do not find structures on either amplitudes or phase shifts and inelasticities.  
As both the $KK$ and $\eta\eta$ channels are already open at the $f_a$ mass, its pole 
occurs in the presence of a background due to a chiral contact term
superimposed to the resonance $f_b$ in which  the mechanism discussed in Sect.\ref{resCOUP} is operating.

\begin{figure}[h!] 
\includegraphics[width=.45\columnwidth,angle=0]{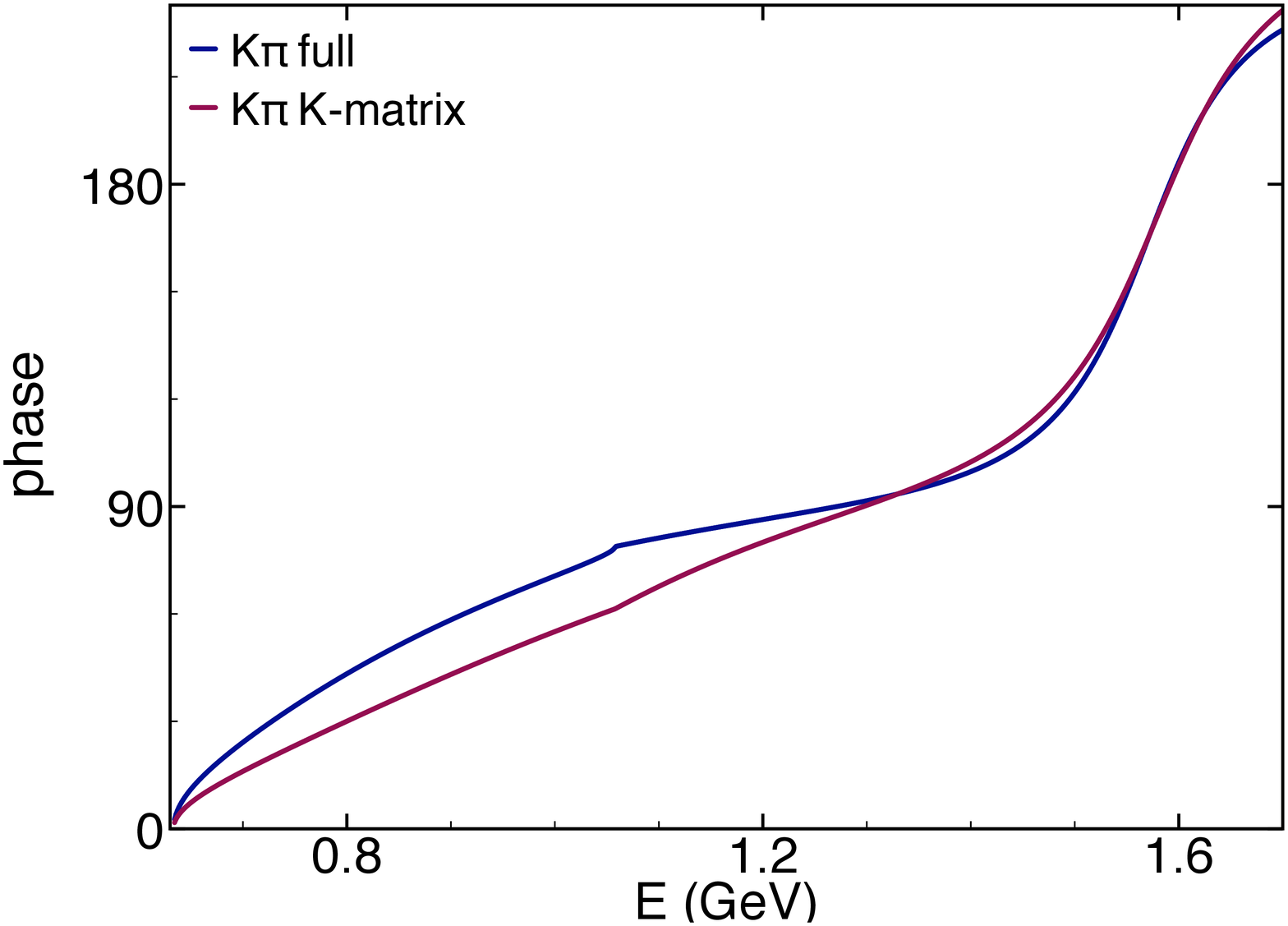}
\includegraphics[width=.45\columnwidth,angle=0]{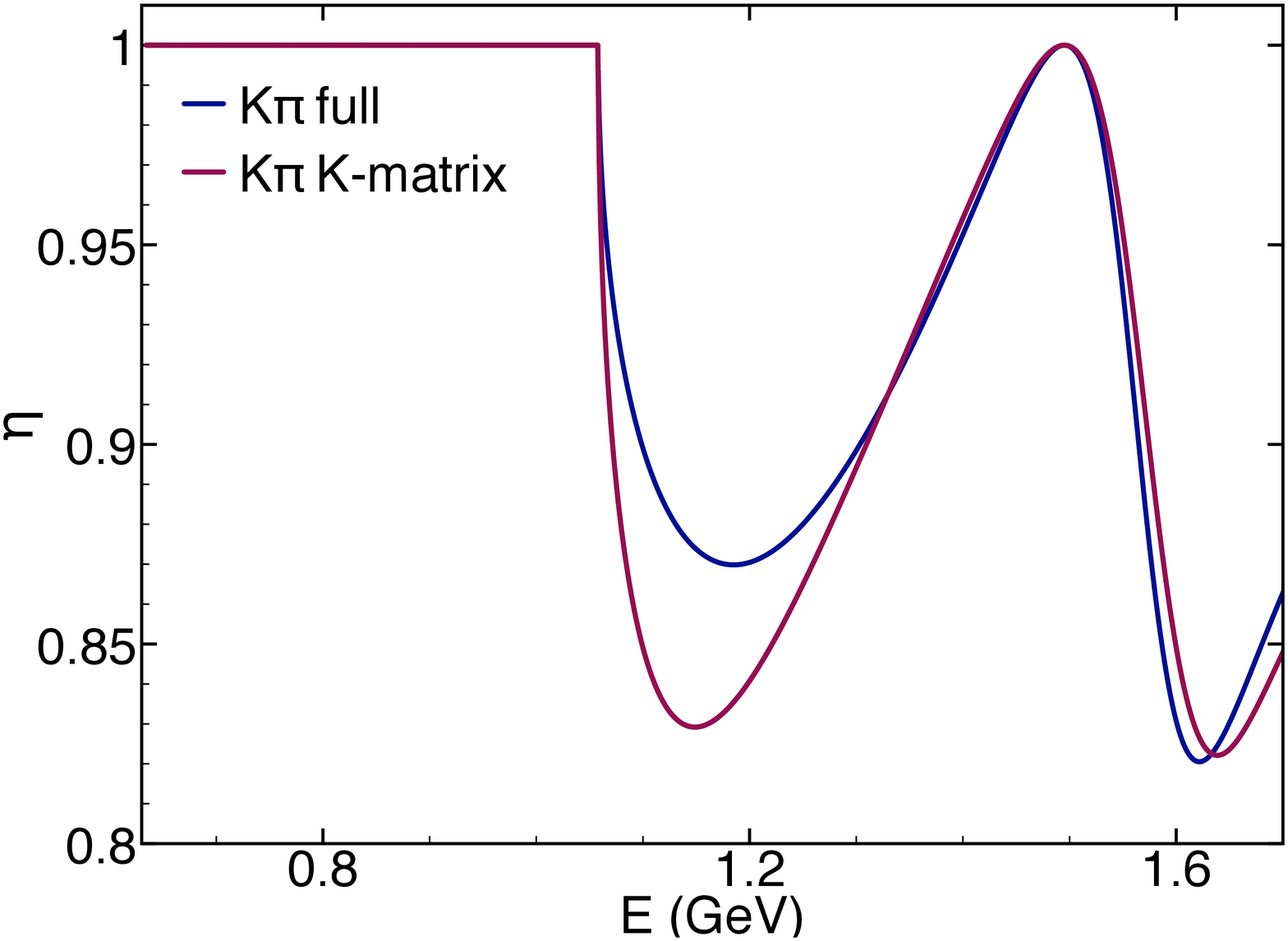}
\\
\caption{ 
Predictions for phase shifts (left) and inelasticity parameters (right)
of  the scalar-isovector $\p K$ amplitude with a resonance $K_0^*$  
with mass $m_{K_0^*}=1.33\,$GeV superimposed  to a 
non-resonant background: (blue)  from the model,
Eq.~(\ref{c.13a}) and (red) the K-matrix.}
\label{Fm-6}
\end{figure}

\begin{figure}[h!] 
\includegraphics[width=.45\columnwidth,angle=0]{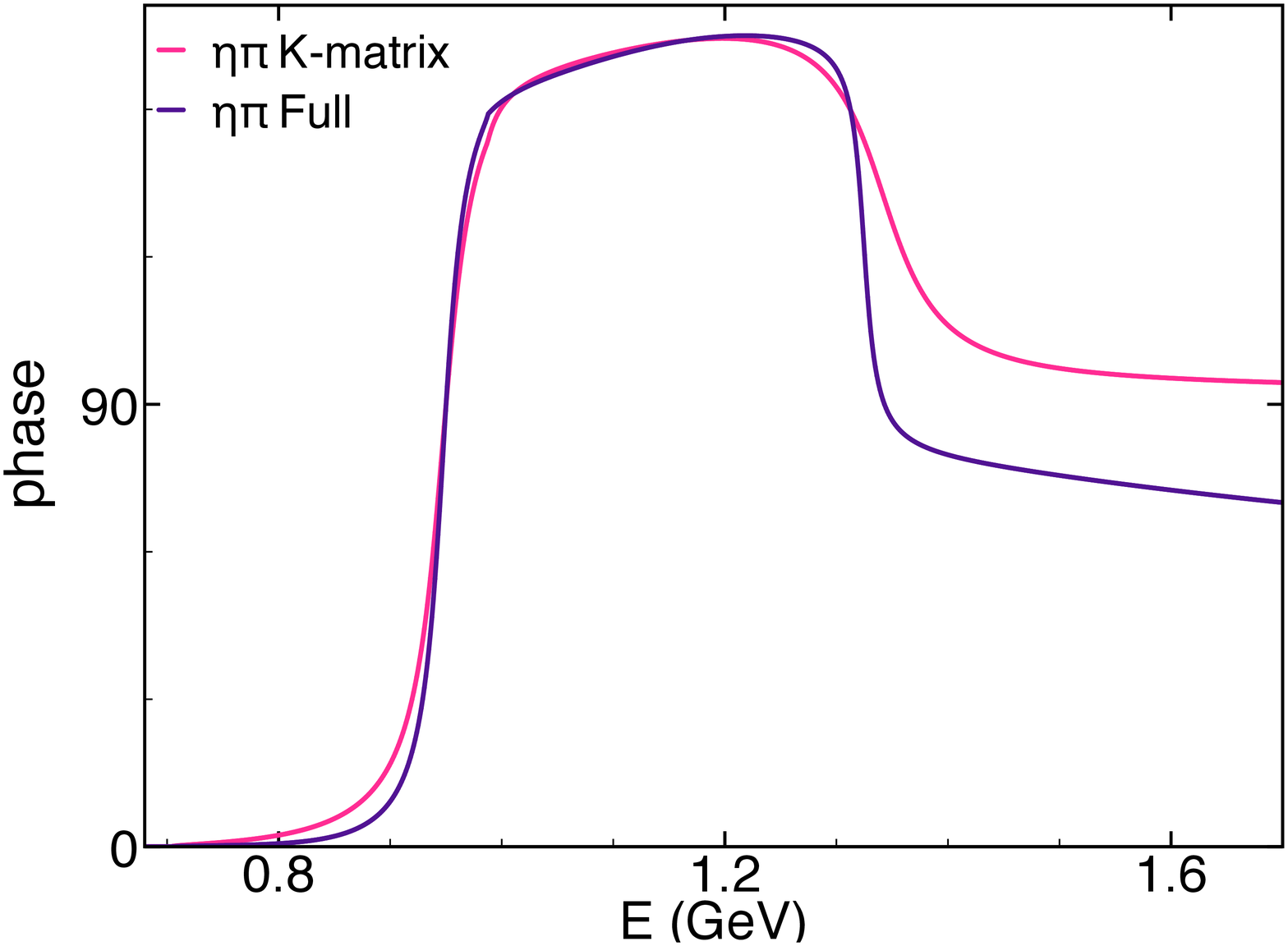}
\includegraphics[width=.45\columnwidth,angle=0]{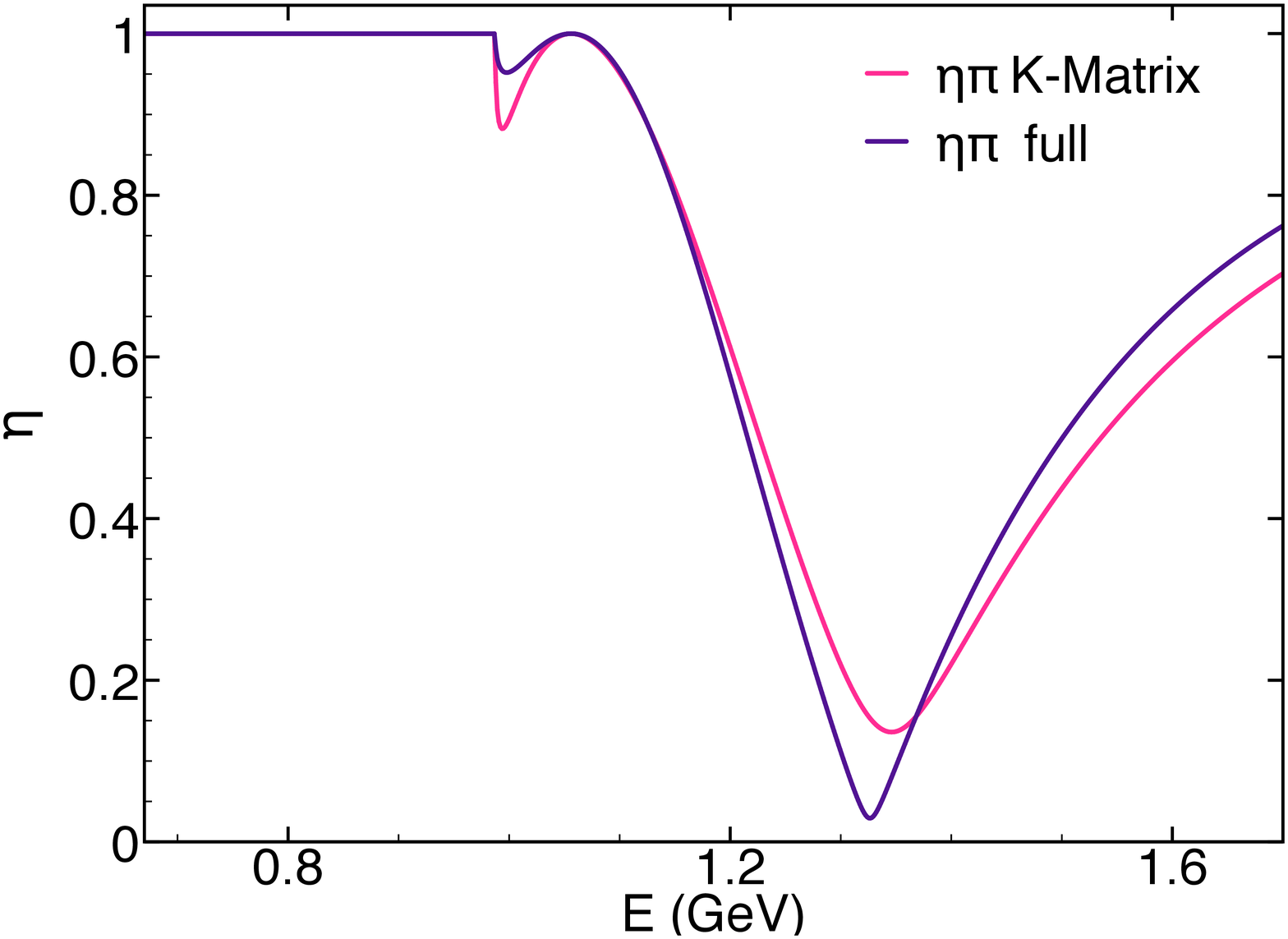}
\\
\caption{ Predictions for phase shifts (left) and inelasticity parameters (right)
of  the scalar-isovector $\pi \eta$ amplitude with
a resonance $a_0$ with mass $m_{a_0}=0.95\,$GeV
superimposed  to a non-resonant background: (blue)  from the model,
Eq.~(\ref{c.12}) and (red) the K-matrix .}
\label{Fm-7}
\end{figure}

For the sake of completeness, in Figs.~\ref{Fm-6} and \ref{Fm-7} we display results 
for phase shifts and inelasticity parameters for scalar $\p K$ and $\p \eta$ 
scatterings,
predicted by Eqs.~(\ref{c.13a}) and (\ref{c.12}).
The $\p K$ process becomes inelastic at the $K \eta$
production threshold and includes a $K_0^*$ with mass $m_{K_0^*}=1.33\,$GeV,
whereas the $\pi \eta$ is coupled to a $K \Kb$ through the $a_0$, with mass $m_{a_0}=0.95\,GeV$. 
Thus, in the $\pi \eta$, the resonance is below threshold and the phase passes through $90^0$
at its mass.
On the other hand, in the $\p K$, the resonance  lies  in the inelastic region and the influence 
of the background in the other channel shows up.
Deviations between the model and the $K$-matrix are noticeable below $1.2$ GeV for the former 
and above that energy for the latter.

While inspecting the results displayed in Figs.~\ref{Fm-2}-\ref{Fm-7}, one should bear
in mind that they rely on the coupling constants precribed in Ref.~\cite{EGPR} and
may change significantly in case other parameters are adopted.

\section{an extra resonance}
 \label{extraR}
The model proposed here allows for the inclusion of any number of resonances. 
In order to illustrate this procedure, we consider the case of an extra resonance $R'$
in each  scalar  channel and begin by resorting to Eq.~(\ref{m6}) in the case
of $\p\p$ scattering and to eq.(\ref{m5}) for $I=1/2$ and $I=0$.
New resonances mean, of course, new masses and coupling constants and, as the number 
of channels is large, one could have, in principle, too many new degrees of freedom to
be fitted by data.
In order to be conservative, we  suggest
 that the same  forms displayed
after the arrows in Eqs.~(\ref{k8})-(\ref{k15}) be  used, with
\bea
\lb \, (c_d \,\mathrm{or} \,\ct_d) \, (s \sm \mathrm{mass}^2) + c_{(R|ab)} \rb \rar 
 \,(c_d \,\mathrm{or} \,\ct_d)\,\lb  \a\,(s \sm \mathrm{mass}^2) + \b_{R'}\,\m^2 \rb\,  \;.
\label{e.1}
\eea
\ni In the case of the $s$-dependent couplings, this preserves the $SU(3)$ structure, 
with a scale given by chiral perturbation theory\cite{EGPR}, $c_d \,= \,0.032\,$GeV and  $\ct_d \,= \, 0.018\,$GeV, whereas $\m=1\,$GeV is just a scale.
These choices allow both  $\a$ and $\b$ to be dimensionless free parameters
and one may guess that their values  will be not far from $-1\leq \a, \b \leq 1$. 

\begin{figure}[h!] 
\includegraphics[width=.45\columnwidth,angle=0]{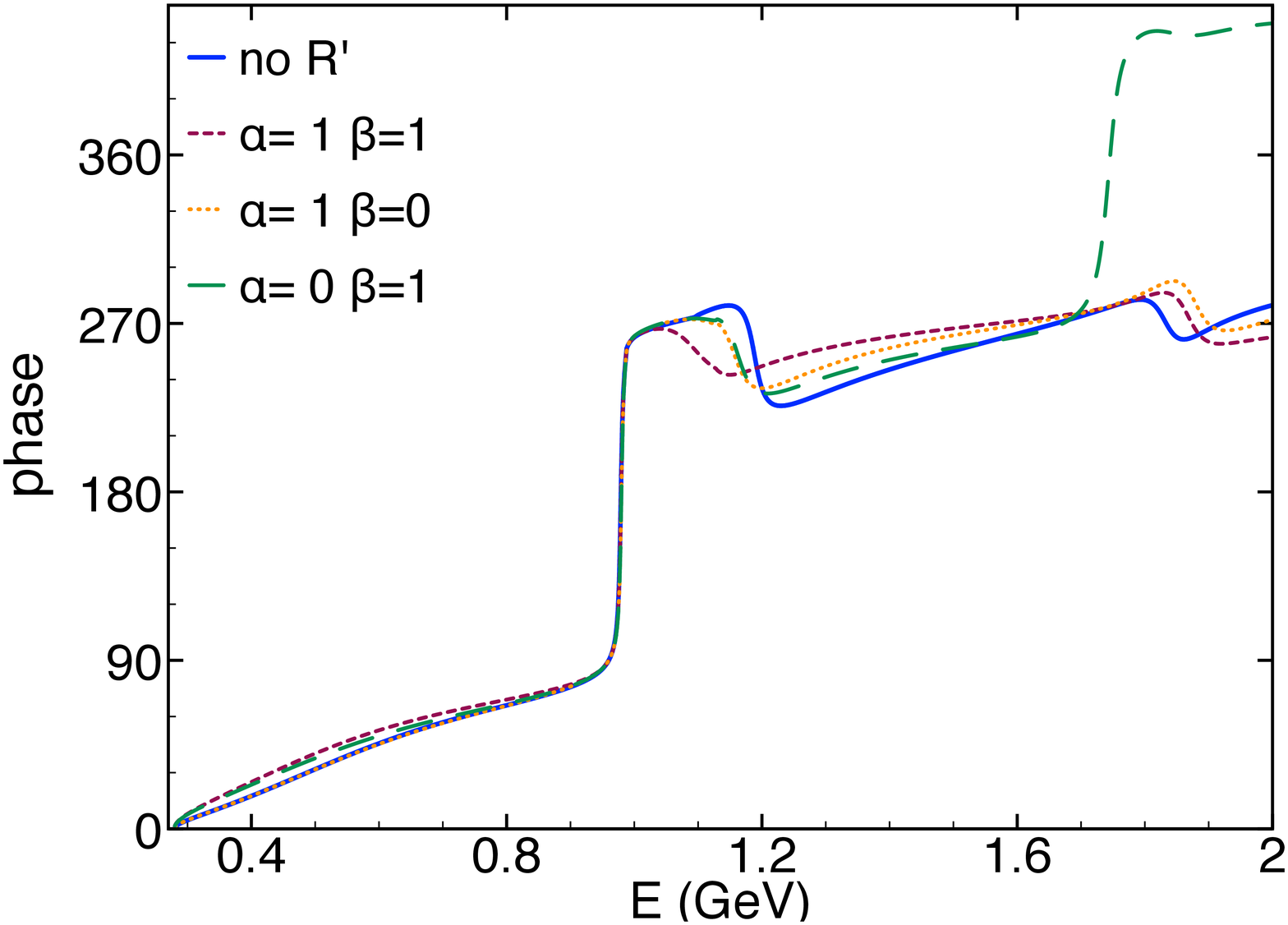}
\includegraphics[width=.45\columnwidth,angle=0]{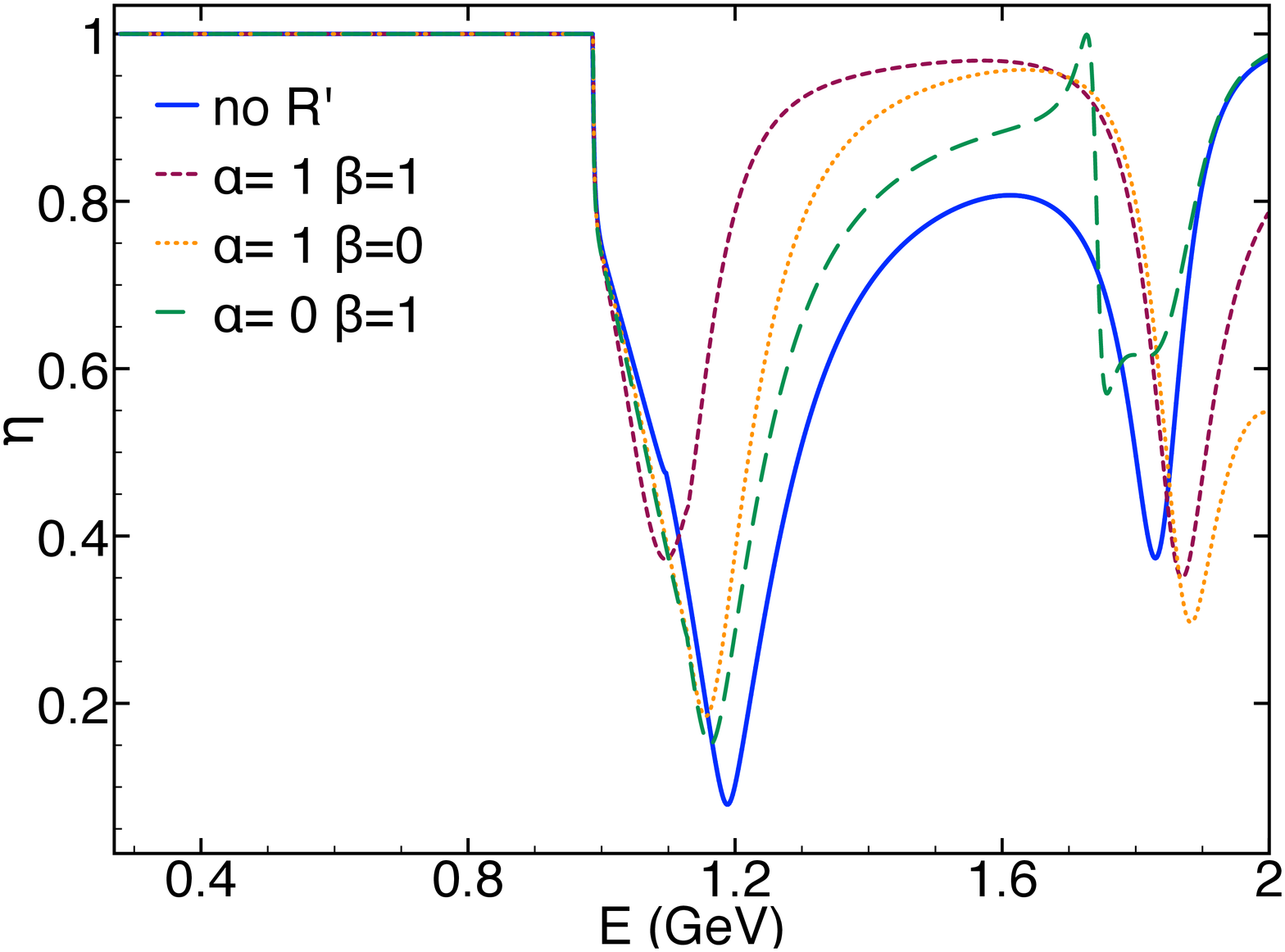}
\caption{ Predictions for phase shifts (left) and inelasticity parameters (right)
 for  the scalar $\pi \pi $ amplitude with
an extra resonance of mass $m_{R'} = m_{f_0}=1.7\,$GeV;
the case \emph{no} $R'$ corresponds to the blue curve of Fig.~\ref{Fm-3}.
}
\label{Fm-8}
\end{figure}

\begin{figure}[h!] 
\includegraphics[width=.45\columnwidth,angle=0]{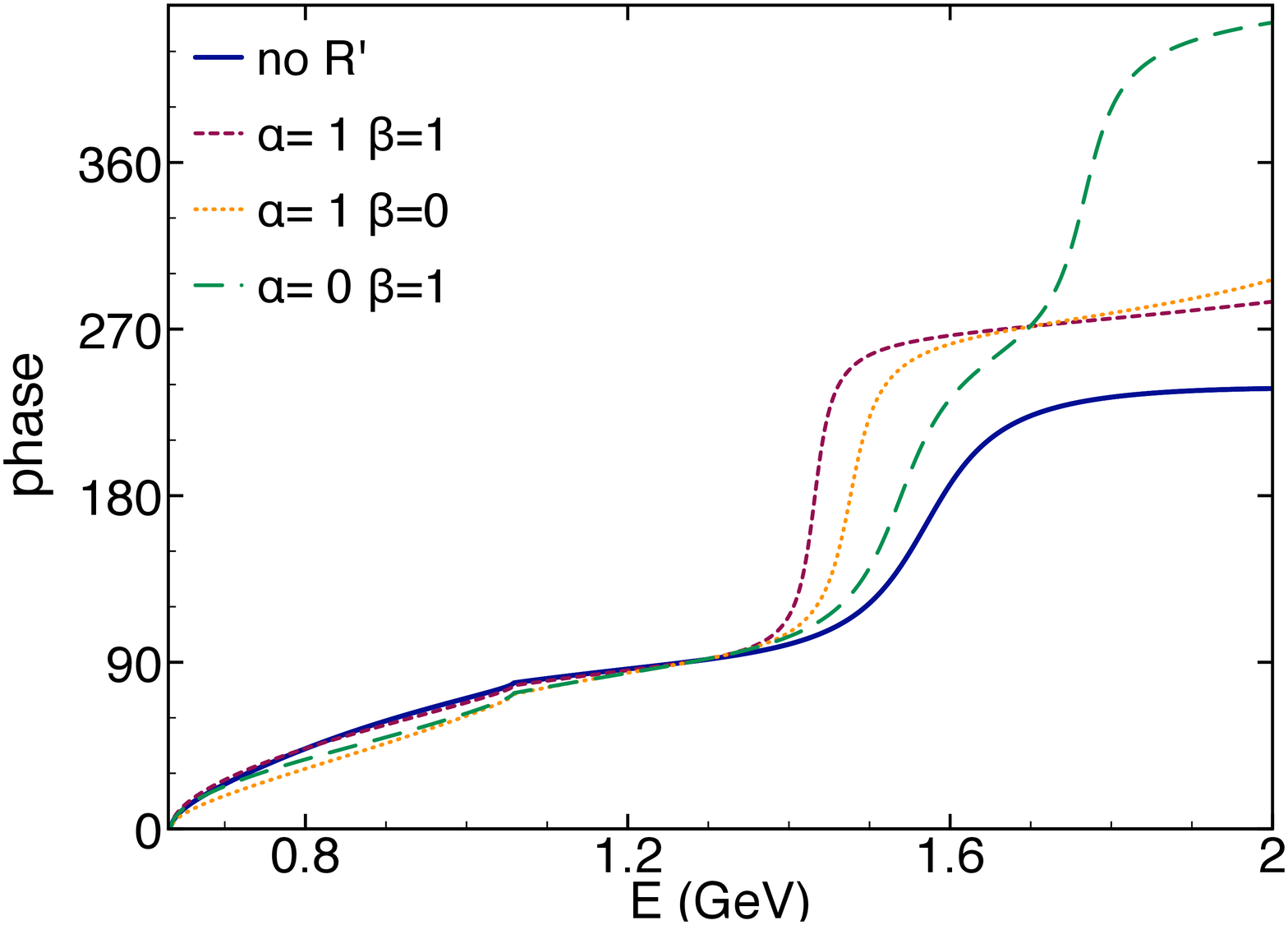}
\includegraphics[width=.45\columnwidth,angle=0]{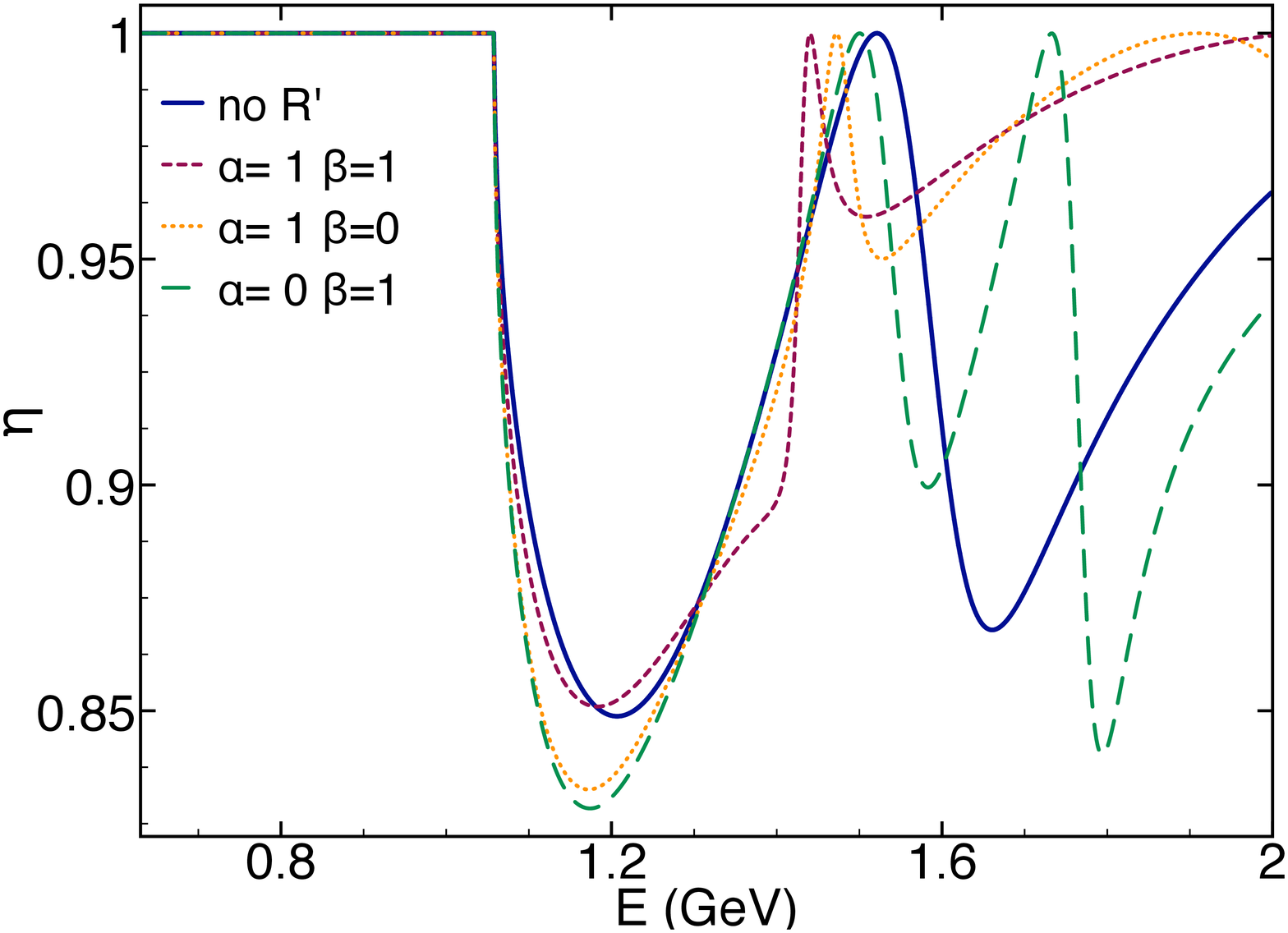}
\\
\caption{ Predictions for phase shifts (left) and inelasticity parameters (right)
for  the scalar-isovector $\pi K$ amplitude with an extra
a resonance of mass $m_{R'} = m_{K_0^*}=1.7\,$GeV;
the case \emph{no} $R'$ corresponds to the dark blue curve of Fig.~\ref{Fm-6}.}
\label{Fm-9}
\end{figure}

\begin{figure}[h!] 
\includegraphics[width=.45\columnwidth,angle=0]{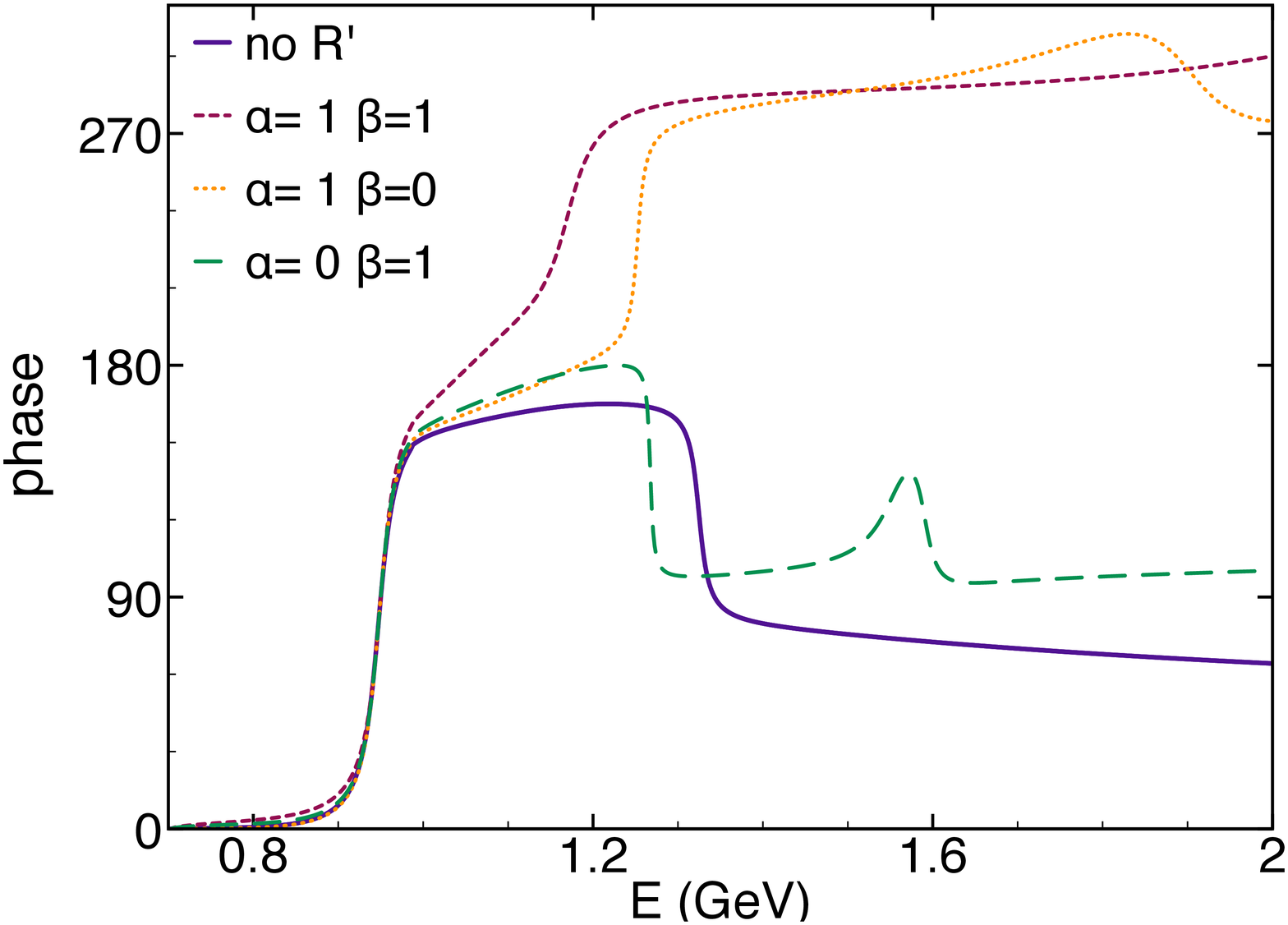}
\includegraphics[width=.45\columnwidth,angle=0]{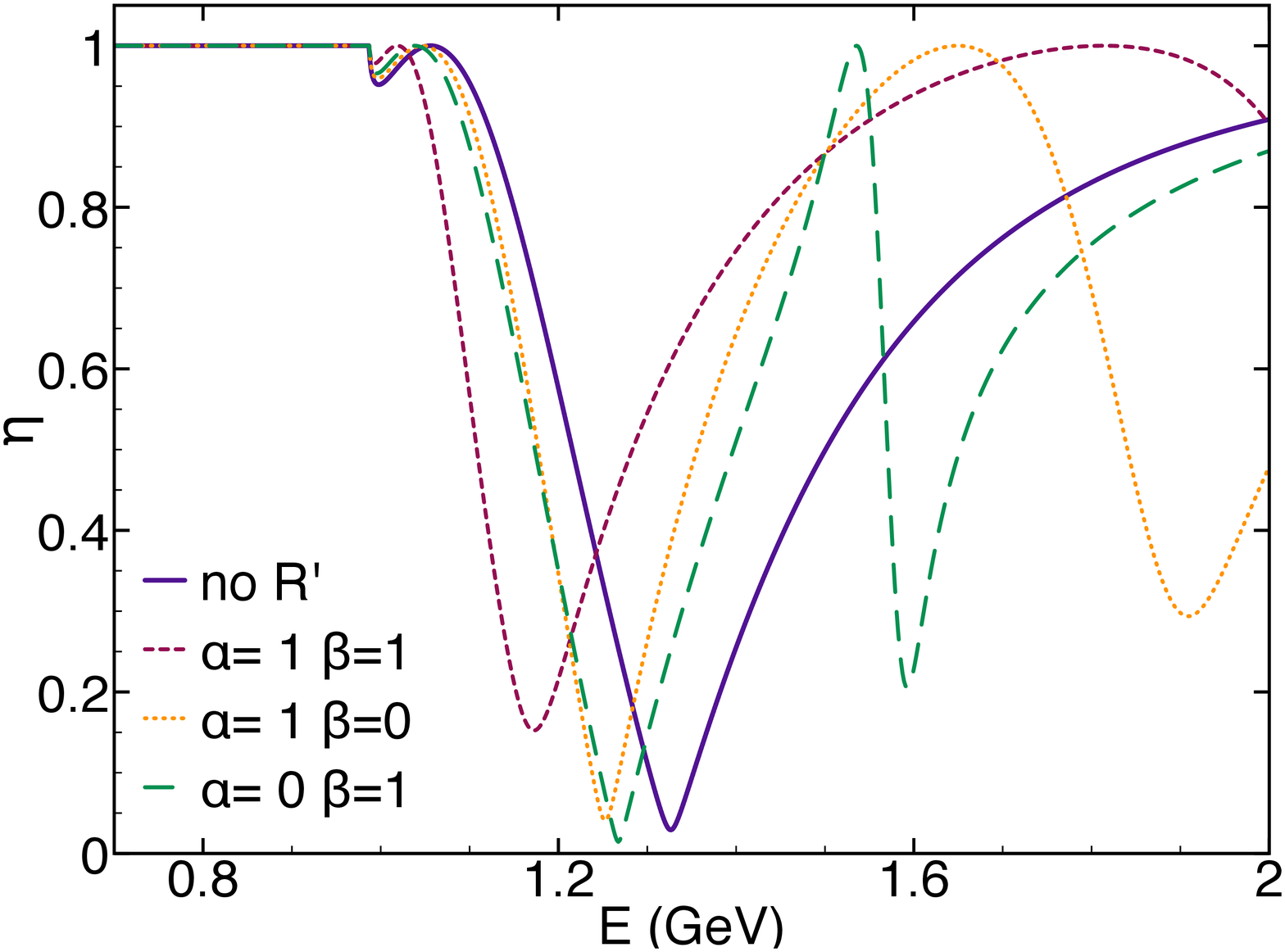}
\\
\caption{ Predictions for phase shifts (left) and inelasticity parameters (right)
 for  the scalar-isovector $\pi \eta$ amplitude with an extra
a resonance of mass $m_{R'} = m_{a_0}=1.5\,$GeV;
the case \emph{no} $R'$ corresponds to the purple curve of Fig.~\ref{Fm-7}.}
\label{Fm-10}
\end{figure}

 As an illustration, in
 Figs.~\ref{Fm-8}-\ref{Fm-10}, we display  phase shifts and inelasticity parameters
for $\p\p$, $\p K$ and $\p \eta$ scatterings including an extra resonance, for a choice
of values of $\a$ and $\b$.
In all cases one notes that results do depend on the the values of $\a$ and $b$ adopted
and also, as expected, that the high energy region of the curves are
more sensitive to the inclusion of the extra resonance. 
In all cases, the extra resonance occurs in the inelastic regime and, 
as discussed in subsection \ref{clopol}, its shape is strongly affected by a background due 
to channel-coupling.

\section{summary and conclusions}
\label{summ}

The standard isobar model (SIM) was produced more than 50 years ago and is still widely used,
in spite of its many limitations.
In the case of heavy-meson decays into three mesons, 
 the model relies on the (2+1) approximation, whereby strong final state interactions involve
just a two-body  interacting system in the presence of a spectator.
The assumption that  meson-meson 
amplitudes are strongly dominated by resonances 
is essential to the model.
We argue that QCD has a strong impact on this picture and that  
the SIM may be reliable for vector mesons in uncoupled channels but 
is not suited to scalar mesons.  
Nowadays a proper description of low-energy meson-meson interactions requires 
contact with chiral perturbation theory,  
which implements 
QCD by means of effective lagrangians.
 Although originally developed for low-energy processes, this theory 
can be reliably extended through the inclusion of resonances and unitarization techniques.  
In Sect. \ref{imUNC} we have shown that the SIM and its post-QCD version give rise 
to rather different predictions for  the scalar $\p\p$ amplitude, 
owing to both dynamics  and unitarity. 
Another problem of the SIM concerns  the coupling of channels.
This effect is compulsory whenever possible and, in subsection \ref{clopol} we 
have shown that resonances cannot be considered as dynamically isolated 
objects beyond coupling thresholds.
This happens because pole dominance in a given channel is contaminated
by background effects occurring elsewhere. 
Therefore,  BW line shapes are unsuited for describing 
resonances in the inelastic regime, as shown in subsection \ref{Kmare}.
 
As an alternative to the versions employed in the SIM
we present, in App.\ref{coupled}, a  set of phenomenological meson-meson 
amplitudes to the $SU(3)$ sector,  which is suitable for amplitude analyses of heavy-meson decays.
Their main features include:
\\
{\bf a. unitarization -}  All amplitudes are automatically unitary for energies
below the first coupling threshold.
\\
{\bf b. coupled channels  -}  The treatment of coupled channels is standard and gives 
rise to the expected inelasticities.
\\
{\bf c. dynamics -}  Interactions are described by chiral lagrangians, which 
include both pure pseudoscalar
 vertices and bare resonances,  with free masses and coupling constants.
This ensures that chiral symmetry is obeyed at low-energies and also give rise 
to fitting parameters with well defined physical meaning.  
\\
{\bf d. model for meson loops -}  Two-meson loops are an important component 
of scattering amplitudes.
In the $s$-channel, they are given by real functions below threshold and acquire 
an imaginary part above it.
The latter is fully determined by  theory whereas the former involve 
unknown renormalization constants.
In Sect.\ref{model} we propose a model for these real parts, which comply with chiral 
symmetry and can accommodate  any number of resonances. 
\\
{\bf e. systematic inclusion of resonances -}  The model can accommodate any number 
of resonances in each given channel.
\\
{\bf f. free parameter have physical meaning -} 
The free parameters of the model are resonance masses and constants describing 
their couplings to pseudoscalar mesons.
Thus,  their conceptual meaning is both rather conventional and process independent,
whereas their  empirical values can be extracted from different reactions.
This allows one to envisage a situation in which one could compare various
sets of values for the {\it same} parameters as determined, for instance, from
chiral perturbation theory, meson-meson scattering up to 2 GeV, 
$D\rar \p\p\p$, $D\rar\p\p K$ and other processes. 
This would definitely promote understanding and, hopefully, much needed progress.
\\

In this constructive approach, all imaginary terms in the amplitudes can be traced 
back to loops, which are also responsible for the finite widths of resonances.
The parameters to be fitted  are just resonance masses and coupling constants, which 
have a rather transparent physical meaning. 
As examples, we have discussed scalar amplitudes, phase shifts and inelasticity parameters
for $\pi \pi $, $\pi K$ and $\pi \eta$ scatterings, employing the low-energy parameters
given in Ref.~\cite{EGPR}. 
In all cases, results from the model for the real parts of the loop functions
were compared with those from the $K$-matrix, where they are absent.
One notices that the main differences occur close to the first inelastic 
 threshold and shows
that the new model provides a clear indication for the 
mechanism responsible for the sharp rise observed in the $\p\p$ phase around
1\, GeV.

\section*{ Acknowledgements} 
We would like to thank Jonas Rademacker and Jos\'e Ramon Pelaez for fruitful discussions. 
PCM work was supported by Marie Currie (MSCA) grant no. 799974.

\newpage
\appendix

\section{two-meson propagators and functions $\Omega$}
\label{omega} 

The conventional expressions presented here are displayed for the sake of completeness 
and rely on results from Ref.~\cite{GL85}.
These integrals do not include symmetry factors, which are accounted for in the main text. 
One deals with both $S$ and $P$ waves and the corresponding two-meson propagators are 
associated with
\bea
&&
\lc I_{ab} ; \,  I_{ab}^{\m \n}  \rc
= \int \frac{d^4  \ell}{(2\p)^4}
\frac{\lc 1; \, \ell^\m \ell^\n \rc} {D_a  D_b },
\label{a.1}\\[2mm]
&& D_a = (\ell \sp p/2)^2 \sm  M_a^2 \;,
\hspace*{6mm} 
D_b = (\ell \sm p/2)^2 \sm  M_b^2 \;,
\label{a.2}
\eea
where $p^2=s$ and both integrals  are evaluated using dimensional regularization  techniques.
The function $I_{ab}$ reads
\bea
I_{ab} &\!=\!&  i\;\frac{1}{16 \p^2} \, \lb \Lambda_{ab} + \Pi_{ab} \rb 
\label{a.3}
\eea
where $\Lambda_{ab}  $ is a function of the renormalization scale $\m $
and the number of dimensions $n\, $, which diverges in the limit $n\rar 4\, $,
whereas $\Pi $ is a regular component, given by
\bea
&& \hspace{-8mm}  s \!<\! (M_a\!-\!M_b)^2  \rar 
\Pi_{ab} = \Pi_{ab}^0 + \frac{\sqrt{\l}}{s}\;
\ln \lb \frac{M_a^2+M_b^2-s + \sqrt{\l}}{2\,M_a\,M_b}\rb 
\label{a.4}\\[4mm]
&& \hspace{-8mm}  (M_a\!-\!M_b)^2 \!<\! s\! < \!(M_a^2\!+\!M_b^2) \rar
\Pi_{ab} =  \Pi_{ab}^0 -\, \frac{\sqrt{- \l}}{s}\;
\tan^{-1} \lb \frac{\sqrt{-\l}} {M_a^2+M_b^2-s}\rb 
\label{a.5}\\[4mm]
&& \hspace{-8mm} (M_a^2\!+\!M_b^2) \!<\! s \! < \! (M_a\!+\!M_b)^2 \rar
\Pi_{ab} =  \Pi_{ab}^0 -\, \frac{\sqrt{- \l}}{s}\;
\lc \tan^{-1} \lb \frac{\sqrt{-\l}}{M_a^2+M_b^2-s}\rb + \p \rc 
\label{a.6}\\[4mm]
&& \hspace{-8mm} s \! > \! (M_a\!+\!M_b)^2 \rar
\Pi_{ab} =  \Pi_{ab}^0  - \,\frac{\sqrt{\l}}{s}\;
\ln \lb \frac{s- M_a^2- M_b^2 + \sqrt{\l}}{2\,M_a\,M_b}\rb
+ \,i\,\p\;\frac{\sqrt{\l}}{s} \;,
\label{a.7}\\[4mm]
&& \Pi_{ab}^0 = 1 + \frac{M_a^2 + M_b^2}{M_a^2 - M_b^2}\;\ln \lb\frac{M_a}{M_b}\rb
- \frac{M_a^2-M_b^2}{s}\;\ln \lb \frac{M_a}{M_b} \rb \;,
\label{a.8}\\[4mm]
&& \l = s^2 - 2\; s\;(M_a^2+M_b^2) + (M_a^2-M_b^2)^2 \;.
\label{a.9}
\eea
For $M_a=M_b$, $\Pi_{aa}^0=2$.
The tensor integral is 
\bea
I_{ab}^{\m\n}  &\!=\!&  i\;\frac{1}{16 \p^2} \, \lc \lb \frac{p^\m \, p^\n}{s} \, \Lambda_{ab}^{pp} 
-  g^{\m\n}\, \Lambda_{ab}^g \rb
+ \lb  \frac{p^\m \, p^\n}{s} -  g^{\m\n}\, \rb \, \frac{\l}{12\, s}  \; \Pi_{ab}  \rc \;,
\label{a.10}
\eea
where $ \Lambda_{ab}^{pp}   $ and $ \Lambda_{ab}^g   $ are divergent quantities.

In the calculation of final state interactions, it is more convenient to use the functions $\O$,
defined from the regular parts of Eqs.~(\ref{a.3}) and (\ref{a.10}) as
\bea
\O_{ab}^S &\! =\! & i\, \lb \mathrm{regular \; part\; of} \, I_{ab} \rb \rar 
\O_{ab}^S = - \, \frac{1}{16\p^2} \, \Pi_{ab} \;,
\label{a.11}\\[4mm]
%
%
\frac{1}{4}\, \lb \frac{p^\m  p^\m }{s} - g^{\m\n}\rb \,\O_{ab}^P 
&\! = \!& i\, \lb \mathrm{regular \; part\; of} \, I_{ab}^{\m\n} \rb \rar 
\O_{ab}^P = -\, \frac{\l}{48\, \p^2\, s} \, \Pi_{ab} \;.
\label{a.12}
\eea 
As indicated in Eqs.~(\ref{a.4}-\ref{a.9}), the functions $\O$ are real below the threshold
at $s_\mathrm{th} = (M_a\!+\!M_b)^2$ and acquire
 an imaginary component above it.
This imaginary part is not affected by infinities and is a well defined prediction of the theory,
needed to implement unitarity.

In the CM the momentum $Q_{ab}$ is given by
\bea 
&& Q_{ab} = \frac{\sqrt{\l}}{2\,\sqrt{s}} 
=\frac{1}{2} \, \sqrt{s - 2\,(M_a^2 + M_b^2) + (M_a^2 - M_b^ 2)^2/s} 
\label{a.15}
\eea 
and the imaginary components read
\bea 
&& [\O_{ab}^S]^I = -\,\frac{1}{8\p} \; \frac{Q_{ab}}{\sqrt{s}} \;
\theta(s \sm (M_a \sp M_b)^2) \;,
\label{a.13}\\[2mm]
&& [\O_{ab}^P]^I = -\,\frac{1}{6\p} \; \frac{Q_{ab}^3}{\sqrt{s}} \;
\theta(s \sm (M_a \sp M_b)^2) \;,
\label{a.14}
\eea 
where $\theta$ is the Heaviside step function.

\section{scattering kernels }
\label{kernel}

We consider scattering amplitudes that
 can  have $SU(3)$ resonances as intermediate states.
They depend on interaction kernels for
 channels with angular momentum $J=1,0 $ and 
isospin $I=1,1/2,0$.
All kernels are written as sums of a leading-order (LO) chiral polynomial and 
next-to-leading-order (NLO) resonance contributions~\cite{EGPR}.
In the resonance sector, we consider the standard $SU(3)$ contributions, supplemented 
by an extra term $R^{(J,I)}$ for each channel, 
with free masses and coupling constants, denoted by a prime.
The usual Mandelstam variables are $s, t, u$ and  the kernels $\cK_{ab\rar cd}^{(J,I)}$
for the process $P_a \, P_b \rar P_c\, P_d$ read
\\[2mm]
{\bf $\bullet$ vector sector - }  In the case $J=1$, kernels are  written without 
a factor $ \lb 2\,t + s - 2(M_a^2\sp M_b^2) + (M_a^2 \sm M_b^2) ^2/s \rb$,
which becomes $(t-u)$ in the case of identical particles and reduces to 
$ [4\, \bQ^2 \cos \theta] $ in the center of mass.
\\
{\bf - isospin  $ \bI$=1} 
\bea
&& \cK_{(\p \p|\p\p)}^{(1,1)} =   \frac{1}{F^2}  - \,  \frac{s\;G_{(\rho|\p\p)}^2}{s-m_\rho^2} 
-\, \frac{s\;G_{(\rho'|\p\p)}^2}{s-m_{\rho'}^2}  \;, 
\label{k1}\\[4mm]
&& \cK_{(\p \p|KK)}^{(1,1)} =  \frac{\rtw}{2\,F^2} - \, \frac{s\; G_{(\rho|\p\p)}\,G_{(\rho|KK)}}{s-m_\rho^2} 
- \, \frac{s\; G_{(\rho'|\p\p)}\,G_{(\rho'|KK)}}{s-m_{\rho'}^2}  \;,
\label{k2}\\[4mm]
&& \cK_{(KK|KK)}^{(1,1)} =   \frac{1}{2\,F^2} -\, \frac{s\, G_{(\rho|KK)}^2}{s-m_\rho^2} 
 -\, \frac{s\, G_{(\rho'|KK)}^2}{s-m_{\rho'}^2}  \;,
\label{k3}\\[4mm]
&& G_{(\rho| \p\p)} = \frac{\rtw\,G_V}{F^2} \;,
\label{k4}\\[4mm]
&& G_{(\rho|KK)} = \frac{G_V}{F^2} \;.
\label{k5}
\eea
In the framework of RChPT, $G_V$ lies in the range $53 \sm 69\,$MeV.
Of special interest is the relationship $G_V=F/\sqrt{2} \simeq 66\,$MeV, 
associated with vector meson dominance~\cite{EGPR}.
\\[2mm]
{\bf - isospin $\bI$=1/2 }
\bea 
&& \cK_{(\p K|\p K)}^{(1,1/2)} =  \frac{3}{8\,F^2}
 - \frac{s\, G_{(K^*|\p K)}^2}{s-m_{K^*}^2} 
  - \frac{s\, G_{(K^{*'}|\p K)}^2}{s-m_{K^{*'}}^2}  \;,
\label{k5a}\\[4mm]
&& \cK_{(\p K|8K)}^{(1,1/2)} = \frac{3}{8\,F^2}
 - \frac{s\, G_{(K^*|\p K)} G_{(K^*|8 K)} }{s-m_{K^*}^2} 
  - \frac{s\, G_{(K^{*'}|\p K)} G_{(K^{*'}|8 K)}}{s-m_{K^{*'}}^2}  \;,
\label{k5b}\\[4mm]
&& \cK_{(K8|K8)}^{(1,1/2)} =  \frac{3}{8\,F^2}
 - \frac{s\, G_{(K^*|8 K)}^2}{s-m_{K^*}^2} 
  - \frac{s\, G_{(K^{*'}|8 K)}^2}{s-m_{K^{*'}}^2}  \;,
\label{k5c}\\[4mm]
&& G_{(K^*|\p K)} = \frac{\sqrt{3}\ G_V}{2\, F^2} \;.
\label{k5d}\\[4mm]
&& G_{(K^*|8K)} =-\, \frac{\sqrt{3}\ G_V}{2\, F^2} \;.
\label{k5e}
\eea
{\bf - isospin $\bI$=0 }
\bea 
&& \cK_{(KK|KK)}^{(1,0)} =  \frac{3}{2\,F^2}
 - \frac{s\, G_{(\phi|KK)}^2}{s-m_\phi^2} 
  - \frac{s\, G_{(\phi'|KK)}^2}{s-m_{\phi'}^2}  \;,
\label{k6}\\[4mm]
&& G_{(\phi|KK)} = \frac{\sqrt{3}\, G_V \, \sin \theta}{F^2} \;.
\label{k7}
\eea
In a previous work~\cite{PatDKKK}, we considered  a dressed $\phi$ propagator,
which accounts for the partial width of the decay $\phi \rar (\rho \p + \p\p\p)$.
This small contribution is  technically involved and here we ignore it for the sake 
of simplicity. 
The partial width for $\phi\rar K \Kb $ yields~\cite{PDG}  $\sin \theta = 0.605$.
\\[2mm]
{\bf $\bullet$ scalar sector} 
\\
Chiral perturbation theory predicts accurately how $SU(3)$ breaking effects,
characterized by pseudoscalar masses, influence low-energy observables.
The couplings of scalar resonances to two pseudoscalars
involve energy dependent factors which conserve $SU(3)$, associated with the constants $c_d$ and $\ct_d$, 
supplemented by symmetry breaking terms, proportional to $c_m$ and $\ct_m$.
In this work we need to extend scattering amplitudes up to energies well beyond the $\rho$-mass,
which is the upper bound for ChPT and, therefore, we keep the $SU(3)$ invariant
parts of scalar-two-pseudoscalar couplings and allow the symmetry breaking
parts to be described by phenomenological parameters $c$.
Below we denote the resonances by $ a_0 \rar (J,I = 0,1)$, 
 $K_0^* \rar (J=0, 1/2)$, 
$So \rar (J,I=0,0)\,$octet, $S1\rar (J, I=0,0)\,$ singlet and
list these couplings
 using the standard RChPT notation~\cite{EGPR} 
before the arrow and our suggested parametrization after it.
\bea
&& G_{(a_0|\p 8)} = \frac{2}{\sqrt{3}\,F^2}  \,
\lb c_d \, (s\sm M_\p^2 \sm M_8^2)  + c_m \, 2M_\p^2 \rb 
\nn\\[2mm]
&& \hspace{30mm} \rar \;\;
\frac{2}{\sqrt{3}\,F^2}  \, \lb c_d \, (s \sm M_\p^2 \sm M_8^2) + c_{(a_0|\p 8)} \rb \;,
\label{k8}\\[4mm]
&& G_{(a_0|KK)} = \frac{\sqrt{2}}{F^2} \,
\lb c_d \, s - (c_d \sm c_m) \, 2M_K^2 \rb \;\; \rar \;\; 
\frac{\sqrt{2}}{F^2} \, \lb c_d \, (s\sm 2M_K^2)  + c_{(a_0|KK)} \rb \;,
\label{k9} \\[4mm]
&& G_{(K_0^*|\p K)} = \frac{\sqrt{3}}{\sqrt{2}\,F^2}  \,
\lb c_d \, s -(c_d\sm c_m)\, (M_\p^2 \sp M_K^2) \rb
\nn\\[2mm]
&& \hspace{30mm} \rar \;\;
\frac{\sqrt{3}}{\sqrt{2}\,F^2}  \, \lb c_d \, (s \sm M_\p^2\sm M_K^2) + c_{(K_0^*|\p 8)} \rb  \;,
\label{k9a}\\[4mm]
&& G_{(K_0^*|K 8)} = -\,  \frac{1}{\sqrt{6}\,F^2}  \,
\lb c_d \, ( s - M_K^2 - M_8^2) +  c_m\,  (-8 M_\p^2 +11  M_K^2 + 3 M_8^2)/3 \rb
\nn\\[2mm]
&& \hspace{30mm} \rar \;\;
- \frac{1}{\sqrt{6}\,F^2}  \, \lb c_d \, (s \sm M_\p^2\sm M_8^2) + c_{(K_0^*|K 8)} \rb  \;,
\label{k9b}\\[4mm]
&& G_{(So|\p\p)} = - \, \frac{\sqrt{2}}{F^2}  \, \lb c_d \, s - (c_d \sm c_m) \, 2M_\p^2 \rb
\;\; \rar \; \; 
- \, \frac{\sqrt{2}}{F^2}  \, \lb c_d \, (s \sm 2M_\p^2)  + c_{(So|\p\p)}  \rb \;,
\label{k10}\\[4mm] 
&& G_{(So|KK)} = \frac{\sqrt{6}}{3\,F^2}  \, \lb c_d \, s - (c_d \sm c_m) \, 2 M_K^2 \rb 
\;\; \rar \; \; 
\frac{\sqrt{6}}{3\,F^2}  \, \lb c_d \, (s\sm 2M_K^2) + c_{(So|KK)} \rb \;,
\label{k11}\\[4mm]
&& G_{(So|88)} = \frac{\sqrt{6}}{3\, F^2}  \,
\lb c_d \, (s\sm 2 M_8^2)  + c_m \, (16M_K^2 \sm 10 M_\p^2)/3\rb
\nn\\[2mm]
&& \hspace{30mm} \rar \;\; 
\frac{\sqrt{6}}{3\, F^2}  \, \lb c_d\, (s\sm 2M_8^2)  + c_{(So|88)} \rb \;,
\label{k12}\\[4mm]
&& G_{(S1| \p\p)} = \frac{2 \sqrt{3}}{F^2} \, \lb \ct_d \, s - (\ct_d \sm \ct_m) \, 2M_\p^2\rb
\;\; \rar \;\; 
\frac{2 \sqrt{3}}{F^2} \, \lb \ct_d \, (s - 2M_\p^2) + c_{(S1|\p\p)} \rb \;,
\label{k13}\\[4mm]
&& G_{(S1|KK)} = \frac{4}{F^2} \,  \lb \ct_d \, s - (\ct_d \sm \ct_m) \, 2M_K^2\rb
\;\; \rar \;\; 
\frac{4}{F^2} \,  \lb \ct_d \, (s - 2M_K^2) + c_{(S1|KK)} \rb \;,
\label{k14}\\[4mm]
&& G_{(S1|88)} = \frac{2}{F^2} \, \lb \ct_d \, s - (\ct_d \sm \ct_m) \, 2M_8^2 \rb 
\;\;  \rar \;\; 
\frac{2}{F^2} \, \lb \ct_d \, (s -2M_8^2) + c_{(S1|88)}  \rb \;.
\label{k15}
\eea
\\
In RChPT~\cite{EGPR}, one has $|c_d|=0.032\,$MeV, $|c_m|=0.042\,$MeV, $|\ct_d|=|c_d|/\sqrt{3}$
and $|\ct_m|=|c_m|/\sqrt{3}$.
\\
{\bf - isospin $\bI$=1 }
\bea 
&& \cK_{(\p 8|\p 8)}^{(0,1)} =   \frac{2 M_\p^2}{3 F^2}   
-\, \frac{G_{(a_0|\p 8)}^2}{s-m_{a_0}^2}
-\, \frac{G_{(a_0'|\p 8)}^2}{s-{m_{a_0'}}^2}\;,
\label{k16}
\\[4mm]
&& \cK_{(\p 8|KK)}^{(0,1)} =  \frac{(3s -4M_K^2)}{\rts\, F^2} 
-\, \frac{G_{(a_0|\p 8)}\, G_{(a_0|KK)}}{s-m_{a_0}^2}
-\, \frac{G_{(a_0'|\p 8)}\, G_{(a_0'|KK)}}{s-{m_{a_0'}}^2} \;,
\label{k17}\\[4mm]
&& \cK_{(KK|KK)}^{(0,1)}  =  \frac{s}{2F^2}   -\frac{G_{(a_0|KK)}^2}{s-m_{a_0}^2}
 -\frac{G_{(a_0'|KK)}^2}{s-{m_{a_0'}}^2} \;.
\label{k18}
\eea
\\
{\bf - isospin $\bI$=1/2 }
\bea 
&& \cK_{(\p K|\p K)}^{(0,1/2)} =   
-\, \frac{1}{8\, F^2} \, \lb 5\,s -2  (M_\p^2\sp M_K^2) + \frac{3(M_\p^2\sp M_K^2)^2}{s}\rb \;,
\nn\\[2mm]
&&  
-\, \frac{G_{(K_0^*|\p K)}^2}{s-m_{K_0^*}^2}
-\, \frac{G_{(K_0^{*'}|\p K)}^2}{s-{m_{K_0^{*'}}}^2} \;.
\label{k18a}\\[4mm]
&& \cK_{(\p K| 8K )}^{(0,1/2)} =   
-\, \frac{1}{24\, F^2} \, \lb 9\,s -16 M_\p^2-  8 M_K^2 + 6 M_8^2
+ \frac{9(M_\p^2\sp M_K^2)^2}{s}\rb \;,
\nn\\[2mm]
&&  
-\, \frac{G_{(K_0^*|\p K)} G_{(K_0^*|8 K )} }{s-m_{K_0^*}^2}
-\, \frac{G_{(K_0^{*'}|\p K)} G_{(K_0^{*'}|8 K)}} {s-{m_{K_0^{*'}}}^2} \;.
\label{k18a}\\[4mm]
&& \cK_{(8K |8K)}^{(0,1/2)} =   
-\, \frac{1}{24\, F^2} \, \lb 9\,s + 4M_\p^2-18 M_K^2 + 3M_8^2
+ \frac{9(M_K^2\sp M_8^2)^2}{s}\rb \;,
\nn\\[2mm]
&&  
-\, \frac{G_{(K_0^*|8 K)}^2}{s-m_{K_0^*}^2}
-\, \frac{G_{(K_0^{*'}|8 K)}^2}{s-{m_{K_0^{*'}}}^2} \;.
\label{k18a}
\eea
\\
{\bf - isospin $\bI$=0  }
\\[2mm]
We allow for the possibility that the two first  observed resonances in this channel,
denoted by $f_a$ and $f_b$, can be mixtures of octet and singlet states $So$ and $S1$.
The mixing angle $\e$ is defined by
\bea
&& | f_a \ra = \cos\e \, | S1 \ra + \sin \e \, |So\ra \;,
\label{k19}
\\
&& | f_b \ra = -\sin\e \, | S1 \ra + \cos \e \, |So\ra \;,
\label{k20} 
\eea
and the kernels read
\bea 
&& \cK_{(\p \p|\p \p)}^{(0,0)}  = \frac{(2 s - M_\p^2)}{ F^2} 
-\, \frac{G_{(f_a| \p\p|\p\p)}}{s-m_{f_a}^2}\,
-\, \frac{G_{(f_b|\p\p|\p\p)}}{s-m_{f_b}^2}\,
- \, \frac{G_{(f'|\p\p)}^2}{s-m_{f'}^2} \;,
\label{k21}\\[4mm]
%
%
%
%
&& \cK_{(\p \p|KK)}^{(0,0)}  =  \frac{\rth\, s}{2 F^2}
-\,\frac{G_{(f_a|\p\p|KK)}}{s-m_{f_a}^2}
-   \frac{G_{(f_b|\p\p|KK)}}{s-m_{f_b}^2} 
- \, \frac{G_{(f'|\p\p)} \, G_{(f'|KK)}}{s-m_{f'}^2} \;,
\label{k22}\\[4mm]
%
%
%
%
&& \cK_{(\p \p|88)}^{(0,0)}  = \frac{\rth\, M_\p^2}{3 F^2}
-\, \frac{G_{(f_a|\p\p|88)}}{s-m_{f_a}^2}
-  \frac{G_{(f_b|\p\p|88)}}{s-m_{f_b}^2}\,
- \frac{G_{(f'|\p\p)} \, G_{(f'|88)}}{s-m_{f'}^2} \;,
\label{k23}\\[4mm]
%
%
%
&& \cK_{(KK|KK)}^{(0,0)}  = \frac{3 s}{2 F^2}
-\, \frac{G_{(f_a|KK|KK)}}{s-m_{f_a}^2} 
-   \frac{G_{(f_b|KK|KK)}}{s-m_{f_b}^2}
- \frac{G_{(f'|KK)}^2}{s-m_{f'}^2}  \;,
\label{k24}\\[4mm]
%
%
%
%
&& \cK_{(KK|88)}^{(0,0)}  =  \frac{(9 s - 8M_K^2)}{6 F^2} 
-\, \frac{G_{(f_a|KK|88)}}{s-m_{f_a}^2}
-  \, \frac{G_{(f_b|KK|88)}}{s-m_{f_b}^2}
- \, \frac{G_{(f'|KK)} \, G_{(f'|88)}}{s-m_{f'}^2} \;,
\label{k25}\\[4mm]
%
%
%
 && \cK_{(88|88)}^{(0,0)}  =  \frac{(-7M_\p^2 +16 M_K^2)}{9 F^2}
 -\, \frac{G_{(f_a|88|88)}}{s-m_{f_a}^2}
-  \, \frac{G_{(f_b|88|88)}}{s-m_{f_b}^2}
- \, \frac{G_{(f'|88)}^2}{s- m_{f'}^2} \;,
\label{k26}
\eea
with
\bea
&& G_{(f_a| \p\p|\p\p)} = \sin^2\!\e \; G_{(So|\p\p)}^2 + \cos^2\!\e \; G_{(S1|\p\p)}^2 \;,
\label{k27}\\[4mm]
&& G_{(f_a|\p\p|KK)} = \sin^2\!\e \; G_{(So|\p\p)}\,G_{(So|KK)} 
+ \cos^2\!\e \; G_{(S1|\p\p)} \, G_{(S1|KK)} \;,
\label{k28}\\[4mm]
&& G_{(f_a|\p\p|88)}= \sin^2\!\e \; G_{(So|\p\p)}\,G_{(So|88)} 
+ \cos^2\!\e \; G_{(S1|\p\p)} \, G_{(S1|88)} \;,
\label{k29}\\[4mm]
&& G_{(f_a|KK|KK)} = \sin^2\!\e \; G_{(So|KK)}^2  + \cos^2\!\e \; G_{(S1|KK)}^2 \;,
\label{k30}\\[4mm] 
&& G_{(f_a|KK|88)} = \sin^2\!\e \; G_{(So|KK)}\,G_{(So|88)} 
+ \cos^2\!\e \; G_{(S1|KK)}^2 \, G_{(S1|88)} \;,
\label{k31}\\[4mm]
&& G_{(f_a|88|88)} = \sin^2\!\e \; G_{(So|88)}^2  + \cos^2\!\e \; G_{(S1|88)}^2 \;,
\label{k32}\\[4mm] 
%
%
&& G_{(f_b| \p\p|\p\p)} = \cos^2\!\e \; G_{(So|\p\p)}^2 + \sin^2\!\e \; G_{(S1|\p\p)}^2 \;,
\label{k33}\\[4mm]
&& G_{(f_b|\p\p|KK)} = \cos^2\!\e \; G_{(So|\p\p)}\,G_{(So|KK)} 
+ \sin^2\!\e \; G_{(S1|\p\p)} \, G_{(S1|KK)} \;,
\label{k34}\\[4mm]
&& G_{(f_b|\p\p|88)} = \cos^2\!\e \; G_{(So|\p\p)}\,G_{(So|88)} 
+ \sin^2\!\e \; G_{(S1|\p\p)} \, G_{(S1|88)} \;,
\label{k35}\\[4mm]
&& G_{(f_b|KK|KK)} = \cos^2\!\e \; G_{(So|KK)}^2  + \sin^2\!\e \; G_{(S1|KK)}^2 \;,
\label{k36}\\[4mm] 
&& G_{(f_b|KK|88)} = \cos^2\!\e \; G_{(So|KK)}\,G_{(So|88)} 
+ \sin^2\!\e \; G_{(S1|KK)} \, G_{(S1|88)} \;,
\label{k37}\\[4mm]
&& G_{(f_b|88|88)} = \cos^2\!\e \; G_{(So|88)}^2  + \sin^2\!\e \; G_{(S1|88)}^2 \;.
\label{k38}
\eea

\section{coupled channel scattering amplitudes}
\label{coupled}

In the discussion of schematic dynamics in the main text, we show that the  scattering 
amplitudes for pseudoscalars 
have the general form given by Eqs.~(\ref{dyn.1})-(\ref{dyn.3}) and reproduced below,
\bea 
&& A =  \cK \times \lb 1 + (\mathrm{loop} \times \cK) + (\mathrm{loop}\times \cK)^2 
+ (\mathrm{loop}\times \cK)^3 + \cdots \rb \,,
\nn\\[2mm]
&& \mathrm{(loop)} = \mathrm{real\,\,part} + i\, \O^I \;,
\nn\\[2mm]
&& \cK= \cK_0 + \cK_1 + \cK_ 2 + \cdots \;,
\nn
\eea
where the functions $(\mathrm{loop})$ involve the $\O$ discused in App.\ref{omega} and the
kernels $\cK_0$ were given in App.\ref{kernel}.
Here we present the scattering amplitudes for the process $P_k \, P_\ell \rar P_a \,P_b$ 
in the coupled channel formalism.
It is important to stress that, although expressed in terms of $\O$ and $\cK$, results displayed 
are quite general and fully independent of the specific forms chosen for these functions. 
They just rely on the well established techniques for dealing with coupled channel problems.

The factor $(\mathrm{loop} \times \cK)$ corresponds to mixing matrix elements
$M^{(J,I)}$, which are given by~\cite{PatDKKK}
\bea
&& M_{11}^{(1,1)} =  - \cK_{(\p\p|\p\p)}^{(1, 1)} \, [ \O_{\p\p}^P /2 ]  \;,
\hspace{10mm}
M_{12}^{(1,1)} =  -  \cK_{(\p\p|KK)}^{(1, 1)}  \, [\O_{KK}^P/2] \;,
\nn \\[2mm]
&& M_{21}^{(1,1)}  = - \cK_{(\p\p|KK)}^{(1, 1)} \, [\O_{\p\p}^P/2] \;,
\hspace{10mm}
M_{22}^{(1,1)} = -  \cK_{(KK|KK)}^{(1, 1)} \, [\O_{KK}^P /2] \;.
\label{c.1}\\[6mm]
&& M_{11}^{(1,1/2)} =  - \cK_{(\p K|\p K)}^{(1, 1/2)} \, [\O_{\p K}^P] \;,
\hspace{10mm} 
M_{12}^{(1,1/2)} =  - \cK_{(\p K |K 8)}^{(1, 1/2)} \, [\O_{K8}^P] \;,
\nn \\[2mm]
&& M_{21}^{(1,1/2)}  = - \cK_{(\p K|K8)}^{(1, 1/2)} \, [\O_{\p K}^P] \;,  
\hspace{10mm}
M_{22}^{(1,1/2)} = - \cK_{(K8|K8)}^{(1, 1/2)} \, [\O_{K8}^P ]\;. 
\label{c.1a}\\[6mm]
&& M^{(1,0)} = - \cK_{(KK|KK)}^{(1, 0)} \, [\O_{KK}^P/2]\;. 
\label{c.2}
\eea
and
\bea
&& M_{11}^{(0,1)} =  - \cK_{(\p 8|\p 8)}^{(0, 1)} \, [\O_{\p 8}^S/2] \;,
\hspace{10mm} 
M_{12}^{(0,1)} =  - \cK_{(\p 8 |KK)}^{(0, 1)} \, [\O_{KK}^S/2] \;,
\nn \\[2mm]
&& M_{21}^{(0,1)}  = - \cK_{(\p 8|KK)}^{(0, 1)} \, [\O_{\p8}^S/2] \;,  
\hspace{10mm}
M_{22}^{(0,1)} = - \cK_{(KK|KK)}^{(0, 1)} \, [\O_{KK}^S/2 ]\;. 
\label{c.3}\\[6mm]
&& M_{11}^{(0,1/2)} =  - \cK_{(\p K|\p K)}^{(1, 1/2)} \, [\O_{\p K}^S] \;,
\hspace{10mm} 
M_{12}^{(0,1/2)} =  - \cK_{(\p K |K 8)}^{(1, 1/2)} \, [\O_{K8}^S] \;,
\nn \\[2mm]
&& M_{21}^{(0,1/2)}  = - \cK_{(\p K|K8)}^{(1, 1/2)} \, [\O_{\p K}^S] \;,  
\hspace{10mm}
M_{22}^{(0,1/2)} = - \cK_{(K8|K8)}^{(1, 1/2)} \, [\O_{K8}^S ]\;. 
\label{c.3a}\\[6mm]
&& M_{11}^{(0,0)} =  - \cK_{(\p\p|\p\p)}^{(0,0)} \, [\O_{\p\p}^S/2] \;, 
\hspace{10mm}
M_{12}^{(0,0)} =  - \cK_{(\p\p|KK)}^{(0,0)} \, [\O_{KK}^S/2] \;,
\nn\\[2mm]
&& M_{13}^{(0,0)} =  - \cK_{(\p\p|88)}^{(0,0)} \, [\O_{88}^S/2] \;,
\hspace{10mm}
M_{21}^{(0,0)} =  - \cK_{(\p\p|KK)}^{(0,0)} \, [\O_{\p\p}^S/2] \;,
\nn\\[2mm]
&& M_{22}^{(0,0)} =  - \cK_{(KK|KK)}^{(0,0)} \, [\O_{KK}^S/2] \;,
\hspace{10mm}
M_{23} ^{(0,0)}=  - \cK_{(KK|88)}^{(0,0)} \, [\O_{88}^S/2] \;,
\nn \\[2mm]
&& M_{31}^{(0,0)} =  - \cK_{(\p\p|88)}^{(0,0)} \, [\O_{\p\p}^S/2] \;,
\hspace{10mm}
M_{32}^{(0,0)} =  - \cK_{(KK|88)}^{(0,0)} \, [\O_{KK}^S/2] \;,
\nn \\[2mm]
&& M_{33}^{(0,0)} =  - \cK_{(88|88)}^{(0,0)} \, [\O_{88}^S/2] \;.
\label{c.4}
\eea
The factor $1/2$ accounts for the symmetry of intermediate states.
It is also present in the functions $M_{11}^{(0,1)}$ and $M_{21}^{(0,1)}$
because we use symmetrized $\pi 8$ intermediate states.

As shown in Eqs.~(\ref{dyn.4}) and (\ref{dyn.5}), the summation of the geometric series
 yields scattering amplitudes based on denominators given schematically by 
$ D = 1- (\mathrm{loop} \times \cK)$.
\bea
&& D^{(1,1)}= 
\lb 1\sm M_{11}^{(1,1)} \rb \,  \lb 1\sm M_{22}^{(1,1)} \rb - M_{12}^{(1,1)}  M_{21}^{(1,1)} \;,
\label{c.5}\\[4mm]
&& D^{(1,1/2)} = 
\lb 1\sm M_{11}^{(1,1/2)} \rb \,  \lb 1\sm M_{22}^{(1,1/2)} \rb - M_{12}^{(1,1/2)}  M_{21}^{(1,1/2)}  \,,
\label{c.5a} \\[4mm] 
&& D^{(0,1)} = 
\lb 1\sm M_{11}^{(0,1)} \rb \,  \lb 1\sm M_{22}^{(0,1)} \rb - M_{12}^{(0,1)}  M_{21}^{(0,1)}  \,,
\label{c.6} \\[4mm] 
&& D^{(0,1/2)} = 
\lb 1\sm M_{11}^{(0,1/2)} \rb \,  \lb 1\sm M_{22}^{(0,1/2)} \rb - M_{12}^{(0,1/2)}  M_{21}^{(0,1/2)}  \,, 
 \label{c.6a}\\[4mm]
&& D^{(0,0)} =
\lb 1\sm M_{11}^{(0,0)}\rb \, \lb 1\sm M_{22}^{(0,0)} \rb \, \lb 1 \sm M_{33}^{(0,0)} \rb 
-  \lb 1 \sm M_{11}^{(0,0)} \rb \,   M_{23}^{(0,0)} \,  M_{32}^{(0,0)}  
\nn\\[2mm]
&& \hspace{10mm} 
- \lb 1 \sm M_{22}^{(0,0)} \rb \,  M_{13}^{(0,0)} \,  M_{31}^{(0,0)}  
- \lb 1 \sm M_{33}^{(0,0)}\rb \,  M_{12}^{(0,0)} \,  M_{21}^{(0,0)} 
\nn\\[2mm]
&& \hspace{10mm}
-  \, M_{12}^{(0,0)} \, M_{23}^{(0,0)} \, M_{31}^{(0,0)}  
- M_{21}^{(0,0)} \,  M_{13}^{(0,0)} \, M_{32}^{(0,0)}  \;.
\label{c.8}
\eea
The scattering amplitudes for the process $P_k \, P_\ell \rar P_a \,P_b$ in the various channels are given by
\\[2mm]
{\bf $\bullet$ vector sector - } 
\\
{\bf - isospin  $ \bI$=1} 
\bea
&& A_{(\p\p|ab)}^{(1,1)} = \frac{1}{D^{(1,1)}}
\lc \lb 1 \sm M_{22}^{(1,1)} \rb \, \cK_{(\p\p|ab)}^{(1,1)} + M_{12}^{(1,1)} \, \cK_{(KK|ab)}^{(1,1)}  \rc \,
(t-u) \;,
\label{c.9}\\[4mm]
&& A_{(KK|ab)}^{(1,1)} = \frac{1}{D^{(1,1)}}
\lc M_{21}^{(1,1)} \, \cK_{(\p\p|ab)}^{(1,1)} + \lb 1 \sm M_{11}^{(1,1)}\rb  \, \cK_{(KK|ab)}^{(1,1)}  \rc 
\, (t-u)\;.
\label{c.10}
\eea
\\
{\bf - isospin $\bI$=1/2 }
\bea
&& A_{(\p K|ab)}^{(1,1/2)} = \frac{1}{D^{(1,1/2)}}\; 
\lc \lb 1 \sm M_{22}^{(1,1/2)} \rb \, \cK_{(\p K|ab)}^{(1,1/2)} + M_{12}^{(1,1/2)} \, \cK_{(K8|ab)}^{(1,1/2)}  \rc 
\nn\\[2mm]
&& \times
 \lb 2\,t + s - 2(M_\p^2\sp M_K^2) + \frac{(M_\p^2 \sm M_K^2) ^2}{s} \rb \;.
\label{c.10a}\\[4mm]
&& A_{( K 8|ab)}^{(1,1/2)} = \frac{1}{D^{(1,1/2)}}\; 
\lc  M_{21}^{(1,1/2)}  \, \cK_{(\p K|ab)}^{(1,1/2)} + \lb 1 \sm  M_{11}^{(1,1/2)}\rb \,\cK_{(K8|ab)}^{(1,1/2)}  \rc 
\nn\\[2mm]
&& \times
 \lb 2\,t + s - 2(M_\p^2\sp M_K^2) + \frac{(M_\p^2 \sm M_K^2) ^2}{s} \rb \;.
 \label{c.10b}
\eea
\\
{\bf - isopin  $ \bI=$0 }
\bea
&& A_{(KK|ab)}^{(1,0)} = \frac{1}{D^{(1,0)}}\; \cK_{(KK|ab)}^{(1,0)} \, (t-u) \;.
\label{c.11}
\eea
\\
{\bf $\bullet$ scalar sector - } 
\\
{\bf - isospin  $ \bI$=1} 
\bea
&& A_{(\p 8|ab)}^{(0,1)} = \frac{1}{D^{(0,1)}}
\lc  \lb 1 \sm M_{22}^{(0,1)} \rb  \, \cK_{(\p 8|ab)}^{(0,1)} + M_{12}^{(0,1)} \, \cK_{(KK|ab)}^{(0,1)} \rc \;,
\label{c.12}\\[4mm]
&& A_{(KK|ab)}^{(0,1)} = \frac{1}{D^{(0,1)}}
\lc M_{21}^{(0,1)} \,  \cK_{(\p 8|ab)}^{(0,1)} +  \lb 1 \sm M_{11}^{(0,1)} \rb \, \cK_{(KK|ab)}^{(0,1)}  \rc \;. 
\label{c.13}
\eea
\\
{\bf - isospin $\bI$=1/2 }
\bea
&& A_{(\p K|ab)}^{(0,1/2)} = \frac{1}{D^{(0,1/2)}}\; 
\lc \lb 1 \sm M_{22}^{(0,1/2)} \rb \, \cK_{(\p K|ab)}^{(0,1/2)} 
+ M_{12}^{(1,1/2)} \, \cK_{(K8|ab)}^{(0,1/2)}  \rc \;.
\label{c.13a}\\[4mm]
&& A_{( K 8|ab)}^{(0,1/2)} = \frac{1}{D^{(0,1/2)}}\; 
\lc  M_{21}^{(0,1/2)}  \, \cK_{(\p K|ab)}^{(0,1/2)} 
+ \lb 1 \sm  M_{11}^{(0,1/2)}\rb \,\cK_{(K8|ab)}^{(0,1/2)}  \rc  \;.
 \label{c.13b}
\eea
\\
{\bf - isopin  $ \bI=$0 }
\bea
A_{(\p\p|ab)}^{(0,0)}  &\!=\!& \frac{1}{D^{(0,0)}}
\lc \lb \lp 1\sm M_{22}^{(0,0)}\rp \lp 1\sm M_{33}^{(0,0)}\rp \sm M_{23}^{(0,0)} \,  M_{32}^{(0,0)} \rb\, 
\cK_{(\p\p|ab)}^{(0,0)}
\right.
\nn\\[2mm]
&\! + \!& \left. \lb M_{12}^{(0,0)} \, \lp 1\sm M_{33}^{(0,0)} \rp \sp M_{13}^{(0,0)} \,  M_{32}^{(0,0)} \rb  \, 
\cK_{(KK|ab)}^{(0,0)}
\right.
\nn\\[2mm]
&\! + \!& \left.
\lb M_{13}^{(0,0)} \, \lp 1\sm M_{22}^{(0,0)} \rp \sp M_{12}^{(0,0)}\,  M_{23}^{(0,0)} \rb \, 
\cK_{(88|ab)}^{(0,0)} \rc  \;,
\label{c.14}\\[4mm]
A_{(KK|ab)}^{(0,0)} &\!=\!& \frac{1}{D^{(0,0)}}
\lc \lb M_{21}^{(0,0)} \, \lp 1\sm M_{33}^{(0,0)} \rp \sp M_{23}^{(0,0)}\, M_{31}^{(0,0)} \rb \, 
\cK_{(\p\p|ab)}^{(0,0)}
\right.
\nn\\[2mm]
&\!+\!& \left.  
\lb \lp 1\sm M_{11}^{(0,0)}\rp \, \lp 1\sm M_{33}^{(0,0)}\rp  \sm M_{13}^{(0,0)} \,  M_{31}^{(0,0)} \rp \, 
\cK_{(KK|ab)}^{(0,0)} 
\right.
\nn\\[2mm]
&\! + \!& \left. 
\lb  M_{23}^{(0,0)} \, \lp 1\sm M_{11}^{(0,0)} \rp  \sp M_{13}^{(0,0)} \,  M_{21}^{(0,0)} \rb \, 
\cK_{(88|ab)}^{(0,0)} \rc 
\label{c.15}\\[4mm] 
A_{(88|ab)}^{(0,0)} &\!=\!& \frac{1}{D^{(0,0)}}
\lc \lb M_{31}^{(0,0)} \, \lp 1\sm M_{22}^{(0,0)} \rp  \sp M_{21}^{(0,0)}\, M_{32}^{(0,0)} \rb \, 
\cK_{(\p\p|ab)}^{(0,0)}
\right.
\nn\\[2mm]
&\!+\!& \left.  
\lb  M_{32}^{(0,0)} \, \lp 1\sm M_{11}^{(0,0)} \rp \sp M_{12}^{(0,0)}\, M_{31}^{(0,0)} \rb \, 
\cK_{(KK|ab)}^{(0,0)}
\right.
\nn\\[2mm]
&\! +\! & \left. 
\lb \lp 1\sm M_{11}^{(0,0)}\rp \, \lb 1\sm M_{22}^{(0,0)} \rp  \sm M_{12}^{(0,0)} \,  M_{21}^{(0,0)} \rb  \, 
\cK_{(88|ab)}^{(0,0)} \rc\,.
\label{c.16}  
\eea

\section{$\p \p$ phase shifts}
\label{nonrelat}

Most examples discussed in the main text refer to $\p \p$ scattering 
and the partial wave expansion of the amplitude for isospin channel $I$ reads
\bea
A_{(\p\p|\p\p)}^I = \frac{32\p}{\rho} \, \sum_{J=0}^\infty \, (2J+1) \, P_J(\cos\theta) \, 
f_{(\p\p|\p\p)}^{(J,I)}(s) \;,
\label{ps.1}
\eea 
where $ f_{(\p\p|\p\p)}^{(J,I)}$ is the non-relativistic scattering amplitude and 
$ \rho = \sqrt{(s- 4\, M_\p^2)/s} $ .
Our amplitudes are written as 
\bea
A_{(\p\p|\p\p)}^I = A_{(\p\p|\p\p)}^{(0,I)} + A_{(\p\p|\p\p)}^{(1, I)} + \cdots 
\label{ps.2}
\eea 
In the CM, one has $ (t-u) = (s-4\,M_\p^2) \, \cos\theta $ and
\bea
A_{(\p\p|\p\p)}^I &\!=\!& A_{(\p\p|\p\p)}^{(0,I)} + [(s-4\,M_\p^2) \, \cos\theta] \, A_{(\p\p|\p\p)}^{(1, I)} + \cdots 
\nn\\[4mm]
&\!=\!&  \frac{32\p}{\rho} \, \lb f_{(\p\p|\p\p)}^{(0,I)}(s) + 3\, \cos\theta \, f_{(\p\p|\p\p)}^{(1,I)}(s)  + \cdots \rb 
\label{ps.3}
\eea
with
\bea
&& f_{(\p\p|\p\p)}^{(0,0)} = \frac{\sqrt{s-4M_\p^2}}{32\,\p\,\sqrt{s}}  \; A_{(\p\p|\p\p)}^{(0,0)} \;,
 \label{ps.4}\\[4mm]
&& f_{(\p\p|\p\p)}^{(1,1)} = \frac{(s-4M_\p^2)^{3/2}}{96\,\p\, \sqrt{s}} \;   A_{(\p\p|\p\p)}^{(1,1)} \;.
\label{ps.5}
\eea
From now on, the formalism independs of $(J,I)$, one drops all subscripts and superscripts 
and expresses the amplitude $f$ in terms of phase shifts $\d$ and inelasticity parameters 
$\eta$ as~\cite{Hyams}
\bea 
f = \frac{1}{2i} \lb \, \eta  \, e^{2\, i \, \d } -1 \rb \;.
\label{ps.6}
\eea
In order to obtain $\d$ and $\eta$ from the $A_{\p\p|\p\p}^{(J,I)}$,
one writes $ f = a + i\, b $,   with $ a =  \mathrm{Re}\lb f \rb , \, b =  \mathrm{Im}\lb f \rb $
and Eq.~(\ref{ps.6}) yields
\bea 
1 + 2\,i\, f &\!=\!& [1 -2\, b] + 2\, i\, a 
= \eta \, \lb \cos 2\d + i\, \sin 2\d \rb \;.
\label{ps.7} 
\eea 
Thus
\bea
&& \eta = \sqrt{ [1\sm 2\, b]^2 + 4\, a^2} \;,
\label{ps.8}\\[4mm]
&& \d = \tan^{-1} \lb \frac{2\,a}{1+\eta-2\,b} \rb \;.
\label{ps.9}
\eea
The alternative form 
\bea
&&\d = \frac{1}{2} \, \tan^{-1} \lb \frac{\sin 2\d}{\cos 2\d} \rb 
\label{ps.10}\\[2mm]
&& \sin 2\d = \frac{2a}{\eta}\;, \hspace{10mm} \cos 2\d = \frac{1-2\,b}{\eta} 
\label{ps.11}
\eea
is more convenient in numerical calculations because, as $\eta >0$, 
the signs of $\sin 2\d $ and $\cos2\d$ in Eq.~(\ref{ps.11}) are well defined
and the quadrant assignment of $2\d$ is unambiguous.
This  yields continuous results in the interval $0\leq \d \leq \p$.

\newpage

\end{document}